\documentclass[%
reprint,
superscriptaddress,
nobibnotes,
amsmath,amssymb,
aps,longbibliography,
prl,
floatfix,
]{revtex4-2}

\usepackage[T1]{fontenc}
\usepackage[utf8]{inputenc}
\usepackage[english]{babel}
\makeatletter
\@ifundefined{selectlanguage}{
}{
  \expandafter\newlanguage\csname l@en\endcsname %
}
\makeatother
\usepackage{graphicx}%
\usepackage{verbatim}
\usepackage[version=4]{mhchem}
\usepackage{xcolor}
\usepackage{physics}
\usepackage{comment}
\usepackage[normalem]{ulem}
\usepackage{pdfpages}
\usepackage{pgffor}
\usepackage{booktabs}
\usepackage{makecell}
\usepackage{float}

\usepackage{multirow}%
\usepackage{longtable}
\usepackage{pdflscape}

\newcommand{\rev}[1]{#1}

\usepackage{etoolbox}                   %
\AtBeginEnvironment{listing}{%
  \raggedright
  \captionsetup{justification=raggedright,singlelinecheck=false}%
}

\usepackage[normalem]{ulem}
\definecolor{tangerine}{rgb}{0.944,0.522,0}
\definecolor{verde}{rgb}{0.,0.6,0}
\definecolor{rosso}{rgb}{0.9,0.0,0.2}
\definecolor{orange}{rgb}{1.0,0.5,0.0}

\newif\ifhighlight

\newcommand{\highlight}{\highlighttrue}
\highlight
\newcommand{\editor}[2]{%
  \expandafter\newcommand\csname #1note\endcsname[1]{%
    \textcolor{#2}{(\textbf{#1note:} \textsc{##1})}}%
  \expandafter\newcommand\csname #1\endcsname[1]{%
    \ifhighlight\textcolor{#2}{##1} \else ##1\fi}%
  \expandafter\newcommand\csname #1cancel\endcsname[1]{%
    \ifhighlight\textcolor{#2}{\sout{##1}}\fi}%
  \expandafter\newcommand\csname #1change\endcsname[2]{%
    \ifhighlight\textcolor{#2}{\sout{##1} ##2}\else ##2\fi}%
  \newenvironment{#1text}{\ifhighlight\color{#2}\fi}{\color{black}}
}

\editor{MC}{red}
\editor{MK}{orange}
\newcommand{\grayout}[1]{}

\newcommand{\shiftmlthree}{ShiftML3}
\newcommand{\shiftmlnew}{ShiftML4}%
\newcommand{\shiftmlnewshort}{sML4}
\newcommand{\shiftmlthreeshort}{sML3}

\makeatletter
\AtBeginDocument{\let\LS@rot\@undefined}
\makeatother

\begin{document}

\title{Machine-Learned NMR Shieldings in Molecular Solids with Built-In Hybrid-Functional Molecular Corrections}

\author{Matthias Kellner}
\affiliation{Laboratory of Computational Science and Modeling, Institut des Mat\'eriaux, \'Ecole Polytechnique F\'ed\'erale de Lausanne, 1015 Lausanne, Switzerland}
\author{Ruben Rodriguez-Madrid}
\affiliation{Laboratory of Magnetic Resonance, Institut des Sciences et Ing\'enierie Chimiques, \'Ecole Polytechnique F\'ed\'erale de Lausanne, 1015 Lausanne, Switzerland}
\author{Jacob B. Holmes}
\affiliation{Laboratory of Magnetic Resonance, Institut des Sciences et Ing\'enierie Chimiques, \'Ecole Polytechnique F\'ed\'erale de Lausanne, 1015 Lausanne, Switzerland}
\author{Pablo~A.~Unzueta}
\affiliation{Department of Chemistry, University of California, Riverside, Riverside, CA 92521, USA}
\author{Gregory J. O. Beran}
\affiliation{Department of Chemistry, University of California, Riverside, Riverside, CA 92521, USA}
\author{Lyndon Emsley}
\email{lyndon.emsley@epfl.ch}
\affiliation{Laboratory of Magnetic Resonance, Institut des Sciences et Ing\'enierie Chimiques, \'Ecole Polytechnique F\'ed\'erale de Lausanne, 1015 Lausanne, Switzerland}
\author{Michele Ceriotti}
\email{michele.ceriotti@epfl.ch}
\affiliation{Laboratory of Computational Science and Modeling, Institut des Mat\'eriaux, \'Ecole Polytechnique F\'ed\'erale de Lausanne, 1015 Lausanne, Switzerland}

\newcommand{\CSM}[1]{{\color{red}#1}}

\date{\today}%

\begin{abstract}
\textbf{Abstract}: 
Fast and accurate chemical shielding estimators are essential for shielding-driven Nuclear Magnetic Resonance (NMR) crystallography. Machine-learning models for shielding predictions have matured significantly and today are primarily limited by the electronic structure reference data they are trained on. 
Here, we introduce ShiftML4, a shielding-tensor model trained directly on monomer-corrected calculations that approximate PBE0, rather than the PBE reference targeted by earlier ShiftML models. ShiftML4 is trained on a diverse set of structures containing 12 of the most common NMR nuclei in molecular organic solids. On experimental benchmark sets, the $^{13}$C isotropic RMSE against experiment is 1.67 ppm, compared with 2.34 ppm for GIPAW-PBE on the same geometries. ShiftML4 gives a similar $^{1}$H prediction RMSE to ShiftML3 (0.5 ppm) and improves the $^{15}$N RMSE from 7.24 to 6.08~ppm. The model also reduces errors in the shielding-tensor anisotropy, with an RMSE of 4.63~ppm on $^{13}$C CSA principal components against 5.85~ppm for GIPAW. The improvements in prediction accuracy are retained on better geometries. Basing shift predictions on structures relaxed with PET-MOLS, a recent machine-learned interatomic potential that reaches approximate hybrid-DFT geometries in seconds, lowers the ShiftML4 errors further to \rev{0.48 ppm ($^1$H),} 1.49~ppm~($^{13}$C) and 3.66~ppm~($^{15}$N).
\end{abstract}

\maketitle

Chemical-shift-driven NMR crystallography determines structure by enumerating candidate geometries and selecting those whose predicted shifts best match experiment.~\cite{emsleySpiersMemorialLecture2025b} Chemical shifts are particularly informative for determining the structure of both crystalline and disordered materials, because the shift of an environment is sensitive to local geometric changes and does not depend strongly on long-range ordering of the lattice.
The resolution of shift-driven NMR crystallography and confidence in the structure determination critically depend on the accuracy of the predicted shifts.~\cite{hofstetterPositionalVarianceNMR2017, engel_bayesian_2019,muellerUniformChisquaredModel2025} 
 For this reason, shifts are usually computed with ab-initio electronic structure calculations at the  DFT level of theory.~\cite{pickardAllelectronMagneticResponse2001b,mauriInitioTheoryNMR1996,bonhommeFirstPrinciplesCalculationNMR2012}
NMR crystallography has determined, or helped determine, structures across a wide range of compositions, including crystalline organic molecular solids,~\cite{salagerPowderCrystallographyCombined2010, baiasNovoDeterminationCrystal2013, baiasPowderCrystallographyPharmaceutical2013,  senkerMicroscopicDescriptionPolyamorphic2005, yatesCombinedFirstPrinciples2005, oshaughnessyPolarTriptyceneBasedNonmetal2025,pindelskaSolidStateNMREffective2015,szeleszczukApplicationCombinedSolidstate2019,daiSolidState1H13And172020,widdifieldFurosemidesOneLittle2016} inorganic materials,~\cite{ ashbrookApplicationNMRCrystallography2020, brouwerGeneralProtocolDetermining2013, 
leeStructureDisorderAmorphous2010, grasCrystallineAmorphousCalcium2016,
moranHuntingHydrogenRandom2016} zeolites,~\cite{brouwerSolidStateNMRMethod2005, brouwerNMRCrystallographyZeolites2008} ceramics,~\cite{romaoZeroThermalExpansion2015, moranEnsembleBasedModelingNMR2019} amorphous drugs,~\cite{ cordovaAtomiclevelStructureDetermination2023b, guestEssentialSynergyMD2025,torodiiDeterminationKeyFunctional2025, torodiiThreeDimensionalAtomicLevelStructure2025} enzymatic sites,~\cite{laiXrayNMRCrystallography2011, holmesImagingActiveSite2022a} cementitious materials,~\cite{rejmak29SiNMRCement2012, walkleySolidstateNuclearMagnetic2019, kunhimohamedAtomicLevelStructureCementitious2020,morales-melgares_atomic-level_2022} and hybrid-perovskites.~\cite{hope_nanoscale_2021, kubickiNMRSpectroscopyProbes2021} The extent of this list of successful structure-determination campaigns motivates continued development of the accompanying computational tools used for chemical shift prediction.

Shielding calculations must also scale to the large models needed to study disordered or amorphous systems. Periodic-DFT GIPAW calculations are ill-suited to study such systems since their computational cost scales cubically with system size. GIPAW is also limited in accuracy, because standard implementations rely on GGA functionals, which have proved insufficient for confident structure selection in challenging polymorphic systems.~\cite{hartmanEnhancedNMRDiscrimination2016} In this letter we address both limitations with \shiftmlnew{}, a transferable machine-learning surrogate for chemical shielding tensors in organic crystals with built-in hybrid-DFT molecular corrections. 

Over recent years numerous Machine Learning (ML) models have been developed to predict chemical shifts directly from geometry,~\cite{ unzuetaReview2025} with materials of interest ranging from organic crystals~\cite{paru+18ncomm, liuMultiresolution3DDenseNetChemical2019,xuUnifiedBenchmarkFramework2025a} to inorganic materials.~\cite{charpentierFirstprinciplesNMROxide2025, cunyInitioQualityNMR2016, venetosMachineLearningFull2023, benmahmoudGraphneuralnetworkPredictionsSolidstate2025, bornesAccurateTensorialModel2026} In a similar spirit, a series of ML models have been developed for other relevant quantities in the nuclear spectroscopies, including zero-field splitting tensors~\cite{zaverkinThermallyAveragedMagnetic2022c} and electric field gradients.~\cite{f.harperTrackingLiAtoms2025a, hussSimulatingQuadrupolarNMR2026} Other ML models have instead aimed to correct low-fidelity DFT calculations, as a post-hoc correction step.~\cite{unzueta_predicting_2021, kleinebuningComputationCCSDTQualityNMR2023} 
For crystalline materials there are no experimental databases comparable to those for proteins or solvated small molecules, so ML models for shielding predictions in condensed matter are typically trained on ab-initio reference calculations instead.
Focusing specifically on transferable models for organic solids, Paruzzo et al. established the ShiftML family of models, trained on a wide corpus of reference GIPAW-DFT calculations.~\cite{paru+18ncomm} The family has developed steadily: from the initial model which was only parametrized to predict isotropic shifts for crystals containing up to four elements, to models equipped with uncertainty estimators and additional coverage of sulfur-containing crystals,~\cite{engel_bayesian_2019} to the inclusion of thermally distorted structures and the extension of ShiftML to 12 elements, ShiftML2,~\cite{cordova_machine_2022} and ShiftML3~\cite{kellnerDeepLearningModel2025b} which predicts shielding tensors directly.

Against experimental benchmark data, the latest iteration \shiftmlthree{} is almost as accurate as its GIPAW-DFT training reference. Recent benchmark studies also indicate that \shiftmlthree{} can be used as a drop-in replacement for many NMR-crystallographic tasks, whether it is used for isotropic chemical shielding  predictions,~\cite{gunagaAccessibleHybridDFTquality} or for chemical shielding anisotropy.~\cite{czernekAssessingReliabilityShiftML32026} The fast evaluation speed of ShiftML also enables the study of systems that require extensive sampling of configurational space, with repeated evaluation of chemical shieldings for an ensemble of structures, either because of thermal and quantum nuclear fluctuations or because of intrinsic structural disorder. This is particularly relevant for computing the shifts of acidic or labile protons, which are inherently ensemble averages, as shown in solids by our recent quantum-nuclei-corrected NMR (QNC-NMR) protocol,~\cite{kellnerQuantumcorrectedNMRCrystallography2026} amorphous structure determination from MD trajectories,~\cite{cordovaAtomiclevelStructureDetermination2023b} and in solution by combining \shiftmlthree{} with machine-learned interatomic potential (MLIP) molecular dynamics.~\cite{sochaQuantitativePredictionExchangeable2026}

The accuracy of ShiftML is now mostly limited by the reference data itself rather than by the model. Improving the electronic-structure reference to a higher level of theory, such as hybrid DFT, is essential to improve prediction accuracies of ShiftML against experiment. However, to the best of our knowledge, periodic hybrid-DFT shielding calculations are not routinely available for molecular solids of the size considered here, which makes hybrid-quality training sets hard to generate.
More practical schemes, which apply chemical-shift computations from molecular DFT codes up to the coupled-cluster level of theory~\cite{gaussCoupledclusterCalculationsNuclear1995} on organic crystals, have therefore focused on fragment based approaches,~\cite{hartmanEnhancedNMRDiscrimination2016, hartmanFragmentbased13CNuclear2015} or on computing the shifts of molecular clusters with point-charge embeddings of the surrounding molecules.~\cite{dittmerComputationNMRShielding2020} 
Other shielding computation schemes focus instead on improving periodic shielding computations by computing correction terms on model structures that have been cut from the bulk. Early efforts by Nakajima and coworkers towards correcting chemical shielding tensor calculations in thymine and inorganic compounds relied on QM/MM ONIOM-type~\cite{humbelIMOMOMethodIntegration1996, dapprichNewONIOMImplementation1999} calculations, cutting fragments from the bulk and capping cut bonds with hydrogen atoms.~\cite{nakajimaExtrapolationSchemeSolidstate2017}
Later, Dračínský and co-workers refined these molecular corrections, by separating entire molecules from the bulk, which reduced shift prediction errors by 50\% and 32\% on $^{13}$C and $^{15}$N experimental benchmarks with respect to reference periodic PBE calculations.~\cite{dracinskyImprovingAccuracySolidstate2019a} In this approach, the hybrid-level crystalline shieldings $\sigma^{\textrm{PBE0}}_{\textrm{cryst}}$ are approximated as:

\begin{equation}
    \hat{\sigma}^{\textrm{PBE0}}_{\textrm{cryst}} \approx \hat{\sigma}^{\text{GIPAW,PBE}}_{\text{cryst}} + (\hat{\sigma}^{\text{GIAO,PBE0}}_{\text{mol}} - \hat{\sigma}^{\text{GIAO,PBE}}_{\text{mol}})
\label{eq:correction}
\end{equation}

where $\sigma^{\text{GIPAW,PBE}}_{\text{cryst}}$ is the shielding tensor of an environment computed by GIPAW-PBE periodic reference calculations. $\sigma^{\text{GIAO,PBE}}_{\text{mol}}$ and $\sigma^{\text{GIAO,PBE0}}_{\text{mol}}$ are the corresponding tensors for the same site in the isolated molecule cut from the crystal, computed with a molecular DFT code using the GGA functional matching GIPAW and the PBE0 hybrid, with the GIAO approach~\cite{ditchfieldSelfconsistentPerturbationTheory1974}, respectively.

Molecular correction schemes were later evaluated on various experimental benchmarks using both the monomer correction in the gas phase,~\cite{dracinskyAccuratePredictionsProton2020, chaloupeckaNMRCrystallographyAmino2024} and using implicit PCM solvation corrections~\cite{hartmanImprovingAccuracyGIPAW2022, iuliucciModelsHybridDensity2023} further increasing $^{13}$C and $^{15}$N prediction accuracies. Iuliucci et al.\ examined how the functional choice influences the prediction accuracy of molecular corrections, finding hybrid functionals sufficient to match the accuracy of double-hybrid ones. \cite{iuliucciModelsHybridDensity2023} Ramos and co-workers then showed that hybrid-DFT-relaxed geometries remove the accuracy limitations imposed by GGA geometries, with double-hybrid corrections slightly improving $^{13}$C predictions over hybrid corrections, achieving isotropic shift prediction RMSEs of  1.1~ppm using hybrid PBE0 corrections and 0.9~ppm using double-hybrid DSD-PBEP86 corrections.~\cite{ramosInterplayDensityFunctional2024}

Molecular correction and fragment based computation schemes have been successfully used to study challenging systems in which GIPAW computations were not accurate enough. 
Hartman et al. demonstrated that a fragment-based approach on the hybrid-DFT level of theory enabled the discrimination of pharmaceutically relevant organic crystal forms,~\cite{hartmanEnhancedNMRDiscrimination2016} including acetaminophen, phenobarbital and testosterone polymorphs. Dračínský and coworkers reported similar gains in discriminating polymorphs using molecular correction schemes and applied molecular corrections to study isocytosine tautomers.~\cite{dracinskyImprovingAccuracySolidstate2019a} Dračínský also used molecular corrections up to coupled cluster CCSD level of theory to investigate pyridinium fumarates.~\cite{dracinskyAnalyzingDiscrepanciesChemicalShift2021} Further applications of molecular corrections or shielding computations from molecular clusters, ranged from determining tautomers in active sites of enzymes,~\cite{holmesImagingActiveSite2022a} to computing corrected shift-tensors,~\cite{holmesChemicalShiftTensors2020} additional verification of shift-driven structure refinement,~\cite{toomeyNMRguidedRefinementCrystal2024} and even shielding computations of $^{51}$V nuclei in molecular crystals.~\cite{mathewsAccurateFragmentbased51V2021}

Chaloupecká and coworkers applied molecular corrections directly to ShiftML predictions to improve their accuracy.~\cite{chaloupeckaDivergingErrorsComparison2025} They concluded that a sufficiently accurate baseline model is needed for the corrections to take full effect. In fact, correcting ShiftML2 reduced the $^{13}$C isotropic shift errors from 3.0 to 2.5~ppm,~\cite{chaloupeckaDivergingErrorsComparison2025} whereas the more accurate ShiftML3 could be corrected from 2.4~ppm to 1.6~ppm.~\cite{chaloupecka2025nmr} Correcting ShiftML predictions this way remains laborious, and it would be preferable if the molecular corrections were absorbed directly into the model predictions.

Here, we present \shiftmlnew{}, a new shielding prediction model trained directly on the monomer-corrected shielding tensors $\sigma^{\textrm{PBE0}}_{\textrm{cryst}}$
 of Equation~\ref{eq:correction}. The model takes the periodic crystal structure as input and directly predicts the corrected shielding tensors. The monomer correction, including implicit cutting of the molecular fragments,  is added directly to the periodic GIPAW-PBE targets, and neither isolated-molecule GIAO calculations nor semiempirical calculations are required during inference. Because \shiftmlnew{} uses the same architecture as \shiftmlthree{}, its evaluation cost and linear scaling with the number of atoms remain unchanged.

Computing molecular corrections for the over 15,000 crystals in the ShiftML2/\shiftmlthree{} training, validation and test sets by hand would be tedious and likely induce biases in separating molecules from the solid. We therefore built a pipeline that takes the crystal structure, cuts out molecules, prepares and runs the low- and high-level molecular DFT calculations, removes outliers, and assembles the training data. We automate the cutting of molecular fragments based on the bond connectivity between atoms in the crystals. For every molecular crystal in the ShiftML2/3 datasets, we compute Mayer-Wiberg bond orders~\cite{wibergApplicationPoplesantrysegalCNDO1968, mayerChargeBondOrder1983}  between all pairs of atoms at the semiempirical GFN2-xTB level of theory.~\cite{bannwarthGFN2xTBAnAccurateBroadly2019} We consider pairs of atoms bonded if their bond order exceeds 0.5. We represent the crystal as a graph, with atoms as nodes and bonded atom pairs as edges. The connected components of this graph define the molecular fragments. Both steps explicitly consider periodic boundary conditions, so molecules crossing the cell boundary are reconstructed across images. After cutting, each fragment is assigned a charge and a spin multiplicity. Atomic graphs and bookkeeping of atomic indices of the molecular fragments are handled with the pymatgen~\cite{ongPythonMaterialsGenomics2013}, Atomic Simulation Environment (ASE)~\cite{hjorthlarsenAtomicSimulationEnvironment2017} and networkx~\cite{hagbergExploringNetworkStructure2008} libraries.
Atomic GFN2-xTB Mulliken charges~\cite{mullikenElectronicPopulationAnalysis1955} from the periodic calculation are rounded to integers, constrained to sum to zero over the cell, and then summed within each fragment to give its net charge. Each fragment is then assigned the lowest spin multiplicity consistent with its resulting electron count. The procedure required handling of selected edge cases, such as spurious bonds between $\pi$-stacked sulfur-containing heterocycles and fractional protonation states, which could be resolved by selectively increasing bond-order thresholds between selected nuclei. GFN2-xTB calculations use the tblite integration.~\cite{TbliteTblite2026} Molecular GIAO shieldings~\cite{stoychevSelfConsistentFieldCalculation2018} are computed with ORCA~6.1.0~\cite{neeseSoftwareUpdateORCA2025} interfaced with the ORCA OPI Python API~\cite{tetenbergORCAMeetsPythonThe2026} to orchestrate shielding computations. We use ORCA input parameters close to the ones selected by Ramos et al.~\cite{ramosInterplayDensityFunctional2024} ORCA DFT input parameters are listed in the SI section~\ref{sec:ORCA_inputs}. Molecular DFT calculations are performed for PBE~\cite{perdewGeneralizedGradientApproximation1996} and PBE0~\cite{adam-baro99jcp} functionals employing the cc-pVTZ basis set~\cite{dunningGaussianBasisSets1989, woonGaussianBasisSets1993}, and the resolution of identity approximation.~\cite{vahtrasIntegralApproximationsLCAOSCF1993, neeseImprovementResolutionIdentity2003} Hybrid computations are performed with the accelerated "chain-of-spheres" algorithm.~\cite{neeseEfficientApproximateParallel2009}
All molecular calculations are performed with the implicit CPCM solvation correction~\cite{baroneQuantumCalculationMolecular1998} using dichloromethane as the medium. 
The molecular corrections are then added to the GIPAW shielding calculations whose parameters are described in the ShiftML2 reference publication.~\cite{cordova_machine_2022}
We perform calculations for the entire ShiftML2/\shiftmlthree{} training set and after discarding unconverged calculations and outliers we obtain a database with a total of 12,613 structures in the training set, 495 structures in the validation set and 1,218 structures in the test set. We discard crystals for which the splitting and spin-state assignment produce doublet fragments. Outliers are removed by discarding entire crystal structures containing any atom with an anomalous absolute shielding or an unusually large molecular correction relative to its element's typical value. 
We then train a committee of eight nanoPET models, a variant of the graph neural network (GNN) Point-Edge-Transformer (PET) architecture, ~\cite{NEURIPS2023_fb4a7e35} on the cleaned training sets.
nanoPET is trained on, and predicts, the shielding tensors in their irreducible spherical tensor (IST) representation: a scalar ($\lambda=0$), a pseudovector ($\lambda=1$) and a proper five-vector ($\lambda=2$), sharing the same rotational behaviour as the corresponding spherical harmonics $Y^{\lambda}$.  
nanoPET converts atomic environments into local representations through one message-passing sequence of unconstrained graph transformer blocks. The IST components are predicted from these representations via linear readouts. The ShiftML implementation converts the IST components back to the Cartesian basis before returning the tensors.
Model training was performed with the \texttt{metatrain} library and using the \texttt{metatomic} integration for model deployment.~\cite{bigiMetatensorMetatomicFoundational2026} The training hyperparameters are identical to those of the ShiftML3 training. Each model is initialized with a different random seed and trained by mini-batch gradient descent with the Adam optimizer~\cite{kingmaAdamMethodStochastic2017a}, a learning rate of $3\cdot10^{-4}$, a batch size of 4, and 1,500 epochs. Note that we remove one model of the committee due to convergence issues and the final ShiftML4 model is made up by an ensemble of 7 nanoPET models. The ShiftML3 paper discusses the benefits of a committee over a single PET model.~\cite{kellnerDeepLearningModel2025b} Importantly, using a committee, higher prediction accuracies are achieved and the spread of the committee predictions can be interpreted as an uncertainty estimate of the model predictions.

We first evaluate \shiftmlnew{} and \shiftmlthree{} against the DFT reference they are trained on:
\shiftmlnew{} against monomer-corrected isotropic shieldings, \shiftmlthree{} against
GIPAW/PBE. On the same hold-out test structures, both reach comparable accuracy. For $^{1}$H \shiftmlnew{} achieves prediction RMSEs of 0.39~ppm and \shiftmlthree{} 0.42~ppm against DFT and for $^{13}$C, 2.20~ppm for \shiftmlnew{} and 2.24~ppm for \shiftmlthree{}. In Table~\ref{tab:cs_tensor_derived_values} we compare prediction error metrics of \shiftmlnew{} and \shiftmlthree{} for isotropic chemical shieldings $\sigma_{\text{iso}}$ on the most commonly studied spin-1/2 nuclei in the test set. We report the full prediction errors against DFT in the SI, Table~\ref{tab:complete_metrics}. Across the remaining elements, both models show similar errors. This suggests that the cutting procedure and the GIAO calculations add little noise to the targets, and that the corrected shieldings can be learned directly from the periodic structure.

\begin{table}[tbp]
\caption{\label{tab:cs_tensor_derived_values} RMSE (in ppm) of \shiftmlthree{} (\shiftmlthreeshort) and \shiftmlnew{} (\shiftmlnewshort) predictions of the isotropic shieldings $\sigma_{\text{iso}}$, the tensor components $\sigma_{ij}$ and the principal tensor components $\sigma_{\text{PAS}}$, evaluated on a hold-out test set. \shiftmlthree{} is evaluated against GIPAW-PBE reference calculations and \shiftmlnew{} against PBE0 molecular-corrected GIPAW-PBE calculations. }
\begin{tabular}{@{}>{\hspace{3 mm}}l @{\hskip 2mm}cc@{\hskip 6mm}cc@{\hskip 6mm}cc}
\toprule
    \multicolumn{1}{c@{\hskip 3mm}} {nucleus} & \multicolumn{2}{c@{\hskip 6mm}}{$\sigma_{\text{iso}}$} & \multicolumn{2}{c@{\hskip 6mm}}{$\sigma_{ij}$} & \multicolumn{2}{c}{$\sigma_{\text{PAS}}$} \\
\midrule
 & \shiftmlthreeshort & \shiftmlnewshort & \shiftmlthreeshort & \shiftmlnewshort & \shiftmlthreeshort & \shiftmlnewshort \\
\midrule
$^{1}\text{H}$ & 0.42 & 0.39 & 0.81 & 0.78 & 0.85 & 0.82 \\
$^{13}\text{C}$ & 2.24 & 2.20 & 3.43 & 3.31 & 3.77 & 3.68 \\
$^{15}\text{N}$ & 10.20 & 11.11 & 11.53 & 11.83 & 15.67 & 16.41 \\
$^{19}\text{F}$ & 5.55 & 4.71 & 6.08 & 5.55 & 7.54 & 6.83 \\
$^{31}\text{P}$ & 18.68 & 16.30 & 19.70 & 19.42 & 23.98 & 21.62 \\
\bottomrule
\end{tabular}
\end{table}

Table~\ref{tab:cs_tensor_derived_values} also compares  \shiftmlnew{} and \shiftmlthree{} prediction errors on tensor components $\sigma_{ij}$ and principal components $\sigma_{\text{PAS}}$. We find that, as for the isotropic shieldings, \shiftmlnew{} and \shiftmlthree{} perform comparably on tensorial quantities. More detailed evaluation metrics of principal components and tensor components are listed in tables~\ref{tab:complete_metrics_PAS} and \ref{tab:complete_metrics_tensor_components}.  Due to the cutting and charge assignment procedure, the alkali and alkaline-earth nuclei ($^{23}$Na, $^{25}$Mg, $^{39}$K, $^{43}$ Ca) are represented by only tens to hundreds of environments in the training set, so their reported metrics rest on limited statistics and predictions of \shiftmlnew{} on metal-ion containing structures should be treated with care. However, including them is important to cover prediction on ionic molecular solids of particular relevance for pharmaceutical salt formulations, for example.

\begin{figure*}
    \centering
    \includegraphics[width=2\columnwidth]{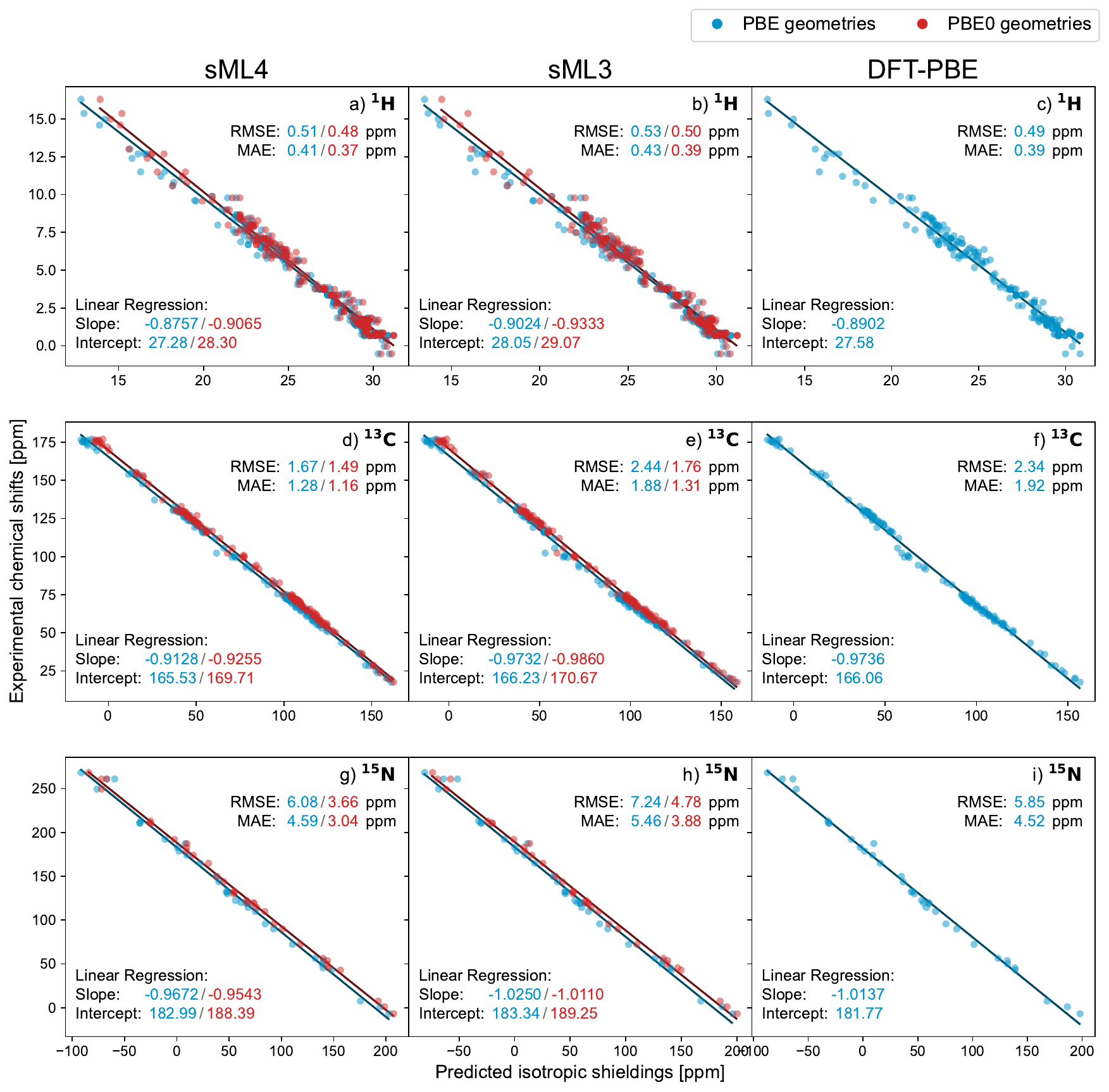}
    \caption{Parity plots between \shiftmlnew{}, \shiftmlthree{} and DFT predicted shifts and experimental reference values, for $^1$H (a,b,c), $^{13}$C (d,e,f), and $^{15}$N (g,h,i). Predictions and GIPAW-PBE calculations (DFT) are performed based on PBE geometries (in blue) from Ramos et al.~\cite{ramosInterplayDensityFunctional2024} (relaxed with PBE-D3(BJ)) and for $^1$H on 13 structures from the ShiftML2 publication (relaxed with PBE-D2)~\cite{cordova_machine_2022} and PBE0 geometries (in red) relaxed using PET-MOLS. The black line shows the linear relation, converting computational isotropic shieldings into shifts. The slope and intercept for each method and nucleus are indicated in the respective subplot.}
    \label{fig:sML_vs_experimental}
\end{figure*}

Next, we evaluate \shiftmlnew{} on representative experimental benchmark sets of organic crystals and assess the prediction accuracy against assigned experimental $^1$H, $^{13}$C and $^{15}$N shifts. We evaluate \shiftmlnew{} and \shiftmlthree{} on published geometries relaxed  with PBE-D2 ($^{1}$H benchmark set) and PBE-D3(BJ) ($^{13}$C and $^{15}$N sets) and compare against GIPAW-PBE reference. In Figure~\ref{fig:sML_vs_experimental} we show parity plots of \shiftmlnew{}, \shiftmlthree{} and GIPAW-DFT predictions against experimental reference values. We find that \shiftmlnew{} prediction errors decrease across all species. On identical PBE-geometries the $^{13}$C RMSE drops from 2.44~ppm (\shiftmlthree{}) to 1.67~ppm, a $\approx$30\% reduction, and a $\approx$20\% reduction for the $^{15}$N RMSE, from 7.24~ppm (\shiftmlthree{}) to 6.08~ppm (\shiftmlnew{}). 
The regression slopes converting shieldings to shifts are systematically smaller in magnitude for \shiftmlnew{} than for \shiftmlthree{} across all three nuclei, and closely match the ab initio PBE0 molecular-corrected slopes of Ramos et al.~\cite{ramosInterplayDensityFunctional2024} on the two nuclei taken from their benchmark using identical geometries (-0.913 against -0.930 for $^{13}$C, -0.967 against -0.975 for $^{15}$N). The deviation from the original slope is inherited from the reference level of theory rather than introduced by the ML model. In line with the observations from Ramos et al.,~\cite{ramosInterplayDensityFunctional2024} molecular corrections and \shiftmlnew{} improve prediction accuracies predominantly for $^{13}$C and $^{15}$N, whose shieldings are dominated by intramolecular interactions, whereas $^{1}$H predictions, which are governed to a much greater extent by intermolecular contributions, change only marginally. In SI  Figure~\ref{fig:parity_tensorcomponents_15N_bench} we plot differences between \shiftmlnew{} and \shiftmlthree{} isotropic shift predictions against experimental reference shifts, for $^1$H, $^{13}$C and $^{15}$N, visualizing the effects of molecular corrections across the observed shift range of the reference nucleus.
Evaluating all 49 benchmark geometries with \shiftmlnew{} took under a minute on a laptop (MacBook Pro M1). The GIPAW reference calculations took 80 hours on a 32-core compute node.

Ramos et al.~\cite{ramosInterplayDensityFunctional2024} recognized that the quality of reference geometries contributes significantly to the chemical shift prediction accuracies. They showed that moving from PBE-D3 to PBE0-D3 geometries removes the bias in bond lengths relative to experiment and improves shift predictions when computing either GIPAW-PBE, or molecular-corrected shieldings from PBE0-D3 geometries.
The steep computational cost of ab-initio geometry relaxations forms another bottleneck in the NMR-crystallography workflow.
We therefore employ the recently developed PET-MOLS machine-learned interatomic potential (MLIP)~\cite{kellnerQuantumcorrectedNMRCrystallography2026} to relax the benchmark crystal structures, cutting the cost for relaxations from CPU-hours to seconds. PET-MOLS covers the same chemical space as \shiftmlnew{} and \shiftmlthree{} and is trained on PBE0+MBD reference data. A similar approach making use of gas-phase molecule trained MLIPs for geometry relaxation in the context of NMR crystallography has been proposed by Gunaga et al.~\cite{gunagaAccessibleHybridDFTquality}. In Fig.~\ref{fig:sML_vs_experimental} we also show parity plots of \shiftmlnew{} and \shiftmlthree{} predictions from PET-MOLS relaxed geometries against experimental reference values. Basing chemical shielding predictions on approximate hybrid-DFT quality geometries further decreases prediction errors of both \shiftmlnew{} and \shiftmlthree{}. On PET-MOLS relaxed geometries the prediction errors of \shiftmlnew{} on $^{13}$C decrease from 1.67~ppm to 1.49~ppm, for $^{15}$N from 6.08~ppm to 3.66~ppm and slightly for $^{1}$H  from 0.51~ppm to 0.48~ppm over the \shiftmlnew{} predictions on PBE-geometries. The $^{13}$C isotropic shift RMSE of \shiftmlthree{} also improves significantly on PET-MOLS geometries, from 2.44~ppm to 1.76~ppm. For comparison, Gunaga and coworkers~\cite{gunagaAccessibleHybridDFTquality} report 1.78~ppm and 1.84~ppm for \shiftmlthree{} predictions on the same crystals relaxed with MACE-POLAR-1~\cite{batatiaMACEPOLAR1PolarisableElectrostatic2026} and UMA-OMOL,~\cite{woodUMAFamilyUniversal2026a} respectively, albeit with looser relaxation convergence criteria. Taken together, moving from \shiftmlthree{} on PBE geometries to \shiftmlnew{} on PET-MOLS geometries reduces the isotropic RMSE against experiment by $39\%$ for $^{13}$C (from $2.44$~ppm to $1.49$~ppm) and by $49\%$ for $^{15}$N (from $7.24$~ppm to $3.66$~ppm). We observe smaller improvements for $^1$H ($0.53$~ppm, to $0.48$~ppm) and note in passing that $^1$H chemical shifts are particularly sensitive to nuclear quantum effects, which are not treated here relying only on static geometry-relaxed structures.~\cite{enge+21jpcl, kellnerQuantumcorrectedNMRCrystallography2026}

\begin{figure*}
    \centering
    \includegraphics[width=2\columnwidth]{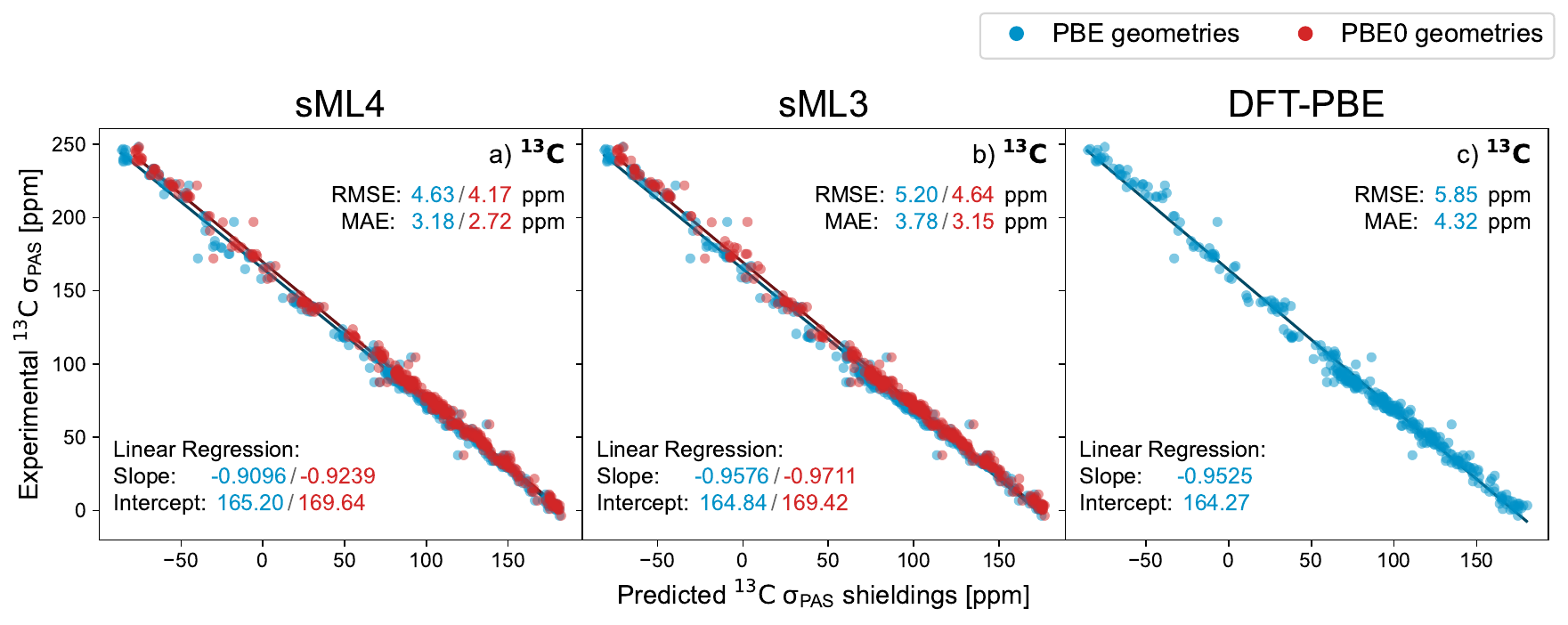}
    \caption{Parity plots between  \shiftmlnew{} (a), \shiftmlthree{} (b), DFT (c) predicted principal components $\sigma_{\text{PAS}}$ of the shielding tensor and experimental reference values. The black line shows the linear relation, converting computational principal components into experimental components. The slope and intercept for each method are indicated in the respective subplot. In blue and red, PBE and PBE0 (PET-MOLS) geometries were taken as starting points for shielding calculations.}
    \label{fig:sML_vs_experimental_csa}
\end{figure*}

Finally, we evaluate \shiftmlnew{} chemical shift anisotropy (CSA) predictions against experiment on a benchmark of $^{13}$C principal tensor components $\sigma_{\text{PAS}}$ of the symmetric shift tensor, from Hartman et al.~\cite{hartmanFragmentbased13CNuclear2015}. In Fig.~\ref{fig:sML_vs_experimental_csa} we show parity plots of GIPAW-DFT, \shiftmlthree{} and \shiftmlnew{} $\sigma_{\text{PAS}}$ predictions against experimental benchmark values.  We find that the gains in isotropic prediction accuracies of \shiftmlnew{} against \shiftmlthree{} also translate to gains in prediction accuracy of $\sigma_{\text{PAS}}$.
\shiftmlnew{} achieves a joint component RMSE of 4.63 ppm against experiment, compared with 5.20 ppm for \shiftmlthree{} and 5.85 ppm for DFT. That \shiftmlthree{} already outperforms its own DFT reference here is likely fortuitous.

In the Supporting Information Section~\ref{sec:15N_PSA_benchmark}, we also present an evaluation of \shiftmlthree{} and \shiftmlnew{} on an experimental benchmark set of  $^{15}$N shielding tensor principal components. Both models achieve comparable prediction RMSEs of 9.49~ppm (\shiftmlthree{}) and 9.22~ppm (\shiftmlnew{}). On PET-MOLS geometries, the prediction accuracies of \shiftmlnew{} and \shiftmlthree{} further improve to 7.89~ppm and 8.12~ppm, respectively, comparable to those of PBE0-fragment based calculations using PBE geometries (7.95~ppm). Kalakewich and coworkers~\cite{kalakewichMonitoringRefinementCrystal2015} report experimental measurement errors of 5.5~ppm for the $^{15}$N shift tensor in single crystalline histidine, which may explain the smaller improvement in prediction accuracy achieved with \shiftmlnew{} compared to $^{13}$C $\sigma_{\text{PAS}}$ predictions.

In summary, here we have introduced \shiftmlnew{}, a transferable deep learning model that predicts chemical shielding tensors in organic solids at near-hybrid-DFT quality, with the monomer corrections absorbed directly into the model so that no molecular DFT calculation is needed at inference. On experimental benchmarks it lowers the isotropic $^{13}$C RMSE from $2.44$~ppm to $1.67$~ppm with respect to \shiftmlthree{}, below the $2.34$~ppm of established GIPAW/PBE workflows. $^1$H shifts, which are dominated by intermolecular contributions, are largely unaffected by the monomer correction, with the \shiftmlnew{}
RMSE only slightly lower than that of \shiftmlthree{} (0.53 ppm to
0.51 ppm), in line with previous
observations.~\cite{dracinskyImprovingAccuracySolidstate2019a, ramosInterplayDensityFunctional2024}  For $^{15}$N the isotropic RMSE is reduced from $7.24$~ppm to $6.08$~ppm, comparable to the $5.85$~ppm of GIPAW-PBE. The improved prediction accuracies extend to the anisotropy, with a $^{13}$C $\sigma_\mathrm{PAS}$ RMSE of $4.63$~ppm against $5.85$~ppm for GIPAW-PBE. Combined with the PET-MOLS MLIP for fast geometry relaxation at approximate hybrid-DFT quality, it brings the $^{13}$C RMSE against experiment to $1.49$~ppm and the $^{15}$N RMSE to $3.66$~ppm, at machine-learning cost throughout. \shiftmlnew{} is available in the ShiftML Python package.

\begin{acknowledgments}
\subsection{ACKNOWLEDGEMENTS}

The authors thank Fr\'ed\'eric Mentink-Vigier for stimulating discussions on higher-level reference corrections for machine-learned chemical shieldings.
This work has been supported by Swiss National Science Foundation Grant No. 200020\_212046, and by the NCCR MARVEL, funded by the Swiss National Science Foundation (grant number 182892). MC also acknowledges funding from the ERC Horizon 2020 Grant No. 101001890-FIAMMA.
GJOB acknowledges funding from the U.S. National Science Foundation (CHE-1955554).
PAU thanks the U.S. Department of Energy, Office of Science, Office of Workforce Development for Teachers and Scientists, Office of Science Graduate Student Research (SCGSR) program. The SCGSR program is administered by the Oak Ridge Institute for Science and Education (ORISE) for the DOE. ORISE is managed by ORAU under contract number DE-SC0014664. All opinions expressed in this paper are the author's and do not necessarily reflect the policies and views of DOE, ORAU, or ORISE.

\subsection{SUPPORTING INFORMATION }

The supporting information is provided below,
containing additional details on the dataset composition, detailed model evaluations and the usage of the ShiftML Python package, which is available under an open-source license at \url{https://github.com/lab-cosmo/shiftml}. \rev{The Python scripts for cutting, evaluation against experimental reference data,} training, validation and test data in the xyz format as well as raw ORCA computations for the molecular computations \rev{can be accessed at the following link DOI:10.5281/zenodo.XXXXXXX [will be completed on acceptance] and is 
available under the CC BY-SA 4.0 (Creative Commons Attribution-ShareAlike 4.0 International) license.}

\end{acknowledgments}

\section{References}
\clearpage

\onecolumngrid
\part*{Supplementary Information}

\setcounter{secnumdepth}{3}           %
\renewcommand{\thesection}{S\arabic{section}}
\setcounter{section}{0}

\renewcommand{\thefigure}{S\arabic{figure}}
\setcounter{figure}{0}

\renewcommand{\thetable}{S\arabic{table}}
\setcounter{table}{0}

\section{Experimental shifts and predicted shieldings for the $^1$H benchmark dataset}

Assigned chemical shift values were used from the previously curated $^1$H benchmark data set. \cite{cordova_machine_2022} Methyl rotational dynamics were accounted for by averaging the chemical shifts of the three proton positions into a single effective value per methyl group, which was assigned a weight of three in the linear regression. The same treatment is applied to other magnetically equivalent protons, such as those in methylene groups. In cases of pairwise ambiguities of the experimental assignments, experimental chemical shifts were resolved by selecting the assignment yielding the lowest shift RMSE. (e.g. H$_a$ and H$_b$ of CH$_2$ groups)

{\renewcommand{\arraystretch}{0.5}
\begin{longtable}{lcccccc}
\caption{Experimental isotropic $^{1}$H chemical shifts and predicted shieldings for the atomic sites in the benchmark dataset. Predicted shieldings are given for PBE and PBE0 (PET-MOLS) geometries, computed with GIPAW-DFT and with the \shiftmlthree{} (\shiftmlthreeshort{}) and \shiftmlnew{} (\shiftmlnewshort{}) machine-learning models.} \label{tab:table_1H_ml} \\
\toprule
\multirow{2}{*}{\textbf{Ref. Code}} &
\multirow{2}{*}{\textbf{Exp. shift [ppm]}} &
\multicolumn{3}{c}{\textbf{PBE geometries}} &
\multicolumn{2}{c}{\textbf{PBE0 geometries}} \\
\cmidrule(lr){3-5} \cmidrule(lr){6-7}
& & \textbf{DFT} & \textbf{\shiftmlthreeshort{}} & \textbf{\shiftmlnewshort{}} & \textbf{\shiftmlthreeshort{}} & \textbf{\shiftmlnewshort{}} \\
\midrule
\endfirsthead
\toprule
\multirow{2}{*}{\textbf{Ref. Code}} &
\multirow{2}{*}{\textbf{Exp. shift [ppm]}} &
\multicolumn{3}{c}{\textbf{PBE geometries}} &
\multicolumn{2}{c}{\textbf{PBE0 geometries}} \\
\cmidrule(lr){3-5} \cmidrule(lr){6-7}
& & \textbf{DFT} & \textbf{\shiftmlthreeshort{}} & \textbf{\shiftmlnewshort{}} & \textbf{\shiftmlthreeshort{}} & \textbf{\shiftmlnewshort{}} \\
\midrule
\endhead
\textbf{Dimethylimidazole}  &     4.80 &    25.91 &    25.91 &    25.75 &    25.92 &    25.66 \\
                            &     0.70 &    29.55 &    29.53 &    29.47 &    29.87 &    29.75 \\
                            &     0.70 &    29.55 &    29.53 &    29.47 &    29.87 &    29.75 \\
                            &     0.70 &    29.55 &    29.53 &    29.47 &    29.87 &    29.75 \\
                            &     1.40 &    29.48 &    29.41 &    29.46 &    29.57 &    29.62 \\
                            &     1.40 &    29.48 &    29.41 &    29.46 &    29.57 &    29.62 \\
                            &     1.40 &    29.48 &    29.41 &    29.46 &    29.57 &    29.62 \\
                            &    13.00 &    15.59 &    16.30 &    15.65 &    16.10 &    15.59 \\
                            &     1.40 &    28.88 &    29.08 &    28.96 &    29.31 &    29.25 \\
                            &     1.40 &    28.88 &    29.08 &    28.96 &    29.31 &    29.25 \\
                            &     1.40 &    28.88 &    29.08 &    28.96 &    29.31 &    29.25 \\
                            &     1.50 &    28.87 &    28.97 &    28.87 &    29.15 &    29.06 \\
                            &     1.50 &    28.87 &    28.97 &    28.87 &    29.15 &    29.06 \\
                            &     1.50 &    28.87 &    28.97 &    28.87 &    29.15 &    29.06 \\
                            &    15.00 &    14.34 &    14.23 &    14.18 &    14.54 &    14.52 \\
                            &     5.20 &    24.73 &    24.83 &    24.82 &    24.96 &    25.05 \\
\midrule
\textbf{AZD5718}            &     1.20 &    29.20 &    29.32 &    29.29 &    29.49 &    29.47 \\
                            &     1.20 &    29.20 &    29.32 &    29.29 &    29.49 &    29.47 \\
                            &     1.20 &    29.20 &    29.32 &    29.29 &    29.49 &    29.47 \\
                            &    10.60 &    18.45 &    18.11 &    18.16 &    18.15 &    18.16 \\
                            &     5.80 &    23.81 &    23.78 &    23.70 &    24.03 &    23.98 \\
                            &     6.90 &    23.17 &    23.88 &    23.82 &    23.99 &    23.93 \\
                            &     7.30 &    23.04 &    23.06 &    22.84 &    23.23 &    23.03 \\
                            &     6.70 &    23.67 &    24.34 &    24.21 &    24.57 &    24.49 \\
                            &     7.00 &    22.51 &    22.78 &    22.66 &    22.94 &    22.95 \\
                            &     3.90 &    26.21 &    27.01 &    26.75 &    27.26 &    27.02 \\
                            &     1.60 &    28.85 &    28.67 &    28.82 &    28.90 &    29.13 \\
                            &     0.00 &    30.34 &    30.35 &    30.62 &    30.52 &    30.79 \\
                            &     1.70 &    28.40 &    28.55 &    28.60 &    28.75 &    28.79 \\
                            &     1.60 &    28.46 &    29.01 &    28.97 &    29.37 &    29.21 \\
                            &     1.60 &    29.05 &    28.90 &    29.05 &    29.06 &    29.37 \\
                            &    -0.50 &    29.96 &    30.02 &    30.26 &    30.19 &    30.43 \\
                            &     0.80 &    29.36 &    29.51 &    29.62 &    29.72 &    29.84 \\
                            &    -0.50 &    30.79 &    30.57 &    30.91 &    30.72 &    31.07 \\
                            &     0.80 &    29.30 &    29.26 &    29.27 &    29.41 &    29.42 \\
                            &     7.70 &    21.33 &    21.36 &    21.72 &    21.72 &    22.14 \\
                            &     7.60 &    22.40 &    22.58 &    22.37 &    22.88 &    22.55 \\
                            &     1.70 &    28.87 &    29.63 &    29.71 &    29.61 &    29.62 \\
                            &     2.70 &    27.49 &    27.80 &    28.08 &    27.94 &    28.19 \\
                            &     6.90 &    22.59 &    22.46 &    22.18 &    23.41 &    23.15 \\
                            &     1.90 &    28.03 &    28.07 &    27.86 &    28.24 &    28.03 \\
                            &     2.70 &    27.23 &    27.50 &    27.59 &    27.64 &    27.79 \\
\midrule
\textbf{Uracil}             &     7.50 &    22.15 &    21.99 &    21.84 &    22.05 &    21.98 \\
                            &    10.80 &    17.95 &    18.32 &    18.23 &    19.06 &    19.01 \\
                            &    11.20 &    17.16 &    17.98 &    17.48 &    19.19 &    18.75 \\
                            &     6.00 &    23.87 &    23.99 &    23.68 &    24.37 &    24.06 \\
\midrule
\textbf{Naproxen}           &     7.00 &    22.77 &    23.09 &    22.93 &    23.29 &    23.18 \\
                            &     6.10 &    24.19 &    24.24 &    24.06 &    24.32 &    24.14 \\
                            &     3.80 &    27.13 &    26.96 &    26.69 &    27.07 &    26.80 \\
                            &     4.50 &    25.77 &    25.69 &    25.58 &    25.97 &    25.86 \\
                            &     4.10 &    26.18 &    25.85 &    25.50 &    25.95 &    25.64 \\
                            &     5.90 &    24.87 &    24.93 &    24.60 &    25.05 &    24.73 \\
                            &     3.20 &    26.76 &    27.73 &    27.68 &    27.79 &    27.78 \\
                            &     1.80 &    28.64 &    29.36 &    29.30 &    29.57 &    29.53 \\
                            &     1.80 &    28.64 &    29.36 &    29.30 &    29.57 &    29.53 \\
                            &     1.80 &    28.64 &    29.36 &    29.30 &    29.57 &    29.53 \\
                            &     2.30 &    28.16 &    28.25 &    28.09 &    28.44 &    28.26 \\
                            &     2.30 &    28.16 &    28.25 &    28.09 &    28.44 &    28.26 \\
                            &     2.30 &    28.16 &    28.25 &    28.09 &    28.44 &    28.26 \\
                            &    11.50 &    15.83 &    16.32 &    16.28 &    17.36 &    17.28 \\
\midrule
\textbf{Furosemide}         &     8.40 &    22.21 &    22.47 &    22.13 &    22.76 &    22.41 \\
                            &    12.70 &    16.65 &    16.73 &    16.50 &    17.10 &    16.95 \\
                            &     6.50 &    23.34 &    23.80 &    23.35 &    24.05 &    23.65 \\
                            &     6.50 &    23.34 &    23.80 &    23.35 &    24.05 &    23.65 \\
                            &     7.90 &    22.71 &    23.26 &    22.90 &    23.26 &    22.94 \\
                            &     8.70 &    21.86 &    22.41 &    21.96 &    22.54 &    22.12 \\
                            &     5.60 &    25.39 &    25.59 &    25.34 &    25.54 &    25.49 \\
                            &     3.80 &    26.58 &    26.57 &    26.53 &    26.60 &    26.57 \\
                            &     6.00 &    24.12 &    24.26 &    24.05 &    24.55 &    24.38 \\
                            &     6.00 &    23.96 &    24.40 &    24.58 &    24.64 &    24.85 \\
                            &     7.70 &    22.13 &    22.47 &    22.52 &    22.64 &    22.67 \\
                            &     8.40 &    21.80 &    22.39 &    22.01 &    22.59 &    22.17 \\
                            &     4.30 &    26.36 &    25.70 &    25.76 &    25.81 &    25.91 \\
                            &    12.70 &    16.27 &    16.43 &    16.12 &    17.13 &    16.90 \\
                            &     6.70 &    22.66 &    22.98 &    22.60 &    23.45 &    23.09 \\
                            &     6.70 &    22.66 &    22.98 &    22.60 &    23.45 &    23.09 \\
                            &     6.20 &    24.96 &    25.23 &    24.63 &    25.47 &    24.85 \\
                            &     8.60 &    21.97 &    22.45 &    22.02 &    22.58 &    22.13 \\
                            &     4.30 &    25.87 &    25.64 &    25.58 &    25.88 &    25.65 \\
                            &     5.70 &    24.86 &    24.64 &    24.65 &    24.91 &    24.94 \\
                            &     6.40 &    23.45 &    23.48 &    23.49 &    23.67 &    23.77 \\
                            &     6.50 &    23.60 &    23.76 &    23.88 &    24.10 &    24.26 \\
\midrule
\textbf{Theophylline}       &    14.60 &    14.22 &    14.32 &    13.86 &    15.46 &    15.10 \\
                            &     7.70 &    22.78 &    22.80 &    22.55 &    23.03 &    22.73 \\
                            &     3.40 &    27.13 &    27.49 &    27.28 &    27.72 &    27.50 \\
                            &     3.40 &    27.13 &    27.49 &    27.28 &    27.72 &    27.50 \\
                            &     3.40 &    27.13 &    27.49 &    27.28 &    27.72 &    27.50 \\
                            &     3.40 &    27.13 &    27.49 &    27.28 &    27.72 &    27.50 \\
                            &     3.40 &    27.13 &    27.49 &    27.28 &    27.72 &    27.50 \\
                            &     3.40 &    27.13 &    27.49 &    27.28 &    27.72 &    27.50 \\
\midrule
\textbf{Flutamide}          &     7.10 &    22.91 &    23.18 &    23.06 &    23.42 &    23.17 \\
                            &     9.90 &    20.49 &    20.66 &    20.47 &    20.63 &    20.51 \\
                            &     8.00 &    22.63 &    23.08 &    22.72 &    23.33 &    22.93 \\
                            &     8.00 &    20.84 &    21.03 &    20.82 &    21.82 &    21.44 \\
                            &     2.00 &    28.30 &    28.39 &    28.59 &    28.53 &    28.71 \\
                            &     1.20 &    29.46 &    29.51 &    29.49 &    29.71 &    29.67 \\
                            &     1.20 &    29.46 &    29.51 &    29.49 &    29.71 &    29.67 \\
                            &     1.20 &    29.46 &    29.51 &    29.49 &    29.71 &    29.67 \\
                            &     1.20 &    29.46 &    29.51 &    29.49 &    29.71 &    29.67 \\
                            &     1.20 &    29.46 &    29.51 &    29.49 &    29.71 &    29.67 \\
                            &     1.20 &    29.46 &    29.51 &    29.49 &    29.71 &    29.67 \\
\midrule
\textbf{Indomethacin-nicotinamide}&     7.30 &    23.55 &    23.20 &    23.16 &    23.39 &    23.33 \\
                            &     5.50 &    25.15 &    25.14 &    24.79 &    25.31 &    24.98 \\
                            &     6.80 &    24.04 &    23.95 &    23.77 &    24.13 &    23.91 \\
                            &     0.90 &    30.15 &    30.04 &    30.04 &    30.15 &    30.15 \\
                            &     0.90 &    30.15 &    30.04 &    30.04 &    30.15 &    30.15 \\
                            &     0.90 &    30.15 &    30.04 &    30.04 &    30.15 &    30.15 \\
                            &     2.90 &    28.16 &    28.02 &    27.64 &    28.19 &    27.84 \\
                            &     2.90 &    28.16 &    28.02 &    27.64 &    28.19 &    27.84 \\
                            &     2.90 &    28.16 &    28.02 &    27.64 &    28.19 &    27.84 \\
                            &     3.40 &    27.44 &    27.69 &    27.67 &    27.83 &    27.91 \\
                            &     3.40 &    27.44 &    27.69 &    27.67 &    27.83 &    27.91 \\
                            &    16.30 &    12.86 &    13.47 &    12.78 &    14.43 &    13.89 \\
                            &     6.40 &    25.28 &    25.07 &    24.58 &    25.28 &    24.76 \\
                            &     6.00 &    23.55 &    23.46 &    23.15 &    23.92 &    23.60 \\
                            &     6.00 &    24.95 &    25.09 &    24.83 &    25.24 &    25.05 \\
                            &     6.40 &    24.69 &    24.61 &    24.60 &    24.78 &    24.76 \\
                            &     9.80 &    21.18 &    21.71 &    21.40 &    21.95 &    21.68 \\
                            &     9.80 &    20.98 &    22.29 &    21.90 &    22.54 &    22.11 \\
                            &     8.30 &    22.37 &    23.27 &    22.83 &    23.59 &    23.18 \\
                            &     7.70 &    23.16 &    23.35 &    23.39 &    23.61 &    23.62 \\
                            &     7.30 &    22.21 &    22.36 &    22.37 &    22.62 &    22.69 \\
                            &     9.00 &    20.97 &    21.13 &    21.06 &    21.58 &    21.52 \\
\midrule
\textbf{Flufenamic acid}    &     8.30 &    22.30 &    22.36 &    22.23 &    22.36 &    22.31 \\
                            &     6.00 &    24.35 &    24.58 &    24.38 &    24.79 &    24.62 \\
                            &     5.40 &    25.19 &    25.46 &    25.00 &    25.65 &    25.22 \\
                            &     6.80 &    23.11 &    23.33 &    23.45 &    23.63 &    23.81 \\
                            &    12.40 &    16.13 &    16.04 &    15.79 &    16.94 &    16.67 \\
                            &     9.60 &    20.02 &    19.59 &    19.54 &    19.97 &    19.92 \\
                            &     6.90 &    23.82 &    23.59 &    23.47 &    23.66 &    23.52 \\
                            &     6.20 &    24.21 &    24.08 &    23.80 &    24.25 &    23.88 \\
                            &     5.90 &    24.63 &    24.62 &    24.62 &    24.93 &    24.86 \\
                            &     7.30 &    23.15 &    23.85 &    23.81 &    24.00 &    24.00 \\
\midrule
\textbf{AZD8329 Form 4}     &     6.92 &    23.71 &    24.13 &    23.76 &    24.35 &    24.01 \\
                            &     8.69 &    21.96 &    22.30 &    21.90 &    22.54 &    22.14 \\
                            &     9.01 &    21.38 &    21.50 &    21.35 &    21.49 &    21.33 \\
                            &     8.47 &    22.29 &    22.80 &    22.61 &    22.84 &    22.79 \\
                            &    15.37 &    12.91 &    13.65 &    12.95 &    15.90 &    15.18 \\
                            &     7.73 &    22.57 &    22.43 &    22.23 &    22.59 &    22.45 \\
                            &     9.64 &    18.92 &    19.60 &    19.46 &    20.65 &    20.52 \\
                            &     2.90 &    28.00 &    28.11 &    28.39 &    28.23 &    28.54 \\
                            &     1.78 &    29.32 &    28.80 &    29.21 &    28.95 &    29.39 \\
                            &     1.88 &    28.88 &    29.47 &    29.67 &    29.69 &    29.88 \\
                            &     1.88 &    28.45 &    28.47 &    28.81 &    28.60 &    28.88 \\
                            &     1.80 &    29.14 &    28.86 &    29.14 &    29.03 &    29.34 \\
                            &     1.60 &    29.29 &    29.58 &    29.89 &    29.86 &    30.18 \\
                            &     0.44 &    30.51 &    30.29 &    30.49 &    30.48 &    30.74 \\
                            &     1.54 &    29.18 &    29.01 &    29.35 &    29.08 &    29.42 \\
                            &     1.88 &    28.20 &    28.62 &    28.83 &    29.02 &    29.20 \\
                            &     1.88 &    28.87 &    29.54 &    29.61 &    29.68 &    29.81 \\
                            &     0.80 &    30.36 &    29.83 &    30.01 &    29.97 &    30.17 \\
                            &     0.80 &    29.54 &    30.11 &    30.27 &    30.18 &    30.34 \\
                            &     1.00 &    30.16 &    29.17 &    29.36 &    29.30 &    29.54 \\
                            &     1.74 &    28.66 &    29.03 &    29.21 &    29.29 &    29.55 \\
                            &     1.74 &    29.66 &    29.75 &    30.09 &    29.92 &    30.29 \\
                            &     0.73 &    29.95 &    30.09 &    30.05 &    30.37 &    30.37 \\
                            &     0.73 &    29.95 &    30.09 &    30.05 &    30.37 &    30.37 \\
                            &     0.73 &    29.95 &    30.09 &    30.05 &    30.37 &    30.37 \\
                            &     0.73 &    29.59 &    29.90 &    29.94 &    30.08 &    30.14 \\
                            &     0.73 &    29.59 &    29.90 &    29.94 &    30.08 &    30.14 \\
                            &     0.73 &    29.59 &    29.90 &    29.94 &    30.08 &    30.14 \\
                            &     0.73 &    30.79 &    30.76 &    30.84 &    31.12 &    31.17 \\
                            &     0.73 &    30.79 &    30.76 &    30.84 &    31.12 &    31.17 \\
                            &     0.73 &    30.79 &    30.76 &    30.84 &    31.12 &    31.17 \\
\midrule
\textbf{Penicillin}         &     4.10 &    26.46 &    26.72 &    26.45 &    26.78 &    26.65 \\
                            &     6.40 &    24.36 &    24.53 &    24.35 &    24.90 &    24.87 \\
                            &     5.70 &    25.63 &    25.45 &    25.55 &    25.38 &    25.36 \\
                            &     1.70 &    29.13 &    29.49 &    29.51 &    29.68 &    29.73 \\
                            &     1.70 &    29.13 &    29.49 &    29.51 &    29.68 &    29.73 \\
                            &     1.70 &    29.13 &    29.49 &    29.51 &    29.68 &    29.73 \\
                            &     0.90 &    30.32 &    30.39 &    30.38 &    30.49 &    30.55 \\
                            &     0.90 &    30.32 &    30.39 &    30.38 &    30.49 &    30.55 \\
                            &     0.90 &    30.32 &    30.39 &    30.38 &    30.49 &    30.55 \\
                            &     6.20 &    24.76 &    24.91 &    24.86 &    25.64 &    25.44 \\
                            &     3.90 &    26.86 &    27.22 &    27.36 &    26.80 &    27.13 \\
                            &     4.70 &    25.79 &    26.32 &    26.28 &    25.96 &    26.05 \\
                            &     7.10 &    23.37 &    23.57 &    23.41 &    23.71 &    23.59 \\
                            &     7.10 &    23.37 &    23.57 &    23.41 &    23.71 &    23.59 \\
                            &     7.10 &    23.37 &    23.57 &    23.41 &    23.71 &    23.59 \\
                            &     7.10 &    23.37 &    23.57 &    23.41 &    23.71 &    23.59 \\
                            &     7.10 &    23.37 &    23.57 &    23.41 &    23.71 &    23.59 \\
\midrule
\textbf{Phenylphosphonic acid}&    12.70 &    16.94 &    16.84 &    16.64 &    17.75 &    17.66 \\
                            &    11.50 &    17.92 &    18.20 &    17.70 &    19.38 &    18.93 \\
                            &     6.50 &    24.49 &    24.74 &    24.32 &    24.94 &    24.54 \\
                            &     6.80 &    24.02 &    24.16 &    23.91 &    24.38 &    24.14 \\
                            &     8.50 &    22.19 &    22.28 &    21.82 &    22.54 &    22.09 \\
                            &     7.10 &    23.96 &    24.11 &    23.55 &    24.32 &    23.76 \\
                            &     6.20 &    24.72 &    24.82 &    24.49 &    24.98 &    24.64 \\
\midrule
\textbf{Cocaine}            &     3.76 &    27.01 &    27.01 &    26.99 &    27.17 &    27.16 \\
                            &     3.78 &    26.75 &    27.24 &    27.45 &    27.47 &    27.70 \\
                            &     5.63 &    25.19 &    25.44 &    25.60 &    25.63 &    25.87 \\
                            &     3.32 &    27.12 &    27.33 &    27.46 &    27.53 &    27.66 \\
                            &     3.06 &    28.20 &    28.32 &    28.65 &    28.44 &    28.79 \\
                            &     3.49 &    27.69 &    28.06 &    28.21 &    28.36 &    28.51 \\
                            &     2.91 &    28.67 &    28.98 &    29.12 &    29.19 &    29.37 \\
                            &     3.38 &    27.64 &    28.33 &    28.15 &    28.60 &    28.41 \\
                            &     2.56 &    28.39 &    28.64 &    28.72 &    28.83 &    28.88 \\
                            &     2.12 &    28.70 &    28.97 &    28.90 &    29.13 &    29.09 \\
                            &     1.04 &    29.58 &    29.47 &    29.37 &    29.62 &    29.54 \\
                            &     1.04 &    29.58 &    29.47 &    29.37 &    29.62 &    29.54 \\
                            &     1.04 &    29.58 &    29.47 &    29.37 &    29.62 &    29.54 \\
                            &     8.01 &    22.55 &    22.53 &    22.53 &    22.75 &    22.71 \\
                            &     8.01 &    22.95 &    22.99 &    22.88 &    23.13 &    23.03 \\
                            &     8.01 &    22.48 &    22.79 &    22.64 &    22.94 &    22.82 \\
                            &     8.01 &    22.33 &    22.73 &    22.41 &    22.91 &    22.60 \\
                            &     8.01 &    22.56 &    22.86 &    22.65 &    23.04 &    22.84 \\
                            &     3.78 &    26.77 &    26.81 &    26.83 &    27.03 &    27.06 \\
                            &     3.78 &    26.77 &    26.81 &    26.83 &    27.03 &    27.06 \\
                            &     3.78 &    26.77 &    26.81 &    26.83 &    27.03 &    27.06 \\
\bottomrule
\end{longtable}
}

\newpage
\section{Experimental shifts and predicted shieldings for the $^{13}$C benchmark dataset}

{\renewcommand{\arraystretch}{0.5}
\begin{longtable}{lcccccc}
\caption{Experimental isotropic $^{13}$C chemical shifts and predicted shieldings for the atomic sites in the benchmark dataset. Predicted shieldings are given for PBE and PBE0 (PET-MOLS) geometries, computed with GIPAW-DFT and with the \shiftmlthree{} (\shiftmlthreeshort{}) and \shiftmlnew{} (\shiftmlnewshort{}) machine-learning models.} \label{tab:table_shieldings} \\
\toprule
\multirow{2}{*}{\textbf{Ref. Code}} &
\multirow{2}{*}{\textbf{Exp. shift [ppm]}} &
\multicolumn{3}{c}{\textbf{PBE geometries}} &
\multicolumn{2}{c}{\textbf{PBE0 geometries}} \\
\cmidrule(lr){3-5} \cmidrule(lr){6-7}
& & \textbf{DFT} & \textbf{\shiftmlthreeshort{}} & \textbf{\shiftmlnewshort{}} & \textbf{\shiftmlthreeshort{}} & \textbf{\shiftmlnewshort{}} \\
\midrule
\endfirsthead
\toprule
\multirow{2}{*}{\textbf{Ref. Code}} &
\multirow{2}{*}{\textbf{Exp. shift [ppm]}} &
\multicolumn{3}{c}{\textbf{PBE geometries}} &
\multicolumn{2}{c}{\textbf{PBE0 geometries}} \\
\cmidrule(lr){3-5} \cmidrule(lr){6-7}
& & \textbf{DFT} & \textbf{\shiftmlthreeshort{}} & \textbf{\shiftmlnewshort{}} & \textbf{\shiftmlthreeshort{}} & \textbf{\shiftmlnewshort{}} \\
\midrule
\endhead
\textbf{ADENOS12} &   154.10 &    12.40 &    13.01 &    11.48 &    17.45 &    16.17 \\
                  &   147.90 &    19.52 &    19.54 &    19.34 &    22.86 &    22.68 \\
                  &   119.30 &    45.57 &    46.12 &    49.15 &    49.73 &    52.79 \\
                  &   154.70 &    15.02 &    15.15 &    13.16 &    18.71 &    16.54 \\
                  &   137.40 &    29.86 &    28.15 &    26.26 &    32.45 &    30.80 \\
                  &    91.90 &    72.07 &    75.88 &    81.40 &    80.10 &    85.54 \\
                  &    74.60 &    91.60 &    92.11 &    99.23 &    96.35 &   103.23 \\
                  &    70.90 &    95.26 &    96.01 &   103.19 &    99.77 &   106.17 \\
                  &    84.50 &    81.46 &    81.59 &    88.57 &    86.75 &    93.17 \\
                  &    62.50 &   106.28 &   106.96 &   114.53 &   110.85 &   118.13 \\
\midrule
\textbf{ASPARM03} &   176.40 &   -12.88 &   -12.11 &   -14.23 &    -4.34 &    -6.04 \\
                  &    51.80 &   119.78 &   119.42 &   123.86 &   122.90 &   127.06 \\
                  &    36.10 &   137.08 &   136.95 &   142.76 &   138.63 &   144.49 \\
                  &   177.10 &    -8.05 &    -7.22 &    -9.61 &    -1.44 &    -3.24 \\
\midrule
\textbf{FRUCTO02} &    65.40 &   102.89 &   102.71 &   109.21 &   106.42 &   112.75 \\
                  &    99.70 &    60.63 &    62.66 &    70.34 &    68.49 &    76.03 \\
                  &    67.20 &    99.94 &    99.59 &   106.42 &   103.60 &   110.12 \\
                  &    69.00 &    99.74 &    99.14 &   106.69 &   102.16 &   109.53 \\
                  &    71.40 &    94.26 &    96.00 &   103.47 &   100.45 &   107.53 \\
                  &    64.90 &   101.98 &   103.54 &   109.72 &   107.97 &   113.85 \\
\midrule
\textbf{GLUTAM01} &   173.00 &    -9.35 &    -9.86 &   -12.01 &    -2.40 &    -3.97 \\
                  &    53.30 &   116.24 &   115.99 &   121.25 &   119.47 &   124.51 \\
                  &    25.50 &   146.35 &   145.36 &   152.14 &   147.31 &   153.97 \\
                  &    28.50 &   143.00 &   143.23 &   148.26 &   145.42 &   150.43 \\
                  &   176.50 &    -9.24 &    -8.84 &   -11.66 &    -3.10 &    -5.33 \\
\midrule
\textbf{GLYCIN29} &   176.20 &   -11.16 &   -10.94 &   -13.59 &    -3.91 &    -6.40 \\
                  &    43.50 &   129.32 &   128.90 &   132.47 &   131.08 &   135.21 \\
\midrule
\textbf{HXACAN09} &   133.10 &    35.89 &    37.49 &    38.12 &    40.95 &    41.59 \\
                  &   123.40 &    45.59 &    46.80 &    47.18 &    48.02 &    47.88 \\
                  &   115.70 &    54.23 &    52.53 &    53.96 &    54.38 &    55.60 \\
                  &   152.30 &    14.69 &    14.47 &    16.93 &    17.48 &    19.81 \\
                  &   116.40 &    52.45 &    51.99 &    53.46 &    54.46 &    55.76 \\
                  &   120.60 &    48.98 &    49.32 &    50.07 &    50.26 &    50.69 \\
                  &   169.80 &    -1.91 &    -4.05 &    -6.57 &     2.63 &     0.48 \\
                  &    23.80 &   148.06 &   148.73 &   153.17 &   150.68 &   155.03 \\
\midrule
\textbf{LALNIN12} &   176.80 &   -14.33 &   -13.47 &   -15.54 &    -6.38 &    -7.88 \\
                  &    50.90 &   119.89 &   119.84 &   124.76 &   123.41 &   128.08 \\
                  &    19.80 &   153.62 &   154.74 &   159.37 &   156.07 &   160.73 \\
\midrule
\textbf{LCYSTN21} &   174.00 &    -8.99 &    -9.71 &   -11.99 &    -1.74 &    -3.73 \\
                  &    56.70 &   114.47 &   114.10 &   118.93 &   117.20 &   121.47 \\
                  &    28.80 &   140.68 &   139.78 &   146.17 &   144.15 &   150.97 \\
\midrule
\textbf{LSERIN01} &   175.10 &   -11.31 &   -10.53 &   -13.15 &    -3.11 &    -5.16 \\
                  &    55.60 &   114.92 &   113.87 &   118.35 &   116.86 &   121.11 \\
                  &    62.90 &   103.47 &   104.28 &   111.81 &   108.32 &   115.99 \\
\midrule
\textbf{LSERMH10} &   175.60 &   -12.30 &   -12.45 &   -14.72 &    -5.20 &    -6.98 \\
                  &    58.30 &   111.17 &   111.98 &   116.55 &   115.28 &   119.59 \\
                  &    61.80 &   106.44 &   106.96 &   114.60 &   110.45 &   118.28 \\
\midrule
\textbf{LTHREO01} &   171.90 &    -7.17 &    -5.76 &    -8.17 &     1.01 &    -0.70 \\
                  &    61.20 &   109.64 &   110.40 &   115.40 &   113.44 &   118.16 \\
                  &    66.80 &    99.50 &    98.49 &   106.50 &   102.68 &   110.32 \\
                  &    20.40 &   153.78 &   152.89 &   157.91 &   153.95 &   158.91 \\
\midrule
\textbf{LTYROS11} &   175.40 &   -11.60 &   -12.34 &   -14.82 &    -5.35 &    -7.29 \\
                  &   130.30 &    37.38 &    36.97 &    36.89 &    39.01 &    38.68 \\
                  &   116.40 &    52.83 &    52.18 &    54.04 &    54.64 &    56.21 \\
                  &    56.40 &   113.77 &   114.49 &   119.20 &   117.87 &   122.58 \\
                  &   131.00 &    38.34 &    36.84 &    36.86 &    38.51 &    38.28 \\
                  &   118.00 &    51.46 &    51.96 &    52.82 &    54.47 &    55.17 \\
                  &   155.60 &    10.41 &    11.30 &    13.30 &    14.05 &    15.64 \\
                  &    36.80 &   134.63 &   134.78 &   142.19 &   136.58 &   143.84 \\
                  &   123.60 &    46.22 &    44.23 &    46.36 &    46.40 &    48.70 \\
\midrule
\textbf{MBDGAL02} &   105.70 &    56.69 &    58.04 &    65.71 &    62.97 &    70.44 \\
                  &    71.20 &    96.65 &    97.59 &   103.78 &   101.15 &   107.15 \\
                  &    72.10 &    94.39 &    95.11 &   101.85 &    98.84 &   105.35 \\
                  &    69.30 &    96.87 &    96.52 &   104.22 &   100.54 &   107.80 \\
                  &    75.60 &    91.87 &    89.06 &    96.09 &    92.57 &    99.27 \\
                  &    62.80 &   105.35 &   106.24 &   113.18 &   109.42 &   116.22 \\
                  &    57.60 &   109.48 &   111.18 &   116.57 &   116.16 &   121.28 \\
\midrule
\textbf{MEMANP11} &    99.60 &    63.11 &    64.94 &    72.54 &    69.83 &    77.06 \\
                  &    71.30 &    95.92 &    95.11 &   103.32 &    98.63 &   106.63 \\
                  &    71.70 &    94.96 &    96.10 &   103.74 &    99.99 &   107.47 \\
                  &    64.80 &   102.74 &   102.86 &   109.12 &   106.82 &   112.68 \\
                  &    71.90 &    94.32 &    96.03 &   101.82 &   100.24 &   105.90 \\
                  &    58.90 &   110.07 &   108.85 &   115.38 &   112.28 &   118.86 \\
                  &    54.90 &   114.21 &   113.93 &   119.83 &   118.07 &   123.65 \\
\midrule
\textbf{MGALPY01} &   100.40 &    62.49 &    63.56 &    71.70 &    69.12 &    77.13 \\
                  &    67.60 &   100.65 &   100.43 &   107.36 &   103.87 &   110.46 \\
                  &    72.60 &    93.84 &    93.10 &   101.16 &    97.32 &   105.24 \\
                  &    70.00 &    97.21 &    98.13 &   105.79 &   101.70 &   108.90 \\
                  &    72.90 &    92.93 &    93.84 &   100.44 &    98.44 &   104.88 \\
                  &    61.40 &   106.83 &   106.83 &   113.68 &   109.98 &   116.85 \\
                  &    55.20 &   114.24 &   113.94 &   120.14 &   118.69 &   124.36 \\
\midrule
\textbf{MGLUCP11} &   101.00 &    62.40 &    63.38 &    71.44 &    68.85 &    76.60 \\
                  &    72.30 &    97.21 &    95.72 &   102.79 &    99.02 &   105.85 \\
                  &    74.60 &    92.68 &    93.19 &   100.98 &    97.36 &   104.82 \\
                  &    72.50 &    94.99 &    93.47 &   101.68 &    97.20 &   105.17 \\
                  &    75.30 &    93.74 &    94.69 &   101.31 &    97.94 &   104.32 \\
                  &    63.80 &   104.75 &   105.34 &   112.91 &   108.94 &   116.48 \\
                  &    56.50 &   112.69 &   114.09 &   120.10 &   118.56 &   124.05 \\
\midrule
\textbf{PERYTO10} &    50.20 &   119.47 &   120.38 &   127.85 &   121.48 &   129.25 \\
                  &    58.40 &   109.67 &   109.49 &   116.50 &   113.24 &   120.02 \\
\midrule
\textbf{RHAMAH12} &    94.50 &    68.08 &    71.12 &    78.89 &    76.54 &    84.12 \\
                  &    72.20 &    94.24 &    95.24 &   102.90 &    99.59 &   106.93 \\
                  &    71.00 &    97.72 &    98.13 &   105.80 &   101.46 &   109.00 \\
                  &    72.50 &    94.58 &    95.38 &   102.57 &    99.44 &   106.29 \\
                  &    69.80 &    96.16 &    98.22 &   104.89 &   102.69 &   109.26 \\
                  &    17.80 &   156.45 &   157.07 &   160.88 &   158.32 &   162.23 \\
\midrule
\textbf{SUCROS04} &    93.30 &    71.00 &    71.48 &    78.15 &    76.75 &    83.17 \\
                  &    66.00 &   104.25 &   105.35 &   112.26 &   107.71 &   114.58 \\
                  &    73.70 &    93.83 &    93.45 &   101.20 &    96.69 &   104.27 \\
                  &   102.40 &    59.07 &    52.83 &    61.52 &    59.47 &    67.82 \\
                  &    72.80 &    94.07 &    93.67 &   101.42 &    98.29 &   105.67 \\
                  &    82.90 &    83.68 &    87.08 &    94.53 &    90.03 &    97.29 \\
                  &    67.90 &    99.86 &    99.91 &   107.29 &   103.09 &   110.18 \\
                  &    71.80 &    96.45 &    96.26 &   104.24 &    99.89 &   107.60 \\
                  &    73.60 &    93.06 &    93.69 &   100.71 &    97.28 &   103.82 \\
                  &    81.80 &    86.36 &    83.50 &    90.78 &    86.94 &    93.82 \\
                  &    60.00 &   109.19 &   108.80 &   116.04 &   112.42 &   119.75 \\
                  &    61.00 &   107.13 &   106.43 &   112.69 &   110.23 &   116.44 \\
\midrule
\textbf{SULAMD06} &   127.10 &    39.42 &    39.82 &    46.29 &    45.98 &    50.38 \\
                  &   129.50 &    40.30 &    39.63 &    39.48 &    41.09 &    40.82 \\
                  &   117.10 &    53.35 &    54.60 &    57.14 &    56.92 &    59.24 \\
                  &   153.40 &    17.47 &    19.71 &    17.76 &    21.62 &    19.76 \\
                  &   112.30 &    57.40 &    57.54 &    60.61 &    60.71 &    63.52 \\
                  &   129.50 &    40.37 &    40.99 &    39.94 &    42.61 &    41.74 \\
\midrule
\textbf{TRIPHE11} &   126.40 &    43.57 &    42.10 &    43.50 &    44.26 &    45.84 \\
                  &   129.50 &    38.42 &    39.33 &    40.69 &    42.49 &    44.00 \\
                  &   124.50 &    43.66 &    44.94 &    45.48 &    48.04 &    48.44 \\
                  &   125.90 &    42.54 &    41.24 &    42.88 &    43.52 &    45.16 \\
                  &   127.50 &    42.20 &    42.78 &    44.06 &    44.38 &    45.55 \\
                  &   122.30 &    47.99 &    48.91 &    49.72 &    50.63 &    50.89 \\
                  &   130.20 &    39.44 &    38.92 &    40.29 &    40.77 &    42.17 \\
                  &   129.50 &    39.75 &    40.70 &    41.71 &    41.57 &    42.85 \\
                  &   120.90 &    49.53 &    50.31 &    50.81 &    51.74 &    52.26 \\
                  &   125.90 &    44.19 &    44.70 &    45.78 &    45.03 &    46.30 \\
                  &   121.70 &    48.88 &    49.19 &    49.90 &    50.25 &    50.80 \\
                  &   129.50 &    40.00 &    40.67 &    41.69 &    42.28 &    43.72 \\
                  &   129.50 &    41.39 &    39.67 &    40.86 &    40.87 &    42.16 \\
                  &   122.30 &    46.72 &    48.50 &    48.92 &    50.08 &    50.53 \\
                  &   126.90 &    41.65 &    41.66 &    42.58 &    43.79 &    44.91 \\
                  &   126.90 &    43.41 &    42.10 &    43.47 &    43.06 &    44.59 \\
                  &   123.80 &    42.98 &    44.47 &    45.17 &    48.65 &    49.11 \\
                  &   129.80 &    40.11 &    40.07 &    41.54 &    41.05 &    42.45 \\
\bottomrule
\end{longtable}
}

\newpage
\section{Experimental shifts and predicted shieldings for the $^{15}$N benchmark dataset}

{\renewcommand{\arraystretch}{0.5}
\begin{longtable}{lcccccc}
\caption{Experimental isotropic $^{15}$N chemical shifts and predicted shieldings for the atomic sites in the benchmark dataset. Predicted shieldings are given for PBE and PBE0 (PET-MOLS) geometries, computed with GIPAW-DFT and with the \shiftmlthree{} (\shiftmlthreeshort{}) and \shiftmlnew{} (\shiftmlnewshort{}) machine-learning models.} \label{tab:table_15N_ml} \\
\toprule
\multirow{2}{*}{\textbf{Ref. Code}} &
\multirow{2}{*}{\textbf{Exp. shift [ppm]}} &
\multicolumn{3}{c}{\textbf{PBE geometries}} &
\multicolumn{2}{c}{\textbf{PBE0 geometries}} \\
\cmidrule(lr){3-5} \cmidrule(lr){6-7}
& & \textbf{DFT} & \textbf{\shiftmlthreeshort{}} & \textbf{\shiftmlnewshort{}} & \textbf{\shiftmlthreeshort{}} & \textbf{\shiftmlnewshort{}} \\
\midrule
\endfirsthead
\toprule
\multirow{2}{*}{\textbf{Ref. Code}} &
\multirow{2}{*}{\textbf{Exp. shift [ppm]}} &
\multicolumn{3}{c}{\textbf{PBE geometries}} &
\multicolumn{2}{c}{\textbf{PBE0 geometries}} \\
\cmidrule(lr){3-5} \cmidrule(lr){6-7}
& & \textbf{DFT} & \textbf{\shiftmlthreeshort{}} & \textbf{\shiftmlnewshort{}} & \textbf{\shiftmlthreeshort{}} & \textbf{\shiftmlnewshort{}} \\
\midrule
\endhead
\textbf{BITZAF}   &   249.50 &   -60.89 &   -68.97 &   -76.47 &   -62.01 &   -71.92 \\
\midrule
\textbf{GEHHEH}   &   187.40 &     9.64 &    11.16 &     7.84 &    13.25 &     9.28 \\
                  &   261.00 &   -64.05 &   -51.89 &   -59.59 &   -58.21 &   -67.06 \\
\midrule
\textbf{GEHHIL}   &   268.50 &   -87.11 &   -81.30 &   -91.49 &   -74.39 &   -84.34 \\
                  &   261.20 &   -73.65 &   -64.26 &   -67.18 &   -69.87 &   -72.16 \\
\midrule
\textbf{LHISTD02} &   210.80 &   -31.50 &   -30.80 &   -35.62 &   -20.79 &   -25.73 \\
                  &   132.60 &    45.11 &    45.81 &    48.15 &    52.48 &    54.85 \\
\midrule
\textbf{LHISTD13} &   210.60 &   -30.87 &   -30.58 &   -35.26 &   -20.50 &   -25.45 \\
                  &   132.40 &    43.87 &    44.66 &    47.30 &    52.03 &    54.64 \\
\midrule
\textbf{TEJWAG}   &   143.90 &    36.52 &    38.03 &    39.93 &    42.86 &    44.69 \\
\midrule
\textbf{GLYCIN03} &    -6.50 &   197.55 &   195.64 &   202.54 &   199.57 &   206.92 \\
\midrule
\textbf{FUSVAQ01} &   183.20 &    -2.80 &    -0.19 &     0.33 &     8.09 &     9.30 \\
                  &   174.20 &     8.67 &     8.87 &     9.52 &    15.23 &    15.71 \\
                  &   192.20 &   -10.82 &    -9.81 &   -11.35 &    -1.04 &    -2.35 \\
                  &   120.20 &    55.15 &    56.21 &    61.13 &    62.60 &    67.27 \\
                  &    50.20 &   131.61 &   131.18 &   138.63 &   136.42 &   143.25 \\
\midrule
\textbf{CYTSIN}   &   110.20 &    65.66 &    65.71 &    74.76 &    74.87 &    83.87 \\
                  &   165.20 &    15.50 &    17.78 &    22.37 &    25.14 &    30.28 \\
                  &    54.20 &   131.20 &   133.13 &   139.80 &   137.10 &   143.64 \\
\midrule
\textbf{THYMIN01} &    90.20 &    85.14 &    83.12 &    91.93 &    92.32 &   100.74 \\
                  &   119.20 &    59.80 &    58.68 &    66.44 &    65.91 &    73.63 \\
\midrule
\textbf{CIMETD}   &   130.50 &    46.58 &    45.40 &    48.17 &    52.01 &    54.25 \\
                  &   213.10 &   -32.07 &   -32.76 &   -34.90 &   -24.66 &   -25.45 \\
                  &    56.60 &   122.91 &   125.63 &   131.98 &   133.37 &   139.45 \\
                  &    43.50 &   138.52 &   140.32 &   147.66 &   149.20 &   156.02 \\
                  &    45.80 &   138.05 &   139.41 &   139.58 &   146.13 &   144.57 \\
                  &   149.90 &    34.88 &    34.30 &    37.45 &    36.37 &    38.23 \\
\midrule
\textbf{BAPLOT01} &   114.70 &    57.48 &    59.89 &    67.85 &    68.39 &    75.90 \\
                  &    72.70 &   100.91 &   101.92 &   110.19 &   109.49 &   117.81 \\
                  &   122.70 &    53.16 &    53.71 &    55.94 &    63.58 &    66.32 \\
                  &   178.70 &     1.47 &     1.61 &     1.76 &     8.72 &     9.10 \\
\midrule
\textbf{LTYRHC10} &     8.00 &   167.71 &   166.90 &   175.27 &   184.36 &   192.27 \\
\midrule
\textbf{CYSCLM}   &     1.50 &   186.09 &   186.34 &   193.96 &   190.64 &   198.19 \\
\midrule
\textbf{URACIL}   &    96.20 &    75.34 &    75.63 &    84.01 &    84.10 &    92.61 \\
                  &   120.20 &    59.55 &    59.00 &    66.51 &    65.38 &    73.07 \\
\bottomrule
\end{longtable}
}
\newpage
\section{Detailed prediction accuracies of \shiftmlnew{} and \shiftmlthree{} on isotropic chemical shielding $\sigma_{\text{iso}}$ DFT reference computations.} \label{sec:detailed_preds_sML5}

\begin{figure}[h!]
    \centering
    \includegraphics[width=\linewidth]{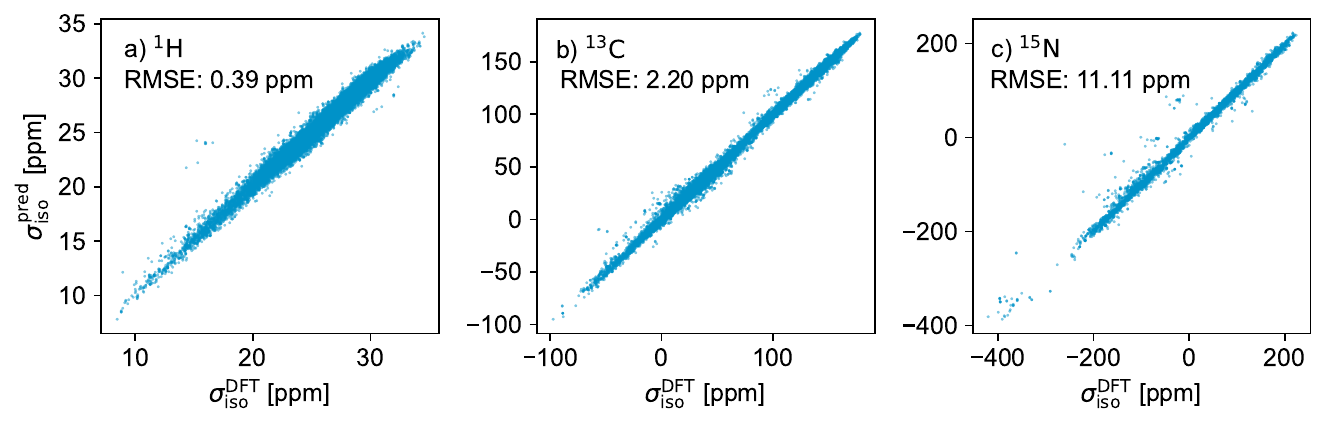}
    \caption{Parity plots of DFT reference isotropic shieldings $\sigma_{\text{iso}}^{\text{DFT}}$ against \shiftmlnew{} predicted isotropic~shieldings~$\sigma_{\text{iso}}^{\text{pred}}$. Subplots show $^{1}$H chemical shieldings (a), $^{13}$C chemical shieldings (b), and $^{15}$N chemical shieldings (c). }
    \label{fig:parity_iso}
\end{figure}

In Table~\ref{tab:complete_metrics} we compare prediction accuracies of isotropic chemical shieldings of \shiftmlnew{} and \shiftmlthree{} against their respective DFT targets on the hold-out test set, for which both GIPAW-PBE labels and molecular corrected labels are available. We evaluate both models on all environments present in the test set, rather than on the farthest-point-sampling subset used to evaluate ShiftML2. We note that our outlier protocol removed a disproportionately large number of structures containing divalent metal cations. The original ShiftML2/\shiftmlthree{} test set contained only a small number of metal cation environments (5-7), and due to the outlier removal protocol, calcium ions are not represented in the validation and test set. For magnesium and sodium we present accuracy numbers computed on the validation set. In Figure~\ref{fig:parity_iso} we show parity plots between \shiftmlnew{} predictions and DFT references. 
The molecular corrections also slightly change the distribution of the species-wise isotropic chemical shieldings and hence, a larger absolute RMSE in Table~\ref{tab:complete_metrics} does not necessarily result in a larger normalized RMSE. 

\begin{table}[h!]
  \centering
  \caption{Predicted accuracy of $\sigma_{\text{iso}}$ of \shiftmlthree{} (\shiftmlthreeshort{}) and \shiftmlnew{} (\shiftmlnewshort{}) on the hold-out test set. In Brackets $()^*$ we report metrics computed on the validation set. Standardized RMSEs (RMSE[\%]) are defined as the nucleus-wise RMSE divided by the nucleus-wise standard deviation of the isotropic shift. \shiftmlthree{} is evaluated against GIPAW-PBE reference calculations and \shiftmlnew{} against PBE0 molecular-corrected GIPAW-PBE calculations.}
    \resizebox{\textwidth}{!}{\begin{tabular}{c|@{\hskip 2mm}c@{\hskip 2mm}c@{\hskip 2mm}c@{\hskip 2mm}|@{\hskip 6mm}cc@{\hskip 6mm}cc@{\hskip 6mm}cc@{\hskip 6mm}cc}
    \toprule
    Nucleus & \hspace{4pt} N$_{\text{train}}$\hspace{4pt}  &  N$_{\text{validation}}$  & \hspace{4pt}N$_{\text{test}}$ \hspace{4pt} & \multicolumn{2}{c@{\hskip 6mm}}{MAE [ppm]} & \multicolumn{2}{c@{\hskip 6mm}}{RMSE [ppm]} & \multicolumn{2}{c@{\hskip 6mm}}{RMSE [\%]} & \multicolumn{2}{c}{R$^2$} \\
    \hline
          &       &       &       & \shiftmlthreeshort{}  & \shiftmlnewshort{}  & \shiftmlthreeshort{}  & \shiftmlnewshort{}  & \shiftmlthreeshort{}  & \shiftmlnewshort{} & \shiftmlthreeshort{}  & \shiftmlnewshort{} \\
\hline
    $^{1}$H & 487,772& 27,242 & 67,964 & 0.31  & 0.29  & 0.42  & 0.39  & 13.2    & 12.1    & 0.98  & 0.99 \\
    $^{13}$C & 392,547& 24,978 & 62,134 & 1.50  & 1.46  & 2.24  & 2.20  & 4.5    & 4.2    & 1.00  & 1.00 \\
    $^{15}$N & 107,576& 2,750 & 6,656 & 4.74  & 5.04  & 10.20  & 11.11  & 9.5    & 9.5    & 0.99  & 0.99 \\
    $^{17}$O & 118,444& 4,902 & 11,506 & 6.91  & 6.95  & 10.45  & 11.23  & 5.9    & 6.0    & 1.00  & 1.00 \\
    $^{19}$F & 31,415& 408 & 804 & 3.87  & 3.41  & 5.55  & 4.71  & 17.6    & 14.4    & 0.97  & 0.98 \\
    $^{33}$S & 26,055& 580 & 1,562 & 13.20  & 13.87  & 21.86  & 24.38  & 14.9    & 15.9    & 0.98  & 0.97 \\
    $^{31}$P & 5,758& 132 & 196 & 8.47  & 8.31  & 18.68  & 16.30  & 57.3    & 48.1    & 0.67  & 0.77 \\
    $^{35}$Cl & 22,900& 276 & 836 & 10.23  & 9.62  & 14.55  & 13.85  & 12.7    & 13.1    & 0.98  & 0.98 \\
    $^{23}$Na & 900& 16 & 0 & (2.13)$^{*}$  & (1.38)$^{*}$  & (2.35)$^{*}$  & (1.56)$^{*}$  & (43.3)$^{*}$    & (28.7)$^{*}$    & (0.81)$^{*}$  & (0.92)$^{*}$ \\
    $^{43}$Ca & 22& 0 & 0 & -  & -  & -  & -  & -    & -    & -  & - \\
    $^{25}$Mg & 30& 2 & 0 & (2.07)$^{*}$  & (2.79)$^{*}$  & (2.08)$^{*}$  & (2.80)$^{*}$  & (-)$^{*}$    & (-)$^{*}$    & (-)$^{*}$  & (-)$^{*}$ \\
    $^{39}$K & 609& 4 & 8 & 4.13  & 3.42  & 4.46  & 3.98  & 55.8    & 49.8    & 0.69  & 0.75 \\
    \bottomrule
    \hline
    \end{tabular}%
    }
  \label{tab:complete_metrics}%
\end{table}

\newpage
\section{Detailed prediction accuracy of \shiftmlnew{} and \shiftmlthree{} on chemical shielding tensor principal components $\sigma_{\text{PAS}}$ DFT reference computations. }

\begin{figure}[h!]
    \centering
    \includegraphics[width=\linewidth]{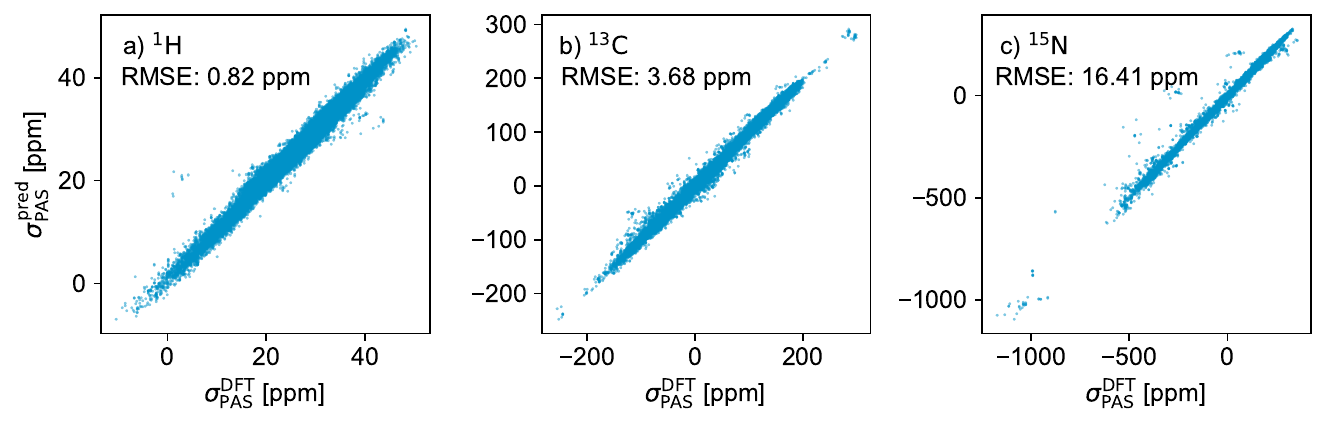}
    \caption{Parity plots of DFT reference chemical shielding tensor principal components $\sigma_{\text{PAS}}$, against \shiftmlnew{} predicted values. Subplots show $^{1}$H principal components (a), $^{13}$C principal components (b), and $^{15}$N principal components (c).}
    \label{fig:parity_psa}
\end{figure}

In Table~\ref{tab:complete_metrics_PAS} we compare prediction accuracies of chemical shielding tensor principal component $\sigma_{\text{PAS}}$ predictions of \shiftmlnew{} against the DFT reference values. In Figure~\ref{fig:parity_psa} we show parity plots between DFT reference chemical shielding tensor principal components $\sigma_{\text{PAS}}$ and \shiftmlnew{} predicted values.

\begin{table}[h!]
  \centering
  \caption{Predicted accuracy of $\sigma_{\text{PAS}}$ of \shiftmlthree{} (\shiftmlthreeshort{}) and \shiftmlnew{} (\shiftmlnewshort{}) on the hold-out test set. In Brackets $()^*$ we report metrics computed on the validation set. Standardized RMSEs (RMSE[\%]) are defined as the nucleus-wise RMSE divided by the nucleus-wise standard deviation of the principal components $\sigma_{\text{PAS}}$. \shiftmlthree{} is evaluated against GIPAW-PBE reference calculations and \shiftmlnew{} against PBE0 molecular-corrected GIPAW-PBE calculations.}
    \resizebox{0.8\textwidth}{!}{\begin{tabular}{c|@{\hskip 6mm}cc@{\hskip 6mm}cc@{\hskip 6mm}cc@{\hskip 6mm}cc}
    \toprule
    Nucleus & \multicolumn{2}{c@{\hskip 6mm}}{MAE [ppm]} & \multicolumn{2}{c@{\hskip 6mm}}{RMSE [ppm]} & \multicolumn{2}{c@{\hskip 6mm}}{RMSE [\%]} & \multicolumn{2}{c}{R$^2$} \\
    \hline
          &    \shiftmlthreeshort{}  & \shiftmlnewshort{}  & \shiftmlthreeshort{}  & \shiftmlnewshort{}  & \shiftmlthreeshort{}  & \shiftmlnewshort{} & \shiftmlthreeshort{}  & \shiftmlnewshort{} \\
\hline
    $^{1}$H &0.64  & 0.61  & 0.85  & 0.82  & 15.2    & 14.1    & 0.98  & 0.98 \\
    $^{13}$C &2.54  & 2.44  & 3.77  & 3.68  & 4.3    & 4.1    & 1.00  & 1.00 \\
    $^{15}$N &6.93  & 7.05  & 15.67  & 16.41  & 9.0    & 8.8    & 0.99  & 0.99 \\
    $^{17}$O &9.61  & 9.44  & 15.63  & 16.25  & 5.6    & 5.6    & 1.00  & 1.00 \\
    $^{19}$F &5.20  & 4.66  & 7.54  & 6.83  & 8.7    & 7.9    & 0.99  & 0.99 \\
    $^{33}$S &20.20  & 20.06  & 34.38  & 36.31  & 15.4    & 15.9    & 0.98  & 0.97 \\
    $^{31}$P &12.64  & 12.16  & 23.98  & 21.62  & 28.7    & 26.9    & 0.92  & 0.93 \\
    $^{35}$Cl &13.65  & 12.97  & 19.92  & 18.83  & 8.1    & 7.9    & 0.99  & 0.99 \\
    $^{23}$Na &(2.95)$^{*}$  & (2.79)$^{*}$  & (3.62)$^{*}$  & (3.37)$^{*}$  & (29.7)$^{*}$    & (27.7)$^{*}$    & (0.91)$^{*}$  & (0.92)$^{*}$ \\
    $^{43}$Ca &-  & -  & -  & -  & -    & -    & -  & - \\
    $^{25}$Mg &(5.52)$^{*}$  & (2.79)$^{*}$  & (6.91)$^{*}$  & (2.92)$^{*}$  & (25.9)$^{*}$    & (10.9)$^{*}$    & (0.93)$^{*}$  & (0.99)$^{*}$ \\
    $^{39}$K &4.33  & 3.84  & 5.07  & 4.95  & 37.7    & 36.7    & 0.86  & 0.87 \\
    \bottomrule
    \hline
    \end{tabular}%
    }
  \label{tab:complete_metrics_PAS}%
\end{table}%

\newpage
\section{Detailed prediction accuracy of \shiftmlnew{}{} and \shiftmlthree{} on chemical shielding tensor components $\sigma_{ij}$ DFT reference computations.}

\begin{figure}[h!]
    \centering
    \includegraphics[width=\linewidth]{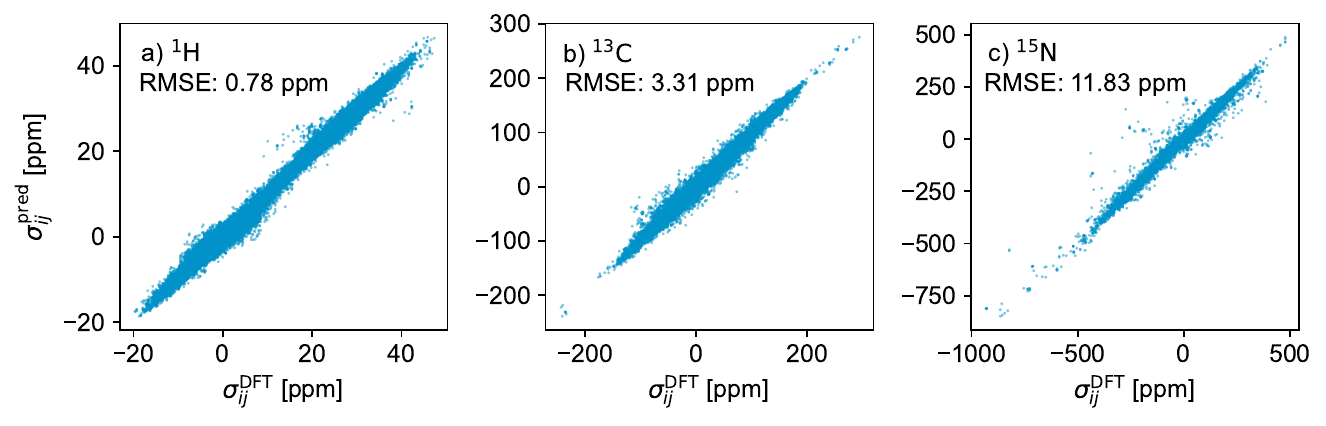}
    \caption{Parity plots of DFT reference chemical shielding tensor components $\sigma_{ij}$ against \shiftmlnew{} predicted values. Subplots show $^1$H shielding tensor components (a), $^{13}$C shielding tensor components (b) and $^{15}$N shielding tensor components (c).}
    \label{fig:parity_tensorcomponents}
\end{figure}

In Table~\ref{tab:complete_metrics_tensor_components} we report prediction accuracies of \shiftmlnew{} on shielding tensor components $\sigma_{ij}$ against DFT reference values. In Figure~\ref{fig:parity_tensorcomponents} we show parity plots of DFT reference shielding tensor components $\sigma_{ij}$ against \shiftmlnew{} predicted values.

\begin{table}[h!]
  \centering
  \caption{Predicted accuracy of $\sigma_{ij}$ of \shiftmlthree{} (\shiftmlthreeshort{}) and \shiftmlnew{} (\shiftmlnewshort{}) on the hold-out test set. In Brackets $()^*$ we report metrics computed on the validation set. Standardized RMSEs (RMSE[\%]) are defined as the nucleus-wise RMSE divided by the nucleus-wise standard deviation of the tensor components $\sigma_{ij}$. \shiftmlthree{} is evaluated against GIPAW-PBE reference calculations and \shiftmlnew{} against PBE0 molecular-corrected GIPAW-PBE calculations.}
    \resizebox{0.8\textwidth}{!}{\begin{tabular}{c|@{\hskip 6mm}cc@{\hskip 6mm}cc@{\hskip 6mm}cc@{\hskip 6mm}cc}
    \toprule
    Nucleus & \multicolumn{2}{c@{\hskip 6mm}}{MAE [ppm]} & \multicolumn{2}{c@{\hskip 6mm}}{RMSE [ppm]} & \multicolumn{2}{c@{\hskip 6mm}}{RMSE [\%]} & \multicolumn{2}{c}{R$^2$} \\
    \hline
          &    \shiftmlthreeshort{}  & \shiftmlnewshort{}  & \shiftmlthreeshort{}  & \shiftmlnewshort{}  & \shiftmlthreeshort{}  & \shiftmlnewshort{} & \shiftmlthreeshort{}  & \shiftmlnewshort{} \\
\hline
    $^{1}$H &0.61  & 0.59  & 0.81  & 0.78  & 6.4    & 6.1    & 1.00  & 1.00 \\
    $^{13}$C &2.30  & 2.22  & 3.43  & 3.31  & 5.9    & 5.5    & 1.00  & 1.00 \\
    $^{15}$N &5.71  & 5.69  & 11.53  & 11.83  & 11.4    & 10.9    & 0.99  & 0.99 \\
    $^{17}$O &8.00  & 7.71  & 12.79  & 12.61  & 7.9    & 7.5    & 0.99  & 0.99 \\
    $^{19}$F &4.12  & 3.80  & 6.08  & 5.55  & 4.9    & 4.2    & 1.00  & 1.00 \\
    $^{33}$S &17.30  & 16.77  & 28.37  & 28.69  & 16.8    & 16.2    & 0.97  & 0.97 \\
    $^{31}$P &10.28  & 9.86  & 19.70  & 19.42  & 14.6    & 13.7    & 0.98  & 0.98 \\
    $^{35}$Cl &10.85  & 10.12  & 16.32  & 15.01  & 5.1    & 4.5    & 1.00  & 1.00 \\
    $^{23}$Na &(1.95)$^{*}$  & (1.92)$^{*}$  & (2.59)$^{*}$  & (2.55)$^{*}$  & (1.0)$^{*}$    & (1.0)$^{*}$    & (1.00)$^{*}$  & (1.00)$^{*}$ \\
    $^{43}$Ca &-  & -  & -  & -  & -    & -    & -  & - \\
    $^{25}$Mg &(2.78)$^{*}$  & (3.23)$^{*}$  & (4.25)$^{*}$  & (3.81)$^{*}$  & (1.5)$^{*}$    & (1.4)$^{*}$    & (1.00)$^{*}$  & (1.00)$^{*}$ \\
    $^{39}$K &3.23  & 3.29  & 4.15  & 4.27  & 0.7    & 0.7    & 1.00  & 1.00 \\
    \bottomrule
    \hline
    \end{tabular}%
    }
  \label{tab:complete_metrics_tensor_components}
\end{table}

\pagebreak
\newpage

\section{Evaluating \shiftmlnew{} and \shiftmlthree{} on $^{15}$N shielding tensor components on an experimental benchmark compared to DFT-PBE0 shielding predictions}
\label{sec:15N_PSA_benchmark}

\begin{figure}[h!]
    \centering
    \includegraphics[width=0.85\linewidth]{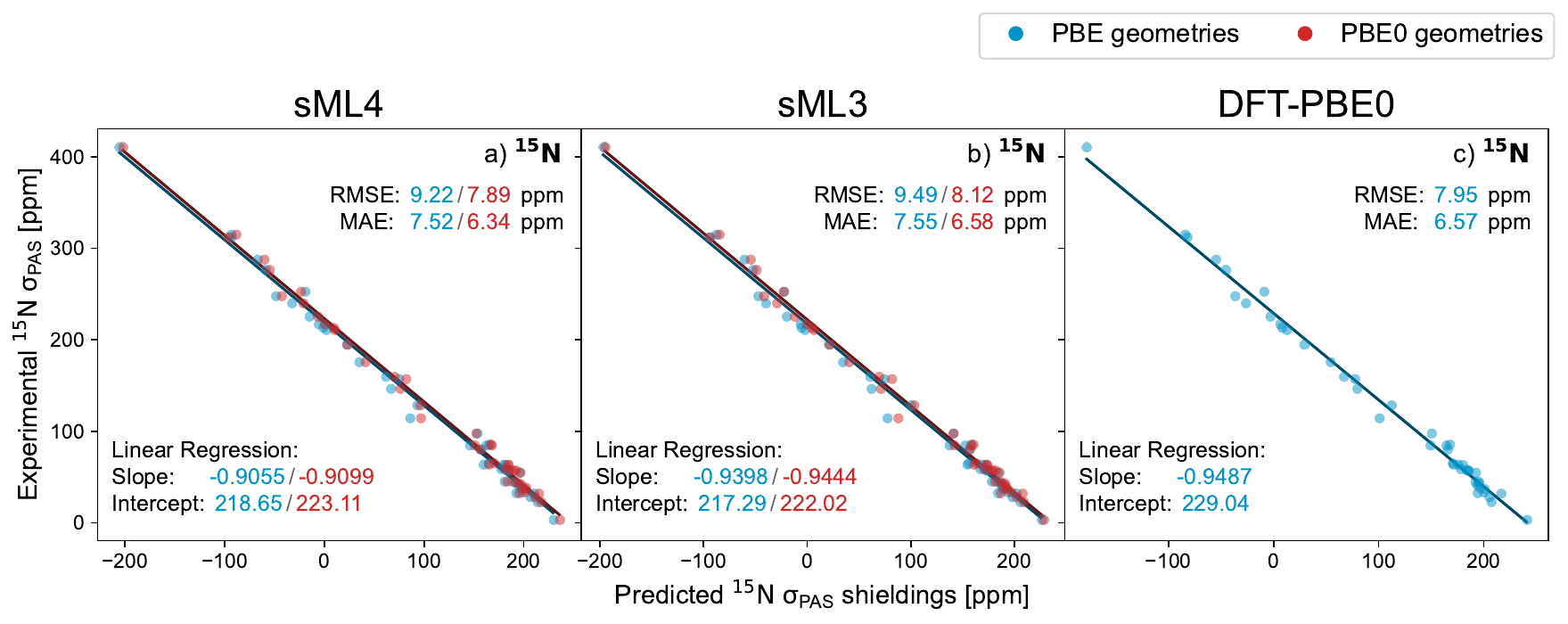}
    \caption{Parity plots between  \shiftmlnew{} (a), \shiftmlthree{} (b), (c) DFT-PBE0 fragment computation $^{15}$ N predicted principal components $\sigma_{\text{PAS}}$ of the shielding tensor and experimental reference values taken from Reference~\citenum{hartmanAccurate13C15N2018}. Subplot (c) shows DFT-PBE0 fragment reference calculations taken directly from Ref.~\citenum{hartmanAccurate13C15N2018}. The black line shows the linear relation, converting computational principal components into experimental components. The slope and intercept for each method are indicated in the respective subplot. In blue and red, PBE and PBE0 (PET-MOLS) geometries were taken as starting points for shielding calculations.}
    \label{fig:parity_delta}
\end{figure}

\section{Evaluating differences of \shiftmlnew{} and \shiftmlthree{} predictions on $^{1}$H, $^{13}$C and $^{15}$N experimental benchmarks}

\begin{figure}[h!]
    \centering
    \includegraphics[width=0.85\linewidth]{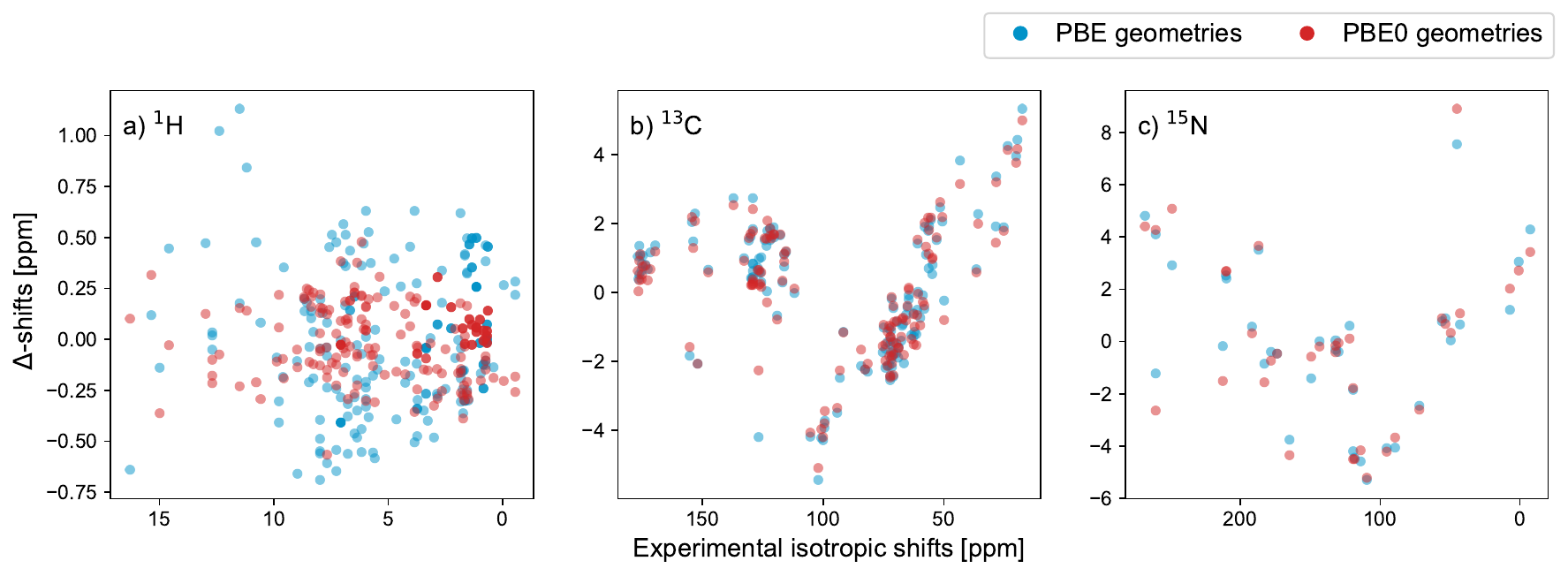}
    \caption{Delta shifts correction evaluated on the $^{1}$H, $^{13}$C and $^{15}$N isotropic chemical shift benchmarks as a function of the experimental chemical shifts. The delta ($\Delta$-shifts) is defined as the difference \shiftmlnew{} - \shiftmlthree{} predicted shifts (i.e. removing constant offsets and skew through linear regression). In blue and red, PBE and PBE0 (PET-MOLS) geometries were taken as starting points for shielding calculations.}
    \label{fig:parity_tensorcomponents_15N_bench}
\end{figure}

\newpage
\section{Evaluating \shiftmlnew{} uncertainty estimates}
\shiftmlnew{} is based on a committee of nanoPET models. The mean $\overline{y}(A_i)$ of the individual model predictions $y^{k}(A_i)$ for a given environment $A_i$, is the \shiftmlnew{} prediction. We can interpret the spread of the committee predictions $ \sigma^{2}_{\text{pred}}(A_i)$ as a measure of the uncertainty of the \shiftmlnew{} predictions:

\begin{equation}
    \label{eq:UQ}
    \sigma^{2}_{\text{pred}}(A_i) = \frac{1}{N_{\text{ens}}-1} \sum_{k=1}^{N_{\text{ens}}} \big[y^{k}(A_i) - \overline{y}(A_i) \big]^2
\end{equation}

Committee based uncertainty estimators tend to be globally miscalibrated, meaning that their uncertainty estimates are globally over- or underconfident.~\cite{guoCalibrationModernNeural2017a, kuleshovAccurateUncertaintiesDeep2018c} This miscalibration can be corrected by computing an input-independent rescaling factor $\alpha$ on a hold-out calibration set, as in Equation~\ref{eq:calibration}.~\cite{musi+19jctc, imbalzanoUncertaintyEstimationMolecular2021} Here $z_i$ are the prediction errors $|\overline{y}(A_i) - y_{\text{ref}}|$ of the ensemble predictions against DFT reference labels, and $\sigma_{i,\text{pred}}^2$ the committee uncertainty. The scaled, or calibrated uncertainty estimates are then obtained by scaling the committee spread $\sigma_{\text{scaled}} = \alpha\sigma_{\text{pred}}$. We use one calibration factor per chemical species in the validation database. We omit calcium, given that there are no calcium environments in the validation dataset. In Table~\ref{tab:alpha} we list the species wise $\alpha$ calibration factors according to Eq.~\ref{eq:calibration}.

\begin{equation}
    \alpha = \sqrt{\frac{1}{N_{\text{samples}}}\sum_{i=1}^{N_{\text{samples}}}\frac{z_i^2}{\sigma_{i,\text{pred}}^2}}
    \label{eq:calibration}
\end{equation}

\begin{figure}[h]
    \centering
    \includegraphics[width=0.9\columnwidth]{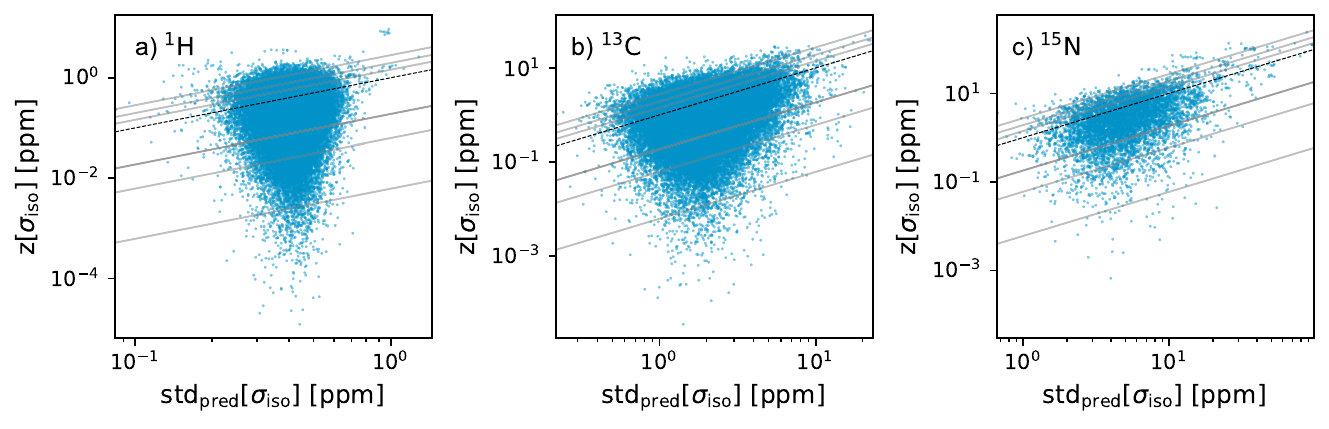}
    \caption{\shiftmlnew{} predicted and calibrated uncertainty estimates of the isotropic shieldings $\text{std}_{\text{pred}}[\sigma_{\text{iso}}]$, and prediction errors against DFT labels $\text{z}[\sigma_{\text{iso}}]$. Subplots show scatter plots for $^1$H predictions (a), $^{13}$C predictions(b) and $^{15}$N predictions (c). The black dashed line indicates the unit line. Grey lines indicate quantiles of the expected log-folded-normal distribution. For example, for any given predicted uncertainty estimate $\text{std}_{\text{pred}}[\sigma_{\text{iso}}]$ 90\% of the samples in the scatter plot should fall between the two outermost grey lines, that correspond to quantile lines of 5\% and 95\% of the expected error distribution.   }
    \label{fig:si_uq_scatter}
\end{figure}

In Figure~\ref{fig:si_uq_scatter} we show double logarithmic scatter plots in which we plot \shiftmlnew{} predicted uncertainties against prediction errors against DFT reference labels. Overall, the \shiftmlnew{} committee uncertainties require only moderate calibration, with values of $\alpha$ ranging from  0.42 to 2.12. 

Note that committee uncertainties can also be used to propagate uncertainties through arbitrarily complex workflows, such as entire NMR crystallography workflows, or evaluating uncertainties on tensor-derived quantities such as the shielding tensor principal components.

\begin{table}[h!]
  \centering
  \caption{Per-nucleus uncertainty calibration factors $\alpha$ of \shiftmlnew{}, determined on the validation set}
    \resizebox{0.5\textwidth}{!}{\begin{tabular}{l|ccccccccccc}
    \toprule
         & $^{1}\text{H}$ & $^{13}\text{C}$ & $^{15}\text{N}$ & $^{17}\text{O}$ & $^{19}\text{F}$ & $^{33}\text{S}$ & $^{31}\text{P}$ & $^{35}\text{Cl}$ & $^{23}\text{Na}$ & $^{25}\text{Mg}$ & $^{39}\text{K}$ \\
    \midrule
        $\alpha$ & 0.42 & 1.29 & 1.44 & 1.54 & 1.25 & 2.12 & 1.00 & 1.52 & 0.65 & 0.82 & 0.87 \\
    \bottomrule
    \end{tabular}}
  \label{tab:alpha}%
\end{table}%

\newpage
\section{Reference ORCA computations}
\label{sec:ORCA_inputs}
In this section we list the ORCA input files, required to reproduce the molecular corrections, computed in this work.
First, we list the input file of the PBE reference computation (as written by the ORCA python API):
\begin{verbatim}
!pbe
!sp
!nmr
!rijcosx
!cc-pvtz
!autoaux
!cpcm(dichloromethane)
!defgrid3
!tightscf
%
    nprocs 12
end

%
    jsonpropfile true
    jsongbwfile true
end
\end{verbatim}

Second we list the input file of the PBE0 reference computation (as written by the ORCA python API):

\begin{verbatim}
!pbe0
!sp
!nmr
!rijcosx
!cc-pvtz
!autoaux
!cpcm(dichloromethane)
!defgrid3
!tightscf
%
    nprocs 12
end

%
    jsonpropfile true
    jsongbwfile true
end
\end{verbatim}

cc-pVTZ~\cite{dunningGaussianBasisSets1989, woonGaussianBasisSets1993} is not parametrized for potassium. To stay consistent with the original work from Ramos et al.~\cite{ramosInterplayDensityFunctional2024} we still use cc-pVTZ throughout, and set the correction for the isolated potassium atoms produced by the cutting procedure to zero, in line with the near-zero corrections found for the other metal ions. This affects only the correction term, so \shiftmlnew{} can still be evaluated meaningfully on potassium-containing structures.

\newpage
\section{Implementation}

\shiftmlnew{} is available via the ShiftML PyPI package.
The package interfaces the ShiftML models via a convenient calculator implementation to the Atomic Simulation Environment (ASE).~\cite{hjorthlarsenAtomicSimulationEnvironment2017}

\begin{verbatim}
from ase.build import bulk
from shiftml.ase import ShiftML
import numpy as np

frame = bulk("C", "diamond", a=3.566)
model = ShiftML("ShiftML4")

Ypred = model.get_cs_tensor(frame)

# Get the committee predictions:
cs_committee_iso = model.get_cs_iso_ensemble(frame)

# Compute uncertainty estimates for the isotropic chemical shieldings
cs_iso_uncertainty = np.std(cs_committee_iso, axis=1, ddof=1)

\end{verbatim}

The package can be easily installed from PyPI via:

\begin{verbatim} pip install shiftml \end{verbatim}


\begin{thebibliography}{109}%
\makeatletter
\providecommand \@ifxundefined [1]{%
 \@ifx{#1\undefined}
}%
\providecommand \@ifnum [1]{%
 \ifnum #1\expandafter \@firstoftwo
 \else \expandafter \@secondoftwo
 \fi
}%
\providecommand \@ifx [1]{%
 \ifx #1\expandafter \@firstoftwo
 \else \expandafter \@secondoftwo
 \fi
}%
\providecommand \natexlab [1]{#1}%
\providecommand \enquote  [1]{``#1''}%
\providecommand \bibnamefont  [1]{#1}%
\providecommand \bibfnamefont [1]{#1}%
\providecommand \citenamefont [1]{#1}%
\providecommand \href@noop [0]{\@secondoftwo}%
\providecommand \href [0]{\begingroup \@sanitize@url \@href}%
\providecommand \@href[1]{\@@startlink{#1}\@@href}%
\providecommand \@@href[1]{\endgroup#1\@@endlink}%
\providecommand \@sanitize@url [0]{\catcode `\\12\catcode `\$12\catcode
  `\&12\catcode `\#12\catcode `\^12\catcode `\_12\catcode `\%12\relax}%
\providecommand \@@startlink[1]{}%
\providecommand \@@endlink[0]{}%
\providecommand \url  [0]{\begingroup\@sanitize@url \@url }%
\providecommand \@url [1]{\endgroup\@href {#1}{\urlprefix }}%
\providecommand \urlprefix  [0]{URL }%
\providecommand \Eprint [0]{\href }%
\providecommand \doibase [0]{https://doi.org/}%
\providecommand \selectlanguage [0]{\@gobble}%
\providecommand \bibinfo  [0]{\@secondoftwo}%
\providecommand \bibfield  [0]{\@secondoftwo}%
\providecommand \translation [1]{[#1]}%
\providecommand \BibitemOpen [0]{}%
\providecommand \bibitemStop [0]{}%
\providecommand \bibitemNoStop [0]{.\EOS\space}%
\providecommand \EOS [0]{\spacefactor3000\relax}%
\providecommand \BibitemShut  [1]{\csname bibitem#1\endcsname}%
\let\auto@bib@innerbib\@empty
\bibitem [{\citenamefont {Emsley}(2025)}]{emsleySpiersMemorialLecture2025b}%
  \BibitemOpen
  \bibfield  {author} {\bibinfo {author} {\bibfnamefont {L.}~\bibnamefont
  {Emsley}},\ }\bibfield  {title} {\bibinfo {title} {Spiers {{Memorial
  Lecture}}: {{NMR}} crystallography},\ }\href
  {https://doi.org/10.1039/d4fd00151f} {\bibfield  {journal} {\bibinfo
  {journal} {Faraday Discussions}\ }\textbf {\bibinfo {volume} {255}},\
  \bibinfo {pages} {9} (\bibinfo {year} {2025})}\BibitemShut {NoStop}%
\bibitem [{\citenamefont {Hofstetter}\ and\ \citenamefont
  {Emsley}(2017)}]{hofstetterPositionalVarianceNMR2017}%
  \BibitemOpen
  \bibfield  {author} {\bibinfo {author} {\bibfnamefont {A.}~\bibnamefont
  {Hofstetter}}\ and\ \bibinfo {author} {\bibfnamefont {L.}~\bibnamefont
  {Emsley}},\ }\bibfield  {title} {\bibinfo {title} {Positional {{Variance}} in
  {{NMR Crystallography}}},\ }\href {https://doi.org/10.1021/jacs.6b12705}
  {\bibfield  {journal} {\bibinfo  {journal} {Journal of the American Chemical
  Society}\ }\textbf {\bibinfo {volume} {139}},\ \bibinfo {pages} {2573}
  (\bibinfo {year} {2017})}\BibitemShut {NoStop}%
\bibitem [{\citenamefont {Engel}\ \emph {et~al.}(2019)\citenamefont {Engel},
  \citenamefont {Anelli}, \citenamefont {Hofstetter}, \citenamefont {Paruzzo},
  \citenamefont {Emsley},\ and\ \citenamefont
  {Ceriotti}}]{engel_bayesian_2019}%
  \BibitemOpen
  \bibfield  {author} {\bibinfo {author} {\bibfnamefont {E.~A.}\ \bibnamefont
  {Engel}}, \bibinfo {author} {\bibfnamefont {A.}~\bibnamefont {Anelli}},
  \bibinfo {author} {\bibfnamefont {A.}~\bibnamefont {Hofstetter}}, \bibinfo
  {author} {\bibfnamefont {F.}~\bibnamefont {Paruzzo}}, \bibinfo {author}
  {\bibfnamefont {L.}~\bibnamefont {Emsley}},\ and\ \bibinfo {author}
  {\bibfnamefont {M.}~\bibnamefont {Ceriotti}},\ }\bibfield  {title}
  {{\selectlanguage {en}\bibinfo {title} {A {Bayesian} approach to {NMR}
  crystal structure determination}},\ }\href
  {https://doi.org/10.1039/C9CP04489B} {\bibfield  {journal} {\bibinfo
  {journal} {Physical Chemistry Chemical Physics}\ }\textbf {\bibinfo {volume}
  {21}},\ \bibinfo {pages} {23385} (\bibinfo {year} {2019})},\ \bibinfo {note}
  {arXiv:1909.00870 [physics]}\BibitemShut {NoStop}%
\bibitem [{\citenamefont {Mueller}(2025)}]{muellerUniformChisquaredModel2025}%
  \BibitemOpen
  \bibfield  {author} {\bibinfo {author} {\bibfnamefont {L.~J.}\ \bibnamefont
  {Mueller}},\ }\bibfield  {title} {\bibinfo {title} {Uniform chi-squared model
  probabilities in {{NMR}} crystallography},\ }\href
  {https://doi.org/10.1039/D4FD00114A} {\bibfield  {journal} {\bibinfo
  {journal} {Faraday Discussions}\ }\textbf {\bibinfo {volume} {255}},\
  \bibinfo {pages} {203} (\bibinfo {year} {2025})}\BibitemShut {NoStop}%
\bibitem [{\citenamefont {Pickard}\ and\ \citenamefont
  {Mauri}(2001)}]{pickardAllelectronMagneticResponse2001b}%
  \BibitemOpen
  \bibfield  {author} {\bibinfo {author} {\bibfnamefont {C.~J.}\ \bibnamefont
  {Pickard}}\ and\ \bibinfo {author} {\bibfnamefont {F.}~\bibnamefont
  {Mauri}},\ }\bibfield  {title} {\bibinfo {title} {All-electron magnetic
  response with pseudopotentials: {{NMR}} chemical shifts},\ }\href
  {https://doi.org/10.1103/PhysRevB.63.245101} {\bibfield  {journal} {\bibinfo
  {journal} {Physical Review B}\ }\textbf {\bibinfo {volume} {63}},\ \bibinfo
  {pages} {245101} (\bibinfo {year} {2001})}\BibitemShut {NoStop}%
\bibitem [{\citenamefont {Mauri}\ \emph {et~al.}(1996)\citenamefont {Mauri},
  \citenamefont {Pfrommer},\ and\ \citenamefont
  {Louie}}]{mauriInitioTheoryNMR1996}%
  \BibitemOpen
  \bibfield  {author} {\bibinfo {author} {\bibfnamefont {F.}~\bibnamefont
  {Mauri}}, \bibinfo {author} {\bibfnamefont {B.~G.}\ \bibnamefont
  {Pfrommer}},\ and\ \bibinfo {author} {\bibfnamefont {S.~G.}\ \bibnamefont
  {Louie}},\ }\bibfield  {title} {\bibinfo {title} {{\emph{Ab
  }}{{{\emph{Initio}}}} {{Theory}} of {{NMR Chemical Shifts}} in {{Solids}} and
  {{Liquids}}},\ }\href {https://doi.org/10.1103/PhysRevLett.77.5300}
  {\bibfield  {journal} {\bibinfo  {journal} {Physical Review Letters}\
  }\textbf {\bibinfo {volume} {77}},\ \bibinfo {pages} {5300} (\bibinfo {year}
  {1996})}\BibitemShut {NoStop}%
\bibitem [{\citenamefont {Bonhomme}\ \emph {et~al.}(2012)\citenamefont
  {Bonhomme}, \citenamefont {Gervais}, \citenamefont {Babonneau}, \citenamefont
  {Coelho}, \citenamefont {Pourpoint}, \citenamefont {Aza{\"i}s}, \citenamefont
  {Ashbrook}, \citenamefont {Griffin}, \citenamefont {Yates}, \citenamefont
  {Mauri},\ and\ \citenamefont
  {Pickard}}]{bonhommeFirstPrinciplesCalculationNMR2012}%
  \BibitemOpen
  \bibfield  {author} {\bibinfo {author} {\bibfnamefont {C.}~\bibnamefont
  {Bonhomme}}, \bibinfo {author} {\bibfnamefont {C.}~\bibnamefont {Gervais}},
  \bibinfo {author} {\bibfnamefont {F.}~\bibnamefont {Babonneau}}, \bibinfo
  {author} {\bibfnamefont {C.}~\bibnamefont {Coelho}}, \bibinfo {author}
  {\bibfnamefont {F.}~\bibnamefont {Pourpoint}}, \bibinfo {author}
  {\bibfnamefont {T.}~\bibnamefont {Aza{\"i}s}}, \bibinfo {author}
  {\bibfnamefont {S.~E.}\ \bibnamefont {Ashbrook}}, \bibinfo {author}
  {\bibfnamefont {J.~M.}\ \bibnamefont {Griffin}}, \bibinfo {author}
  {\bibfnamefont {J.~R.}\ \bibnamefont {Yates}}, \bibinfo {author}
  {\bibfnamefont {F.}~\bibnamefont {Mauri}},\ and\ \bibinfo {author}
  {\bibfnamefont {C.~J.}\ \bibnamefont {Pickard}},\ }\bibfield  {title}
  {\bibinfo {title} {First-{{Principles Calculation}} of {{NMR Parameters
  Using}} the {{Gauge Including Projector Augmented Wave Method}}: {{A
  Chemist}}'s {{Point}} of {{View}}},\ }\href
  {https://doi.org/10.1021/cr300108a} {\bibfield  {journal} {\bibinfo
  {journal} {Chemical Reviews}\ }\textbf {\bibinfo {volume} {112}},\ \bibinfo
  {pages} {5733} (\bibinfo {year} {2012})}\BibitemShut {NoStop}%
\bibitem [{\citenamefont {Salager}\ \emph {et~al.}(2010)\citenamefont
  {Salager}, \citenamefont {Day}, \citenamefont {Stein}, \citenamefont
  {Pickard}, \citenamefont {Elena},\ and\ \citenamefont
  {Emsley}}]{salagerPowderCrystallographyCombined2010}%
  \BibitemOpen
  \bibfield  {author} {\bibinfo {author} {\bibfnamefont {E.}~\bibnamefont
  {Salager}}, \bibinfo {author} {\bibfnamefont {G.~M.}\ \bibnamefont {Day}},
  \bibinfo {author} {\bibfnamefont {R.~S.}\ \bibnamefont {Stein}}, \bibinfo
  {author} {\bibfnamefont {C.~J.}\ \bibnamefont {Pickard}}, \bibinfo {author}
  {\bibfnamefont {B.}~\bibnamefont {Elena}},\ and\ \bibinfo {author}
  {\bibfnamefont {L.}~\bibnamefont {Emsley}},\ }\bibfield  {title} {\bibinfo
  {title} {Powder {{Crystallography}} by {{Combined Crystal Structure
  Prediction}} and {{High-Resolution}} {\textsuperscript{1}} {{H Solid-State
  NMR Spectroscopy}}},\ }\href {https://doi.org/10.1021/ja909449k} {\bibfield
  {journal} {\bibinfo  {journal} {Journal of the American Chemical Society}\
  }\textbf {\bibinfo {volume} {132}},\ \bibinfo {pages} {2564} (\bibinfo {year}
  {2010})}\BibitemShut {NoStop}%
\bibitem [{\citenamefont {Baias}\ \emph
  {et~al.}(2013{\natexlab{a}})\citenamefont {Baias}, \citenamefont {Dumez},
  \citenamefont {Svensson}, \citenamefont {Schantz}, \citenamefont {Day},\ and\
  \citenamefont {Emsley}}]{baiasNovoDeterminationCrystal2013}%
  \BibitemOpen
  \bibfield  {author} {\bibinfo {author} {\bibfnamefont {M.}~\bibnamefont
  {Baias}}, \bibinfo {author} {\bibfnamefont {J.-N.}\ \bibnamefont {Dumez}},
  \bibinfo {author} {\bibfnamefont {P.~H.}\ \bibnamefont {Svensson}}, \bibinfo
  {author} {\bibfnamefont {S.}~\bibnamefont {Schantz}}, \bibinfo {author}
  {\bibfnamefont {G.~M.}\ \bibnamefont {Day}},\ and\ \bibinfo {author}
  {\bibfnamefont {L.}~\bibnamefont {Emsley}},\ }\bibfield  {title} {\bibinfo
  {title} {{\emph{De }}{{{\emph{Novo}}}} {{Determination}} of the {{Crystal
  Structure}} of a {{Large Drug Molecule}} by {{Crystal Structure
  Prediction-Based Powder NMR Crystallography}}},\ }\href
  {https://doi.org/10.1021/ja4088874} {\bibfield  {journal} {\bibinfo
  {journal} {Journal of the American Chemical Society}\ }\textbf {\bibinfo
  {volume} {135}},\ \bibinfo {pages} {17501} (\bibinfo {year}
  {2013}{\natexlab{a}})}\BibitemShut {NoStop}%
\bibitem [{\citenamefont {Baias}\ \emph
  {et~al.}(2013{\natexlab{b}})\citenamefont {Baias}, \citenamefont
  {Widdifield}, \citenamefont {Dumez}, \citenamefont {Thompson}, \citenamefont
  {Cooper}, \citenamefont {Salager}, \citenamefont {Bassil}, \citenamefont
  {Stein}, \citenamefont {Lesage}, \citenamefont {Day},\ and\ \citenamefont
  {Emsley}}]{baiasPowderCrystallographyPharmaceutical2013}%
  \BibitemOpen
  \bibfield  {author} {\bibinfo {author} {\bibfnamefont {M.}~\bibnamefont
  {Baias}}, \bibinfo {author} {\bibfnamefont {C.~M.}\ \bibnamefont
  {Widdifield}}, \bibinfo {author} {\bibfnamefont {J.-N.}\ \bibnamefont
  {Dumez}}, \bibinfo {author} {\bibfnamefont {H.~P.~G.}\ \bibnamefont
  {Thompson}}, \bibinfo {author} {\bibfnamefont {T.~G.}\ \bibnamefont
  {Cooper}}, \bibinfo {author} {\bibfnamefont {E.}~\bibnamefont {Salager}},
  \bibinfo {author} {\bibfnamefont {S.}~\bibnamefont {Bassil}}, \bibinfo
  {author} {\bibfnamefont {R.~S.}\ \bibnamefont {Stein}}, \bibinfo {author}
  {\bibfnamefont {A.}~\bibnamefont {Lesage}}, \bibinfo {author} {\bibfnamefont
  {G.~M.}\ \bibnamefont {Day}},\ and\ \bibinfo {author} {\bibfnamefont
  {L.}~\bibnamefont {Emsley}},\ }\bibfield  {title} {\bibinfo {title} {Powder
  crystallography of pharmaceutical materials by combined crystal structure
  prediction and solid-state {{1H NMR}} spectroscopy},\ }\href
  {https://doi.org/10.1039/c3cp41095a} {\bibfield  {journal} {\bibinfo
  {journal} {Physical Chemistry Chemical Physics}\ }\textbf {\bibinfo {volume}
  {15}},\ \bibinfo {pages} {8069} (\bibinfo {year}
  {2013}{\natexlab{b}})}\BibitemShut {NoStop}%
\bibitem [{\citenamefont {Senker}\ \emph {et~al.}(2005)\citenamefont {Senker},
  \citenamefont {Sehnert},\ and\ \citenamefont
  {Correll}}]{senkerMicroscopicDescriptionPolyamorphic2005}%
  \BibitemOpen
  \bibfield  {author} {\bibinfo {author} {\bibfnamefont {J.}~\bibnamefont
  {Senker}}, \bibinfo {author} {\bibfnamefont {J.}~\bibnamefont {Sehnert}},\
  and\ \bibinfo {author} {\bibfnamefont {S.}~\bibnamefont {Correll}},\
  }\bibfield  {title} {\bibinfo {title} {Microscopic {{Description}} of the
  {{Polyamorphic Phases}} of {{Triphenyl Phosphite}} by {{Means}} of
  {{Multidimensional Solid-State NMR Spectroscopy}}},\ }\href
  {https://doi.org/10.1021/ja046602q} {\bibfield  {journal} {\bibinfo
  {journal} {Journal of the American Chemical Society}\ }\textbf {\bibinfo
  {volume} {127}},\ \bibinfo {pages} {337} (\bibinfo {year}
  {2005})}\BibitemShut {NoStop}%
\bibitem [{\citenamefont {Yates}\ \emph {et~al.}(2005)\citenamefont {Yates},
  \citenamefont {Dobbins}, \citenamefont {Pickard}, \citenamefont {Mauri},
  \citenamefont {Ghi},\ and\ \citenamefont
  {Harris}}]{yatesCombinedFirstPrinciples2005}%
  \BibitemOpen
  \bibfield  {author} {\bibinfo {author} {\bibfnamefont {J.~R.}\ \bibnamefont
  {Yates}}, \bibinfo {author} {\bibfnamefont {S.~E.}\ \bibnamefont {Dobbins}},
  \bibinfo {author} {\bibfnamefont {C.~J.}\ \bibnamefont {Pickard}}, \bibinfo
  {author} {\bibfnamefont {F.}~\bibnamefont {Mauri}}, \bibinfo {author}
  {\bibfnamefont {P.~Y.}\ \bibnamefont {Ghi}},\ and\ \bibinfo {author}
  {\bibfnamefont {R.~K.}\ \bibnamefont {Harris}},\ }\bibfield  {title}
  {\bibinfo {title} {A combined first principles computational and solid-state
  {{NMR}} study of a molecular crystal: Flurbiprofen},\ }\href
  {https://doi.org/10.1039/b500674k} {\bibfield  {journal} {\bibinfo  {journal}
  {Physical Chemistry Chemical Physics}\ }\textbf {\bibinfo {volume} {7}},\
  \bibinfo {pages} {1402} (\bibinfo {year} {2005})}\BibitemShut {NoStop}%
\bibitem [{\citenamefont {O'Shaughnessy}\ \emph {et~al.}(2025)\citenamefont
  {O'Shaughnessy}, \citenamefont {Qu}, \citenamefont {Wang}, \citenamefont
  {Holmes}, \citenamefont {Emsley}, \citenamefont {Glover}, \citenamefont
  {Hafizi}, \citenamefont {Day},\ and\ \citenamefont
  {Cooper}}]{oshaughnessyPolarTriptyceneBasedNonmetal2025}%
  \BibitemOpen
  \bibfield  {author} {\bibinfo {author} {\bibfnamefont {M.}~\bibnamefont
  {O'Shaughnessy}}, \bibinfo {author} {\bibfnamefont {H.}~\bibnamefont {Qu}},
  \bibinfo {author} {\bibfnamefont {X.}~\bibnamefont {Wang}}, \bibinfo {author}
  {\bibfnamefont {J.~B.}\ \bibnamefont {Holmes}}, \bibinfo {author}
  {\bibfnamefont {L.}~\bibnamefont {Emsley}}, \bibinfo {author} {\bibfnamefont
  {J.}~\bibnamefont {Glover}}, \bibinfo {author} {\bibfnamefont
  {R.}~\bibnamefont {Hafizi}}, \bibinfo {author} {\bibfnamefont {G.~M.}\
  \bibnamefont {Day}},\ and\ \bibinfo {author} {\bibfnamefont {A.~I.}\
  \bibnamefont {Cooper}},\ }\bibfield  {title} {\bibinfo {title} {Polar
  {{Triptycene-Based Nonmetal Organic Frameworks Show Enhanced Hydrogen
  Adsorption}}},\ }\href {https://doi.org/10.1021/jacs.5c11317} {\bibfield
  {journal} {\bibinfo  {journal} {Journal of the American Chemical Society}\
  }\textbf {\bibinfo {volume} {147}},\ \bibinfo {pages} {39351} (\bibinfo
  {year} {2025})}\BibitemShut {NoStop}%
\bibitem [{\citenamefont {Pindelska}\ \emph {et~al.}(2015)\citenamefont
  {Pindelska}, \citenamefont {Szeleszczuk}, \citenamefont {Pisklak},
  \citenamefont {Mazurek},\ and\ \citenamefont
  {Kolodziejski}}]{pindelskaSolidStateNMREffective2015}%
  \BibitemOpen
  \bibfield  {author} {\bibinfo {author} {\bibfnamefont {E.}~\bibnamefont
  {Pindelska}}, \bibinfo {author} {\bibfnamefont {L.}~\bibnamefont
  {Szeleszczuk}}, \bibinfo {author} {\bibfnamefont {D.~M.}\ \bibnamefont
  {Pisklak}}, \bibinfo {author} {\bibfnamefont {A.}~\bibnamefont {Mazurek}},\
  and\ \bibinfo {author} {\bibfnamefont {W.}~\bibnamefont {Kolodziejski}},\
  }\bibfield  {title} {\bibinfo {title} {Solid-{{State NMR}} as an {{Effective
  Method}} of {{Polymorphic Analysis}}: {{Solid Dosage Forms}} of {{Clopidogrel
  Hydrogensulfate}}},\ }\href {https://doi.org/10.1002/jps.24249} {\bibfield
  {journal} {\bibinfo  {journal} {Journal of Pharmaceutical Sciences}\ }\textbf
  {\bibinfo {volume} {104}},\ \bibinfo {pages} {106} (\bibinfo {year}
  {2015})}\BibitemShut {NoStop}%
\bibitem [{\citenamefont {Szeleszczuk}\ \emph {et~al.}(2019)\citenamefont
  {Szeleszczuk}, \citenamefont {Pisklak}, \citenamefont {Gubica}, \citenamefont
  {Matjakowska}, \citenamefont {Ka{\'z}mierski},\ and\ \citenamefont
  {{Zieli{\'n}ska-Pisklak}}}]{szeleszczukApplicationCombinedSolidstate2019}%
  \BibitemOpen
  \bibfield  {author} {\bibinfo {author} {\bibfnamefont {{\L}.}~\bibnamefont
  {Szeleszczuk}}, \bibinfo {author} {\bibfnamefont {D.~M.}\ \bibnamefont
  {Pisklak}}, \bibinfo {author} {\bibfnamefont {T.}~\bibnamefont {Gubica}},
  \bibinfo {author} {\bibfnamefont {K.}~\bibnamefont {Matjakowska}}, \bibinfo
  {author} {\bibfnamefont {S.}~\bibnamefont {Ka{\'z}mierski}},\ and\ \bibinfo
  {author} {\bibfnamefont {M.}~\bibnamefont {{Zieli{\'n}ska-Pisklak}}},\
  }\bibfield  {title} {\bibinfo {title} {Application of combined solid-state
  {{NMR}} and {{DFT}} calculations for the study of piracetam polymorphism},\
  }\href {https://doi.org/10.1016/j.ssnmr.2018.11.002} {\bibfield  {journal}
  {\bibinfo  {journal} {Solid State Nuclear Magnetic Resonance}\ }\textbf
  {\bibinfo {volume} {97}},\ \bibinfo {pages} {17} (\bibinfo {year}
  {2019})}\BibitemShut {NoStop}%
\bibitem [{\citenamefont {Dai}\ \emph {et~al.}(2020)\citenamefont {Dai},
  \citenamefont {Terskikh}, \citenamefont {Brinmkmann},\ and\ \citenamefont
  {Wu}}]{daiSolidState1H13And172020}%
  \BibitemOpen
  \bibfield  {author} {\bibinfo {author} {\bibfnamefont {Y.}~\bibnamefont
  {Dai}}, \bibinfo {author} {\bibfnamefont {V.}~\bibnamefont {Terskikh}},
  \bibinfo {author} {\bibfnamefont {A.}~\bibnamefont {Brinmkmann}},\ and\
  \bibinfo {author} {\bibfnamefont {G.}~\bibnamefont {Wu}},\ }\bibfield
  {title} {\bibinfo {title} {Solid-{{State}}{\textsuperscript{1}}
  {{H}},{\textsuperscript{13}} {{C}}, and{\textsuperscript{17}} {{O NMR
  Characterization}} of the {{Two Uncommon Polymorphs}} of {{Curcumin}}},\
  }\href {https://doi.org/10.1021/acs.cgd.0c01164} {\bibfield  {journal}
  {\bibinfo  {journal} {Crystal Growth \& Design}\ }\textbf {\bibinfo {volume}
  {20}},\ \bibinfo {pages} {7484} (\bibinfo {year} {2020})}\BibitemShut
  {NoStop}%
\bibitem [{\citenamefont {Widdifield}\ \emph {et~al.}(2016)\citenamefont
  {Widdifield}, \citenamefont {Robson},\ and\ \citenamefont
  {Hodgkinson}}]{widdifieldFurosemidesOneLittle2016}%
  \BibitemOpen
  \bibfield  {author} {\bibinfo {author} {\bibfnamefont {C.~M.}\ \bibnamefont
  {Widdifield}}, \bibinfo {author} {\bibfnamefont {H.}~\bibnamefont {Robson}},\
  and\ \bibinfo {author} {\bibfnamefont {P.}~\bibnamefont {Hodgkinson}},\
  }\bibfield  {title} {\bibinfo {title} {Furosemide's one little hydrogen atom:
  {{NMR}} crystallography structure verification of powdered molecular
  organics},\ }\href {https://doi.org/10.1039/C6CC02171A} {\bibfield  {journal}
  {\bibinfo  {journal} {Chemical Communications}\ }\textbf {\bibinfo {volume}
  {52}},\ \bibinfo {pages} {6685} (\bibinfo {year} {2016})}\BibitemShut
  {NoStop}%
\bibitem [{\citenamefont {Ashbrook}\ \emph {et~al.}(2020)\citenamefont
  {Ashbrook}, \citenamefont {Dawson}, \citenamefont {Gan}, \citenamefont
  {Hooper}, \citenamefont {Hung}, \citenamefont {Macfarlane}, \citenamefont
  {McKay}, \citenamefont {McLeod},\ and\ \citenamefont
  {Walton}}]{ashbrookApplicationNMRCrystallography2020}%
  \BibitemOpen
  \bibfield  {author} {\bibinfo {author} {\bibfnamefont {S.~E.}\ \bibnamefont
  {Ashbrook}}, \bibinfo {author} {\bibfnamefont {D.~M.}\ \bibnamefont
  {Dawson}}, \bibinfo {author} {\bibfnamefont {Z.}~\bibnamefont {Gan}},
  \bibinfo {author} {\bibfnamefont {J.~E.}\ \bibnamefont {Hooper}}, \bibinfo
  {author} {\bibfnamefont {I.}~\bibnamefont {Hung}}, \bibinfo {author}
  {\bibfnamefont {L.~E.}\ \bibnamefont {Macfarlane}}, \bibinfo {author}
  {\bibfnamefont {D.}~\bibnamefont {McKay}}, \bibinfo {author} {\bibfnamefont
  {L.~K.}\ \bibnamefont {McLeod}},\ and\ \bibinfo {author} {\bibfnamefont
  {R.~I.}\ \bibnamefont {Walton}},\ }\bibfield  {title} {\bibinfo {title}
  {Application of {{NMR Crystallography}} to {{Highly Disordered Templated
  Materials}}: {{Extensive Local Structural Disorder}} in the {{Gallophosphate
  GaPO-34A}}},\ }\href {https://doi.org/10.1021/acs.inorgchem.0c01450}
  {\bibfield  {journal} {\bibinfo  {journal} {Inorganic Chemistry}\ }\textbf
  {\bibinfo {volume} {59}},\ \bibinfo {pages} {11616} (\bibinfo {year}
  {2020})}\BibitemShut {NoStop}%
\bibitem [{\citenamefont {Brouwer}\ \emph {et~al.}(2013)\citenamefont
  {Brouwer}, \citenamefont {Cadars}, \citenamefont {Eckert}, \citenamefont
  {Liu}, \citenamefont {Terasaki},\ and\ \citenamefont
  {Chmelka}}]{brouwerGeneralProtocolDetermining2013}%
  \BibitemOpen
  \bibfield  {author} {\bibinfo {author} {\bibfnamefont {D.~H.}\ \bibnamefont
  {Brouwer}}, \bibinfo {author} {\bibfnamefont {S.}~\bibnamefont {Cadars}},
  \bibinfo {author} {\bibfnamefont {J.}~\bibnamefont {Eckert}}, \bibinfo
  {author} {\bibfnamefont {Z.}~\bibnamefont {Liu}}, \bibinfo {author}
  {\bibfnamefont {O.}~\bibnamefont {Terasaki}},\ and\ \bibinfo {author}
  {\bibfnamefont {B.~F.}\ \bibnamefont {Chmelka}},\ }\bibfield  {title}
  {\bibinfo {title} {A {{General Protocol}} for {{Determining}} the
  {{Structures}} of {{Molecularly Ordered}} but {{Noncrystalline Silicate
  Frameworks}}},\ }\href {https://doi.org/10.1021/ja311649m} {\bibfield
  {journal} {\bibinfo  {journal} {Journal of the American Chemical Society}\
  }\textbf {\bibinfo {volume} {135}},\ \bibinfo {pages} {5641} (\bibinfo {year}
  {2013})}\BibitemShut {NoStop}%
\bibitem [{\citenamefont {Lee}\ \emph {et~al.}(2010)\citenamefont {Lee},
  \citenamefont {Park}, \citenamefont {Yi},\ and\ \citenamefont
  {Moon}}]{leeStructureDisorderAmorphous2010}%
  \BibitemOpen
  \bibfield  {author} {\bibinfo {author} {\bibfnamefont {S.~K.}\ \bibnamefont
  {Lee}}, \bibinfo {author} {\bibfnamefont {S.~Y.}\ \bibnamefont {Park}},
  \bibinfo {author} {\bibfnamefont {Y.~S.}\ \bibnamefont {Yi}},\ and\ \bibinfo
  {author} {\bibfnamefont {J.}~\bibnamefont {Moon}},\ }\bibfield  {title}
  {\bibinfo {title} {Structure and {{Disorder}} in {{Amorphous Alumina Thin
  Films}}: {{Insights}} from {{High-Resolution Solid-State NMR}}},\ }\href
  {https://doi.org/10.1021/jp105306r} {\bibfield  {journal} {\bibinfo
  {journal} {The Journal of Physical Chemistry C}\ }\textbf {\bibinfo {volume}
  {114}},\ \bibinfo {pages} {13890} (\bibinfo {year} {2010})}\BibitemShut
  {NoStop}%
\bibitem [{\citenamefont {Gras}\ \emph {et~al.}(2016)\citenamefont {Gras},
  \citenamefont {Baker}, \citenamefont {Combes}, \citenamefont {Rey},
  \citenamefont {Sarda}, \citenamefont {Wright}, \citenamefont {Smith},
  \citenamefont {Hanna}, \citenamefont {Gervais}, \citenamefont {Laurencin},\
  and\ \citenamefont {Bonhomme}}]{grasCrystallineAmorphousCalcium2016}%
  \BibitemOpen
  \bibfield  {author} {\bibinfo {author} {\bibfnamefont {P.}~\bibnamefont
  {Gras}}, \bibinfo {author} {\bibfnamefont {A.}~\bibnamefont {Baker}},
  \bibinfo {author} {\bibfnamefont {C.}~\bibnamefont {Combes}}, \bibinfo
  {author} {\bibfnamefont {C.}~\bibnamefont {Rey}}, \bibinfo {author}
  {\bibfnamefont {S.}~\bibnamefont {Sarda}}, \bibinfo {author} {\bibfnamefont
  {A.~J.}\ \bibnamefont {Wright}}, \bibinfo {author} {\bibfnamefont {M.~E.}\
  \bibnamefont {Smith}}, \bibinfo {author} {\bibfnamefont {J.~V.}\ \bibnamefont
  {Hanna}}, \bibinfo {author} {\bibfnamefont {C.}~\bibnamefont {Gervais}},
  \bibinfo {author} {\bibfnamefont {D.}~\bibnamefont {Laurencin}},\ and\
  \bibinfo {author} {\bibfnamefont {C.}~\bibnamefont {Bonhomme}},\ }\bibfield
  {title} {\bibinfo {title} {From crystalline to amorphous calcium
  pyrophosphates: {{A}} solid state {{Nuclear Magnetic Resonance}}
  perspective},\ }\href {https://doi.org/10.1016/j.actbio.2015.10.016}
  {\bibfield  {journal} {\bibinfo  {journal} {Acta Biomaterialia}\ }\textbf
  {\bibinfo {volume} {31}},\ \bibinfo {pages} {348} (\bibinfo {year}
  {2016})}\BibitemShut {NoStop}%
\bibitem [{\citenamefont {Moran}\ \emph {et~al.}(2016)\citenamefont {Moran},
  \citenamefont {McKay}, \citenamefont {Pickard}, \citenamefont {Berry},
  \citenamefont {Griffin},\ and\ \citenamefont
  {Ashbrook}}]{moranHuntingHydrogenRandom2016}%
  \BibitemOpen
  \bibfield  {author} {\bibinfo {author} {\bibfnamefont {R.~F.}\ \bibnamefont
  {Moran}}, \bibinfo {author} {\bibfnamefont {D.}~\bibnamefont {McKay}},
  \bibinfo {author} {\bibfnamefont {C.~J.}\ \bibnamefont {Pickard}}, \bibinfo
  {author} {\bibfnamefont {A.~J.}\ \bibnamefont {Berry}}, \bibinfo {author}
  {\bibfnamefont {J.~M.}\ \bibnamefont {Griffin}},\ and\ \bibinfo {author}
  {\bibfnamefont {S.~E.}\ \bibnamefont {Ashbrook}},\ }\bibfield  {title}
  {\bibinfo {title} {Hunting for hydrogen: Random structure searching and
  prediction of {{NMR}} parameters of hydrous wadsleyite},\ }\href
  {https://doi.org/10.1039/c6cp01529h} {\bibfield  {journal} {\bibinfo
  {journal} {Physical Chemistry Chemical Physics}\ }\textbf {\bibinfo {volume}
  {18}},\ \bibinfo {pages} {10173} (\bibinfo {year} {2016})}\BibitemShut
  {NoStop}%
\bibitem [{\citenamefont {Brouwer}\ \emph {et~al.}(2005)\citenamefont
  {Brouwer}, \citenamefont {Darton}, \citenamefont {Morris},\ and\
  \citenamefont {Levitt}}]{brouwerSolidStateNMRMethod2005}%
  \BibitemOpen
  \bibfield  {author} {\bibinfo {author} {\bibfnamefont {D.~H.}\ \bibnamefont
  {Brouwer}}, \bibinfo {author} {\bibfnamefont {R.~J.}\ \bibnamefont {Darton}},
  \bibinfo {author} {\bibfnamefont {R.~E.}\ \bibnamefont {Morris}},\ and\
  \bibinfo {author} {\bibfnamefont {M.~H.}\ \bibnamefont {Levitt}},\ }\bibfield
   {title} {\bibinfo {title} {A {{Solid-State NMR Method}} for {{Solution}} of
  {{Zeolite Crystal Structures}}},\ }\href {https://doi.org/10.1021/ja052306h}
  {\bibfield  {journal} {\bibinfo  {journal} {Journal of the American Chemical
  Society}\ }\textbf {\bibinfo {volume} {127}},\ \bibinfo {pages} {10365}
  (\bibinfo {year} {2005})}\BibitemShut {NoStop}%
\bibitem [{\citenamefont
  {Brouwer}(2008)}]{brouwerNMRCrystallographyZeolites2008}%
  \BibitemOpen
  \bibfield  {author} {\bibinfo {author} {\bibfnamefont {D.~H.}\ \bibnamefont
  {Brouwer}},\ }\bibfield  {title} {\bibinfo {title} {{{NMR Crystallography}}
  of {{Zeolites}}: {{Refinement}} of an {{NMR-Solved Crystal Structure Using}}
  ab {{Initio Calculations}} of {{29Si Chemical Shift Tensors}}},\ }\href
  {https://doi.org/10.1021/ja800227f} {\bibfield  {journal} {\bibinfo
  {journal} {Journal of the American Chemical Society}\ }\textbf {\bibinfo
  {volume} {130}},\ \bibinfo {pages} {6306} (\bibinfo {year}
  {2008})}\BibitemShut {NoStop}%
\bibitem [{\citenamefont {Romao}\ \emph {et~al.}(2015)\citenamefont {Romao},
  \citenamefont {Perras}, \citenamefont {{Werner-Zwanziger}}, \citenamefont
  {Lussier}, \citenamefont {Miller}, \citenamefont {Calahoo}, \citenamefont
  {Zwanziger}, \citenamefont {Bieringer}, \citenamefont {Marinkovic},
  \citenamefont {Bryce},\ and\ \citenamefont
  {White}}]{romaoZeroThermalExpansion2015}%
  \BibitemOpen
  \bibfield  {author} {\bibinfo {author} {\bibfnamefont {C.~P.}\ \bibnamefont
  {Romao}}, \bibinfo {author} {\bibfnamefont {F.~A.}\ \bibnamefont {Perras}},
  \bibinfo {author} {\bibfnamefont {U.}~\bibnamefont {{Werner-Zwanziger}}},
  \bibinfo {author} {\bibfnamefont {J.~A.}\ \bibnamefont {Lussier}}, \bibinfo
  {author} {\bibfnamefont {K.~J.}\ \bibnamefont {Miller}}, \bibinfo {author}
  {\bibfnamefont {C.~M.}\ \bibnamefont {Calahoo}}, \bibinfo {author}
  {\bibfnamefont {J.~W.}\ \bibnamefont {Zwanziger}}, \bibinfo {author}
  {\bibfnamefont {M.}~\bibnamefont {Bieringer}}, \bibinfo {author}
  {\bibfnamefont {B.~A.}\ \bibnamefont {Marinkovic}}, \bibinfo {author}
  {\bibfnamefont {D.~L.}\ \bibnamefont {Bryce}},\ and\ \bibinfo {author}
  {\bibfnamefont {M.~A.}\ \bibnamefont {White}},\ }\bibfield  {title} {\bibinfo
  {title} {Zero {{Thermal Expansion}} in {{ZrMgMo3O12}}: {{NMR Crystallography
  Reveals Origins}} of {{Thermoelastic Properties}}},\ }\href
  {https://doi.org/10.1021/acs.chemmater.5b00429} {\bibfield  {journal}
  {\bibinfo  {journal} {Chemistry of Materials}\ }\textbf {\bibinfo {volume}
  {27}},\ \bibinfo {pages} {2633} (\bibinfo {year} {2015})}\BibitemShut
  {NoStop}%
\bibitem [{\citenamefont {Moran}\ \emph {et~al.}(2019)\citenamefont {Moran},
  \citenamefont {McKay}, \citenamefont {Tornstrom}, \citenamefont {Aziz},
  \citenamefont {Fernandes}, \citenamefont {{Grau-Crespo}},\ and\ \citenamefont
  {Ashbrook}}]{moranEnsembleBasedModelingNMR2019}%
  \BibitemOpen
  \bibfield  {author} {\bibinfo {author} {\bibfnamefont {R.~F.}\ \bibnamefont
  {Moran}}, \bibinfo {author} {\bibfnamefont {D.}~\bibnamefont {McKay}},
  \bibinfo {author} {\bibfnamefont {P.~C.}\ \bibnamefont {Tornstrom}}, \bibinfo
  {author} {\bibfnamefont {A.}~\bibnamefont {Aziz}}, \bibinfo {author}
  {\bibfnamefont {A.}~\bibnamefont {Fernandes}}, \bibinfo {author}
  {\bibfnamefont {R.}~\bibnamefont {{Grau-Crespo}}},\ and\ \bibinfo {author}
  {\bibfnamefont {S.~E.}\ \bibnamefont {Ashbrook}},\ }\bibfield  {title}
  {\bibinfo {title} {Ensemble-{{Based Modeling}} of the {{NMR Spectra}} of
  {{Solid Solutions}}: {{Cation Disorder}} in {{Y2}}({{Sn}},{{Ti}}){{2O7}}},\
  }\href {https://doi.org/10.1021/jacs.9b09036} {\bibfield  {journal} {\bibinfo
   {journal} {Journal of the American Chemical Society}\ }\textbf {\bibinfo
  {volume} {141}},\ \bibinfo {pages} {17838} (\bibinfo {year}
  {2019})}\BibitemShut {NoStop}%
\bibitem [{\citenamefont {Cordova}\ \emph {et~al.}(2023)\citenamefont
  {Cordova}, \citenamefont {Moutzouri}, \citenamefont {Nilsson~Lill},
  \citenamefont {Cousen}, \citenamefont {Kearns}, \citenamefont {Norberg},
  \citenamefont {Svensk~Ankarberg}, \citenamefont {McCabe}, \citenamefont
  {Pinon}, \citenamefont {Schantz},\ and\ \citenamefont
  {Emsley}}]{cordovaAtomiclevelStructureDetermination2023b}%
  \BibitemOpen
  \bibfield  {author} {\bibinfo {author} {\bibfnamefont {M.}~\bibnamefont
  {Cordova}}, \bibinfo {author} {\bibfnamefont {P.}~\bibnamefont {Moutzouri}},
  \bibinfo {author} {\bibfnamefont {S.~O.}\ \bibnamefont {Nilsson~Lill}},
  \bibinfo {author} {\bibfnamefont {A.}~\bibnamefont {Cousen}}, \bibinfo
  {author} {\bibfnamefont {M.}~\bibnamefont {Kearns}}, \bibinfo {author}
  {\bibfnamefont {S.~T.}\ \bibnamefont {Norberg}}, \bibinfo {author}
  {\bibfnamefont {A.}~\bibnamefont {Svensk~Ankarberg}}, \bibinfo {author}
  {\bibfnamefont {J.}~\bibnamefont {McCabe}}, \bibinfo {author} {\bibfnamefont
  {A.~C.}\ \bibnamefont {Pinon}}, \bibinfo {author} {\bibfnamefont
  {S.}~\bibnamefont {Schantz}},\ and\ \bibinfo {author} {\bibfnamefont
  {L.}~\bibnamefont {Emsley}},\ }\bibfield  {title} {\bibinfo {title}
  {Atomic-level structure determination of amorphous molecular solids by
  {{NMR}}},\ }\href {https://doi.org/10.1038/s41467-023-40853-2} {\bibfield
  {journal} {\bibinfo  {journal} {Nature Communications}\ }\textbf {\bibinfo
  {volume} {14}},\ \bibinfo {pages} {5138} (\bibinfo {year}
  {2023})}\BibitemShut {NoStop}%
\bibitem [{\citenamefont {Guest}\ \emph {et~al.}(2025)\citenamefont {Guest},
  \citenamefont {Bourne}, \citenamefont {Screen}, \citenamefont {Wilson},
  \citenamefont {Pham},\ and\ \citenamefont
  {Hodgkinson}}]{guestEssentialSynergyMD2025}%
  \BibitemOpen
  \bibfield  {author} {\bibinfo {author} {\bibfnamefont {J.~L.}\ \bibnamefont
  {Guest}}, \bibinfo {author} {\bibfnamefont {E.~A.~E.}\ \bibnamefont
  {Bourne}}, \bibinfo {author} {\bibfnamefont {M.~A.}\ \bibnamefont {Screen}},
  \bibinfo {author} {\bibfnamefont {M.~R.}\ \bibnamefont {Wilson}}, \bibinfo
  {author} {\bibfnamefont {T.~N.}\ \bibnamefont {Pham}},\ and\ \bibinfo
  {author} {\bibfnamefont {P.}~\bibnamefont {Hodgkinson}},\ }\bibfield  {title}
  {\bibinfo {title} {The essential synergy of {{MD}} simulation and {{NMR}} in
  understanding amorphous drug forms},\ }\href
  {https://doi.org/10.1039/D4FD00097H} {\bibfield  {journal} {\bibinfo
  {journal} {Faraday Discussions}\ }\textbf {\bibinfo {volume} {255}},\
  \bibinfo {pages} {325} (\bibinfo {year} {2025})}\BibitemShut {NoStop}%
\bibitem [{\citenamefont {Torodii}\ \emph
  {et~al.}(2025{\natexlab{a}})\citenamefont {Torodii}, \citenamefont {Holmes},
  \citenamefont {Cordova}, \citenamefont {Moutzouri}, \citenamefont {{van
  Beek}}, \citenamefont {Edfeldt}, \citenamefont {Malmerberg}, \citenamefont
  {Friis}, \citenamefont {Johansson}, \citenamefont {Milbradt}, \citenamefont
  {Nilsson~Lill}, \citenamefont {Malfait}, \citenamefont {Schantz},\ and\
  \citenamefont {Emsley}}]{torodiiDeterminationKeyFunctional2025}%
  \BibitemOpen
  \bibfield  {author} {\bibinfo {author} {\bibfnamefont {D.}~\bibnamefont
  {Torodii}}, \bibinfo {author} {\bibfnamefont {J.~B.}\ \bibnamefont {Holmes}},
  \bibinfo {author} {\bibfnamefont {M.}~\bibnamefont {Cordova}}, \bibinfo
  {author} {\bibfnamefont {P.}~\bibnamefont {Moutzouri}}, \bibinfo {author}
  {\bibfnamefont {L.}~\bibnamefont {{van Beek}}}, \bibinfo {author}
  {\bibfnamefont {F.}~\bibnamefont {Edfeldt}}, \bibinfo {author} {\bibfnamefont
  {E.}~\bibnamefont {Malmerberg}}, \bibinfo {author} {\bibfnamefont {S.~D.}\
  \bibnamefont {Friis}}, \bibinfo {author} {\bibfnamefont {J.~R.}\ \bibnamefont
  {Johansson}}, \bibinfo {author} {\bibfnamefont {A.~G.}\ \bibnamefont
  {Milbradt}}, \bibinfo {author} {\bibfnamefont {S.~O.}\ \bibnamefont
  {Nilsson~Lill}}, \bibinfo {author} {\bibfnamefont {B.}~\bibnamefont
  {Malfait}}, \bibinfo {author} {\bibfnamefont {S.}~\bibnamefont {Schantz}},\
  and\ \bibinfo {author} {\bibfnamefont {L.}~\bibnamefont {Emsley}},\
  }\bibfield  {title} {\bibinfo {title} {Determination of key functional
  structures of an amorphous {{VHL-based SMARCA2 PROTAC}}},\ }\href
  {https://doi.org/10.1038/s41467-025-65478-5} {\bibfield  {journal} {\bibinfo
  {journal} {Nature Communications}\ }\textbf {\bibinfo {volume} {16}},\
  \bibinfo {pages} {9694} (\bibinfo {year} {2025}{\natexlab{a}})}\BibitemShut
  {NoStop}%
\bibitem [{\citenamefont {Torodii}\ \emph
  {et~al.}(2025{\natexlab{b}})\citenamefont {Torodii}, \citenamefont {Cordova},
  \citenamefont {Holmes}, \citenamefont {Moutzouri}, \citenamefont {Casalini},
  \citenamefont {Nilsson~Lill}, \citenamefont {Pinon}, \citenamefont {Knee},
  \citenamefont {Svensk~Ankarberg}, \citenamefont {Putra}, \citenamefont
  {Schantz},\ and\ \citenamefont
  {Emsley}}]{torodiiThreeDimensionalAtomicLevelStructure2025}%
  \BibitemOpen
  \bibfield  {author} {\bibinfo {author} {\bibfnamefont {D.}~\bibnamefont
  {Torodii}}, \bibinfo {author} {\bibfnamefont {M.}~\bibnamefont {Cordova}},
  \bibinfo {author} {\bibfnamefont {J.~B.}\ \bibnamefont {Holmes}}, \bibinfo
  {author} {\bibfnamefont {P.}~\bibnamefont {Moutzouri}}, \bibinfo {author}
  {\bibfnamefont {T.}~\bibnamefont {Casalini}}, \bibinfo {author}
  {\bibfnamefont {S.~O.}\ \bibnamefont {Nilsson~Lill}}, \bibinfo {author}
  {\bibfnamefont {A.~C.}\ \bibnamefont {Pinon}}, \bibinfo {author}
  {\bibfnamefont {C.~S.}\ \bibnamefont {Knee}}, \bibinfo {author}
  {\bibfnamefont {A.}~\bibnamefont {Svensk~Ankarberg}}, \bibinfo {author}
  {\bibfnamefont {O.~D.}\ \bibnamefont {Putra}}, \bibinfo {author}
  {\bibfnamefont {S.}~\bibnamefont {Schantz}},\ and\ \bibinfo {author}
  {\bibfnamefont {L.}~\bibnamefont {Emsley}},\ }\bibfield  {title} {\bibinfo
  {title} {Three-{{Dimensional Atomic-Level Structure}} of an {{Amorphous
  Glucagon-Like Peptide-1 Receptor Agonist}}},\ }\href
  {https://doi.org/10.1021/jacs.5c01925} {\bibfield  {journal} {\bibinfo
  {journal} {Journal of the American Chemical Society}\ }\textbf {\bibinfo
  {volume} {147}},\ \bibinfo {pages} {17077} (\bibinfo {year}
  {2025}{\natexlab{b}})}\BibitemShut {NoStop}%
\bibitem [{\citenamefont {Lai}\ \emph {et~al.}(2011)\citenamefont {Lai},
  \citenamefont {Niks}, \citenamefont {Wang}, \citenamefont {Domratcheva},
  \citenamefont {Barends}, \citenamefont {Schwarz}, \citenamefont {Olsen},
  \citenamefont {Elliott}, \citenamefont {Fatmi}, \citenamefont {Chang},
  \citenamefont {Schlichting}, \citenamefont {Dunn},\ and\ \citenamefont
  {Mueller}}]{laiXrayNMRCrystallography2011}%
  \BibitemOpen
  \bibfield  {author} {\bibinfo {author} {\bibfnamefont {J.}~\bibnamefont
  {Lai}}, \bibinfo {author} {\bibfnamefont {D.}~\bibnamefont {Niks}}, \bibinfo
  {author} {\bibfnamefont {Y.}~\bibnamefont {Wang}}, \bibinfo {author}
  {\bibfnamefont {T.}~\bibnamefont {Domratcheva}}, \bibinfo {author}
  {\bibfnamefont {T.~R.~M.}\ \bibnamefont {Barends}}, \bibinfo {author}
  {\bibfnamefont {F.}~\bibnamefont {Schwarz}}, \bibinfo {author} {\bibfnamefont
  {R.~A.}\ \bibnamefont {Olsen}}, \bibinfo {author} {\bibfnamefont {D.~W.}\
  \bibnamefont {Elliott}}, \bibinfo {author} {\bibfnamefont {M.~Q.}\
  \bibnamefont {Fatmi}}, \bibinfo {author} {\bibfnamefont {C.-e.~A.}\
  \bibnamefont {Chang}}, \bibinfo {author} {\bibfnamefont {I.}~\bibnamefont
  {Schlichting}}, \bibinfo {author} {\bibfnamefont {M.~F.}\ \bibnamefont
  {Dunn}},\ and\ \bibinfo {author} {\bibfnamefont {L.~J.}\ \bibnamefont
  {Mueller}},\ }\bibfield  {title} {\bibinfo {title} {X-ray and {{NMR
  Crystallography}} in an {{Enzyme Active Site}}: {{The Indoline Quinonoid
  Intermediate}} in {{Tryptophan Synthase}}},\ }\href
  {https://doi.org/10.1021/ja106555c} {\bibfield  {journal} {\bibinfo
  {journal} {Journal of the American Chemical Society}\ }\textbf {\bibinfo
  {volume} {133}},\ \bibinfo {pages} {4} (\bibinfo {year} {2011})}\BibitemShut
  {NoStop}%
\bibitem [{\citenamefont {Holmes}\ \emph {et~al.}(2022)\citenamefont {Holmes},
  \citenamefont {Liu}, \citenamefont {Caulkins}, \citenamefont {Hilario},
  \citenamefont {Ghosh}, \citenamefont {Drago}, \citenamefont {Young},
  \citenamefont {Romero}, \citenamefont {Gill}, \citenamefont {Bogie},
  \citenamefont {Paulino}, \citenamefont {Wang}, \citenamefont {Riviere},
  \citenamefont {Bosken}, \citenamefont {Struppe}, \citenamefont {Hassan},
  \citenamefont {Guidoulianov}, \citenamefont {Perrone}, \citenamefont
  {{Mentink-Vigier}}, \citenamefont {Chang}, \citenamefont {Long},
  \citenamefont {Hooley}, \citenamefont {Mueser}, \citenamefont {Dunn},\ and\
  \citenamefont {Mueller}}]{holmesImagingActiveSite2022a}%
  \BibitemOpen
  \bibfield  {author} {\bibinfo {author} {\bibfnamefont {J.~B.}\ \bibnamefont
  {Holmes}}, \bibinfo {author} {\bibfnamefont {V.}~\bibnamefont {Liu}},
  \bibinfo {author} {\bibfnamefont {B.~G.}\ \bibnamefont {Caulkins}}, \bibinfo
  {author} {\bibfnamefont {E.}~\bibnamefont {Hilario}}, \bibinfo {author}
  {\bibfnamefont {R.~K.}\ \bibnamefont {Ghosh}}, \bibinfo {author}
  {\bibfnamefont {V.~N.}\ \bibnamefont {Drago}}, \bibinfo {author}
  {\bibfnamefont {R.~P.}\ \bibnamefont {Young}}, \bibinfo {author}
  {\bibfnamefont {J.~A.}\ \bibnamefont {Romero}}, \bibinfo {author}
  {\bibfnamefont {A.~D.}\ \bibnamefont {Gill}}, \bibinfo {author}
  {\bibfnamefont {P.~M.}\ \bibnamefont {Bogie}}, \bibinfo {author}
  {\bibfnamefont {J.}~\bibnamefont {Paulino}}, \bibinfo {author} {\bibfnamefont
  {X.}~\bibnamefont {Wang}}, \bibinfo {author} {\bibfnamefont {G.}~\bibnamefont
  {Riviere}}, \bibinfo {author} {\bibfnamefont {Y.~K.}\ \bibnamefont {Bosken}},
  \bibinfo {author} {\bibfnamefont {J.}~\bibnamefont {Struppe}}, \bibinfo
  {author} {\bibfnamefont {A.}~\bibnamefont {Hassan}}, \bibinfo {author}
  {\bibfnamefont {J.}~\bibnamefont {Guidoulianov}}, \bibinfo {author}
  {\bibfnamefont {B.}~\bibnamefont {Perrone}}, \bibinfo {author} {\bibfnamefont
  {F.}~\bibnamefont {{Mentink-Vigier}}}, \bibinfo {author} {\bibfnamefont
  {C.-e.~A.}\ \bibnamefont {Chang}}, \bibinfo {author} {\bibfnamefont {J.~R.}\
  \bibnamefont {Long}}, \bibinfo {author} {\bibfnamefont {R.~J.}\ \bibnamefont
  {Hooley}}, \bibinfo {author} {\bibfnamefont {T.~C.}\ \bibnamefont {Mueser}},
  \bibinfo {author} {\bibfnamefont {M.~F.}\ \bibnamefont {Dunn}},\ and\
  \bibinfo {author} {\bibfnamefont {L.~J.}\ \bibnamefont {Mueller}},\
  }\bibfield  {title} {\bibinfo {title} {Imaging active site chemistry and
  protonation states: {{NMR}} crystallography of the tryptophan synthase
  {$\alpha$}-aminoacrylate intermediate},\ }\href
  {https://doi.org/10.1073/pnas.2109235119} {\bibfield  {journal} {\bibinfo
  {journal} {Proceedings of the National Academy of Sciences}\ }\textbf
  {\bibinfo {volume} {119}},\ \bibinfo {pages} {e2109235119} (\bibinfo {year}
  {2022})}\BibitemShut {NoStop}%
\bibitem [{\citenamefont {Rejmak}\ \emph {et~al.}(2012)\citenamefont {Rejmak},
  \citenamefont {Dolado}, \citenamefont {Stott},\ and\ \citenamefont
  {Ayuela}}]{rejmak29SiNMRCement2012}%
  \BibitemOpen
  \bibfield  {author} {\bibinfo {author} {\bibfnamefont {P.}~\bibnamefont
  {Rejmak}}, \bibinfo {author} {\bibfnamefont {J.~S.}\ \bibnamefont {Dolado}},
  \bibinfo {author} {\bibfnamefont {M.~J.}\ \bibnamefont {Stott}},\ and\
  \bibinfo {author} {\bibfnamefont {A.}~\bibnamefont {Ayuela}},\ }\bibfield
  {title} {\bibinfo {title} {{{29Si NMR}} in {{Cement}}: {{A Theoretical
  Study}} on {{Calcium Silicate Hydrates}}},\ }\href
  {https://doi.org/10.1021/jp302218j} {\bibfield  {journal} {\bibinfo
  {journal} {The Journal of Physical Chemistry C}\ }\textbf {\bibinfo {volume}
  {116}},\ \bibinfo {pages} {9755} (\bibinfo {year} {2012})}\BibitemShut
  {NoStop}%
\bibitem [{\citenamefont {Walkley}\ and\ \citenamefont
  {Provis}(2019)}]{walkleySolidstateNuclearMagnetic2019}%
  \BibitemOpen
  \bibfield  {author} {\bibinfo {author} {\bibfnamefont {B.}~\bibnamefont
  {Walkley}}\ and\ \bibinfo {author} {\bibfnamefont {J.}~\bibnamefont
  {Provis}},\ }\bibfield  {title} {\bibinfo {title} {Solid-state nuclear
  magnetic resonance spectroscopy of cements},\ }\href
  {https://doi.org/10.1016/j.mtadv.2019.100007} {\bibfield  {journal} {\bibinfo
   {journal} {Materials Today Advances}\ }\textbf {\bibinfo {volume} {1}},\
  \bibinfo {pages} {100007} (\bibinfo {year} {2019})}\BibitemShut {NoStop}%
\bibitem [{\citenamefont {Kunhi~Mohamed}\ \emph {et~al.}(2020)\citenamefont
  {Kunhi~Mohamed}, \citenamefont {Moutzouri}, \citenamefont {Berruyer},
  \citenamefont {Walder}, \citenamefont {Siramanont}, \citenamefont {Harris},
  \citenamefont {Negroni}, \citenamefont {Galmarini}, \citenamefont {Parker},
  \citenamefont {Scrivener}, \citenamefont {Emsley},\ and\ \citenamefont
  {Bowen}}]{kunhimohamedAtomicLevelStructureCementitious2020}%
  \BibitemOpen
  \bibfield  {author} {\bibinfo {author} {\bibfnamefont {A.}~\bibnamefont
  {Kunhi~Mohamed}}, \bibinfo {author} {\bibfnamefont {P.}~\bibnamefont
  {Moutzouri}}, \bibinfo {author} {\bibfnamefont {P.}~\bibnamefont {Berruyer}},
  \bibinfo {author} {\bibfnamefont {B.~J.}\ \bibnamefont {Walder}}, \bibinfo
  {author} {\bibfnamefont {J.}~\bibnamefont {Siramanont}}, \bibinfo {author}
  {\bibfnamefont {M.}~\bibnamefont {Harris}}, \bibinfo {author} {\bibfnamefont
  {M.}~\bibnamefont {Negroni}}, \bibinfo {author} {\bibfnamefont {S.~C.}\
  \bibnamefont {Galmarini}}, \bibinfo {author} {\bibfnamefont {S.~C.}\
  \bibnamefont {Parker}}, \bibinfo {author} {\bibfnamefont {K.~L.}\
  \bibnamefont {Scrivener}}, \bibinfo {author} {\bibfnamefont {L.}~\bibnamefont
  {Emsley}},\ and\ \bibinfo {author} {\bibfnamefont {P.}~\bibnamefont
  {Bowen}},\ }\bibfield  {title} {\bibinfo {title} {The {{Atomic-Level
  Structure}} of {{Cementitious Calcium Aluminate Silicate Hydrate}}},\ }\href
  {https://doi.org/10.1021/jacs.0c02988} {\bibfield  {journal} {\bibinfo
  {journal} {Journal of the American Chemical Society}\ }\textbf {\bibinfo
  {volume} {142}},\ \bibinfo {pages} {11060} (\bibinfo {year}
  {2020})}\BibitemShut {NoStop}%
\bibitem [{\citenamefont {Morales-Melgares}\ \emph {et~al.}(2022)\citenamefont
  {Morales-Melgares}, \citenamefont {Casar}, \citenamefont {Moutzouri},
  \citenamefont {Venkatesh}, \citenamefont {Cordova}, \citenamefont
  {Kunhi~Mohamed}, \citenamefont {Scrivener}, \citenamefont {Bowen},\ and\
  \citenamefont {Emsley}}]{morales-melgares_atomic-level_2022}%
  \BibitemOpen
  \bibfield  {author} {\bibinfo {author} {\bibfnamefont {A.}~\bibnamefont
  {Morales-Melgares}}, \bibinfo {author} {\bibfnamefont {Z.}~\bibnamefont
  {Casar}}, \bibinfo {author} {\bibfnamefont {P.}~\bibnamefont {Moutzouri}},
  \bibinfo {author} {\bibfnamefont {A.}~\bibnamefont {Venkatesh}}, \bibinfo
  {author} {\bibfnamefont {M.}~\bibnamefont {Cordova}}, \bibinfo {author}
  {\bibfnamefont {A.}~\bibnamefont {Kunhi~Mohamed}}, \bibinfo {author}
  {\bibfnamefont {K.~L.}\ \bibnamefont {Scrivener}}, \bibinfo {author}
  {\bibfnamefont {P.}~\bibnamefont {Bowen}},\ and\ \bibinfo {author}
  {\bibfnamefont {L.}~\bibnamefont {Emsley}},\ }\bibfield  {title} {\bibinfo
  {title} {Atomic-{Level} {Structure} of {Zinc}-{Modified} {Cementitious}
  {Calcium} {Silicate} {Hydrate}},\ }\href
  {https://doi.org/10.1021/jacs.2c06749} {\bibfield  {journal} {\bibinfo
  {journal} {Journal of the American Chemical Society}\ }\textbf {\bibinfo
  {volume} {144}},\ \bibinfo {pages} {22915} (\bibinfo {year}
  {2022})}\BibitemShut {NoStop}%
\bibitem [{\citenamefont {Hope}\ \emph {et~al.}(2021)\citenamefont {Hope},
  \citenamefont {Nakamura}, \citenamefont {Ahlawat}, \citenamefont {Mishra},
  \citenamefont {Cordova}, \citenamefont {Jahanbakhshi}, \citenamefont
  {Mladenović}, \citenamefont {Runjhun}, \citenamefont {Merten}, \citenamefont
  {Hinderhofer}, \citenamefont {Carlsen}, \citenamefont {Kubicki},
  \citenamefont {Gershoni-Poranne}, \citenamefont {Schneeberger}, \citenamefont
  {Carbone}, \citenamefont {Liu}, \citenamefont {Zakeeruddin}, \citenamefont
  {Lewinski}, \citenamefont {Hagfeldt}, \citenamefont {Schreiber},
  \citenamefont {Rothlisberger}, \citenamefont {Grätzel}, \citenamefont
  {Milić},\ and\ \citenamefont {Emsley}}]{hope_nanoscale_2021}%
  \BibitemOpen
  \bibfield  {author} {\bibinfo {author} {\bibfnamefont {M.~A.}\ \bibnamefont
  {Hope}}, \bibinfo {author} {\bibfnamefont {T.}~\bibnamefont {Nakamura}},
  \bibinfo {author} {\bibfnamefont {P.}~\bibnamefont {Ahlawat}}, \bibinfo
  {author} {\bibfnamefont {A.}~\bibnamefont {Mishra}}, \bibinfo {author}
  {\bibfnamefont {M.}~\bibnamefont {Cordova}}, \bibinfo {author} {\bibfnamefont
  {F.}~\bibnamefont {Jahanbakhshi}}, \bibinfo {author} {\bibfnamefont
  {M.}~\bibnamefont {Mladenović}}, \bibinfo {author} {\bibfnamefont
  {R.}~\bibnamefont {Runjhun}}, \bibinfo {author} {\bibfnamefont
  {L.}~\bibnamefont {Merten}}, \bibinfo {author} {\bibfnamefont
  {A.}~\bibnamefont {Hinderhofer}}, \bibinfo {author} {\bibfnamefont {B.~I.}\
  \bibnamefont {Carlsen}}, \bibinfo {author} {\bibfnamefont {D.~J.}\
  \bibnamefont {Kubicki}}, \bibinfo {author} {\bibfnamefont {R.}~\bibnamefont
  {Gershoni-Poranne}}, \bibinfo {author} {\bibfnamefont {T.}~\bibnamefont
  {Schneeberger}}, \bibinfo {author} {\bibfnamefont {L.~C.}\ \bibnamefont
  {Carbone}}, \bibinfo {author} {\bibfnamefont {Y.}~\bibnamefont {Liu}},
  \bibinfo {author} {\bibfnamefont {S.~M.}\ \bibnamefont {Zakeeruddin}},
  \bibinfo {author} {\bibfnamefont {J.}~\bibnamefont {Lewinski}}, \bibinfo
  {author} {\bibfnamefont {A.}~\bibnamefont {Hagfeldt}}, \bibinfo {author}
  {\bibfnamefont {F.}~\bibnamefont {Schreiber}}, \bibinfo {author}
  {\bibfnamefont {U.}~\bibnamefont {Rothlisberger}}, \bibinfo {author}
  {\bibfnamefont {M.}~\bibnamefont {Grätzel}}, \bibinfo {author}
  {\bibfnamefont {J.~V.}\ \bibnamefont {Milić}},\ and\ \bibinfo {author}
  {\bibfnamefont {L.}~\bibnamefont {Emsley}},\ }\bibfield  {title} {\bibinfo
  {title} {Nanoscale {Phase} {Segregation} in {Supramolecular}
  $\pi$-{Templating} for {Hybrid} {Perovskite} {Photovoltaics} from {NMR}
  {Crystallography}},\ }\href {https://doi.org/10.1021/jacs.0c11563} {\bibfield
   {journal} {\bibinfo  {journal} {Journal of the American Chemical Society}\
  }\textbf {\bibinfo {volume} {143}},\ \bibinfo {pages} {1529} (\bibinfo {year}
  {2021})}\BibitemShut {NoStop}%
\bibitem [{\citenamefont {Kubicki}\ \emph {et~al.}(2021)\citenamefont
  {Kubicki}, \citenamefont {Stranks}, \citenamefont {Grey},\ and\ \citenamefont
  {Emsley}}]{kubickiNMRSpectroscopyProbes2021}%
  \BibitemOpen
  \bibfield  {author} {\bibinfo {author} {\bibfnamefont {D.~J.}\ \bibnamefont
  {Kubicki}}, \bibinfo {author} {\bibfnamefont {S.~D.}\ \bibnamefont
  {Stranks}}, \bibinfo {author} {\bibfnamefont {C.~P.}\ \bibnamefont {Grey}},\
  and\ \bibinfo {author} {\bibfnamefont {L.}~\bibnamefont {Emsley}},\
  }\bibfield  {title} {\bibinfo {title} {{{NMR}} spectroscopy probes
  microstructure, dynamics and doping of metal halide perovskites},\ }\href
  {https://doi.org/10.1038/s41570-021-00309-x} {\bibfield  {journal} {\bibinfo
  {journal} {Nature Reviews Chemistry}\ }\textbf {\bibinfo {volume} {5}},\
  \bibinfo {pages} {624} (\bibinfo {year} {2021})}\BibitemShut {NoStop}%
\bibitem [{\citenamefont {Hartman}\ \emph {et~al.}(2016)\citenamefont
  {Hartman}, \citenamefont {Day},\ and\ \citenamefont
  {Beran}}]{hartmanEnhancedNMRDiscrimination2016}%
  \BibitemOpen
  \bibfield  {author} {\bibinfo {author} {\bibfnamefont {J.~D.}\ \bibnamefont
  {Hartman}}, \bibinfo {author} {\bibfnamefont {G.~M.}\ \bibnamefont {Day}},\
  and\ \bibinfo {author} {\bibfnamefont {G.~J.~O.}\ \bibnamefont {Beran}},\
  }\bibfield  {title} {\bibinfo {title} {Enhanced {{NMR Discrimination}} of
  {{Pharmaceutically Relevant Molecular Crystal Forms}} through
  {{Fragment-Based Ab Initio Chemical Shift Predictions}}},\ }\href
  {https://doi.org/10.1021/acs.cgd.6b01157} {\bibfield  {journal} {\bibinfo
  {journal} {Crystal Growth \& Design}\ }\textbf {\bibinfo {volume} {16}},\
  \bibinfo {pages} {6479} (\bibinfo {year} {2016})}\BibitemShut {NoStop}%
\bibitem [{\citenamefont {Unzueta}\ and\ \citenamefont
  {Beran}(2025)}]{unzuetaReview2025}%
  \BibitemOpen
  \bibfield  {author} {\bibinfo {author} {\bibfnamefont {P.~A.}\ \bibnamefont
  {Unzueta}}\ and\ \bibinfo {author} {\bibfnamefont {G.~J.~O.}\ \bibnamefont
  {Beran}},\ }\bibfield  {title} {\bibinfo {title} {Predicting solid-state nmr
  observables via machine learning},\ }in\ \href
  {https://doi.org/10.1039/9781837673179-00224} {\emph {\bibinfo {booktitle}
  {Modern NMR Crystallography: Concepts and Applications}}},\ \bibinfo {editor}
  {edited by\ \bibinfo {editor} {\bibfnamefont {D.~L.}\ \bibnamefont {Bryce}}}\
  (\bibinfo  {publisher} {Royal Society of Chemistry},\ \bibinfo {year}
  {2025})\BibitemShut {NoStop}%
\bibitem [{\citenamefont {Paruzzo}\ \emph {et~al.}(2018)\citenamefont
  {Paruzzo}, \citenamefont {Hofstetter}, \citenamefont {Musil}, \citenamefont
  {De}, \citenamefont {Ceriotti},\ and\ \citenamefont {Emsley}}]{paru+18ncomm}%
  \BibitemOpen
  \bibfield  {author} {\bibinfo {author} {\bibfnamefont {F.~M.}\ \bibnamefont
  {Paruzzo}}, \bibinfo {author} {\bibfnamefont {A.}~\bibnamefont {Hofstetter}},
  \bibinfo {author} {\bibfnamefont {F.}~\bibnamefont {Musil}}, \bibinfo
  {author} {\bibfnamefont {S.}~\bibnamefont {De}}, \bibinfo {author}
  {\bibfnamefont {M.}~\bibnamefont {Ceriotti}},\ and\ \bibinfo {author}
  {\bibfnamefont {L.}~\bibnamefont {Emsley}},\ }\bibfield  {title} {\bibinfo
  {title} {{Chemical shifts in molecular solids by machine learning}},\
  }\href@noop {} {\bibfield  {journal} {\bibinfo  {journal} {Nature Comm.}\
  }\textbf {\bibinfo {volume} {9}},\ \bibinfo {pages} {4501} (\bibinfo {year}
  {2018})}\BibitemShut {NoStop}%
\bibitem [{\citenamefont {Liu}\ \emph {et~al.}(2019)\citenamefont {Liu},
  \citenamefont {Li}, \citenamefont {Bennett}, \citenamefont {Ganoe},
  \citenamefont {Stauch}, \citenamefont {{Head-Gordon}}, \citenamefont
  {Hexemer}, \citenamefont {Ushizima},\ and\ \citenamefont
  {{Head-Gordon}}}]{liuMultiresolution3DDenseNetChemical2019}%
  \BibitemOpen
  \bibfield  {author} {\bibinfo {author} {\bibfnamefont {S.}~\bibnamefont
  {Liu}}, \bibinfo {author} {\bibfnamefont {J.}~\bibnamefont {Li}}, \bibinfo
  {author} {\bibfnamefont {K.~C.}\ \bibnamefont {Bennett}}, \bibinfo {author}
  {\bibfnamefont {B.}~\bibnamefont {Ganoe}}, \bibinfo {author} {\bibfnamefont
  {T.}~\bibnamefont {Stauch}}, \bibinfo {author} {\bibfnamefont
  {M.}~\bibnamefont {{Head-Gordon}}}, \bibinfo {author} {\bibfnamefont
  {A.}~\bibnamefont {Hexemer}}, \bibinfo {author} {\bibfnamefont
  {D.}~\bibnamefont {Ushizima}},\ and\ \bibinfo {author} {\bibfnamefont
  {T.}~\bibnamefont {{Head-Gordon}}},\ }\bibfield  {title} {\bibinfo {title}
  {Multiresolution {{3D-DenseNet}} for {{Chemical Shift Prediction}} in {{NMR
  Crystallography}}},\ }\href {https://doi.org/10.1021/acs.jpclett.9b01570}
  {\bibfield  {journal} {\bibinfo  {journal} {The Journal of Physical Chemistry
  Letters}\ }\textbf {\bibinfo {volume} {10}},\ \bibinfo {pages} {4558}
  (\bibinfo {year} {2019})}\BibitemShut {NoStop}%
\bibitem [{\citenamefont {Xu}\ \emph {et~al.}(2025)\citenamefont {Xu},
  \citenamefont {Guo}, \citenamefont {Wang}, \citenamefont {Yao}, \citenamefont
  {Wang}, \citenamefont {Tang}, \citenamefont {Gao}, \citenamefont {Zhang},
  \citenamefont {E}, \citenamefont {Tian},\ and\ \citenamefont
  {Cheng}}]{xuUnifiedBenchmarkFramework2025a}%
  \BibitemOpen
  \bibfield  {author} {\bibinfo {author} {\bibfnamefont {F.}~\bibnamefont
  {Xu}}, \bibinfo {author} {\bibfnamefont {W.}~\bibnamefont {Guo}}, \bibinfo
  {author} {\bibfnamefont {F.}~\bibnamefont {Wang}}, \bibinfo {author}
  {\bibfnamefont {L.}~\bibnamefont {Yao}}, \bibinfo {author} {\bibfnamefont
  {H.}~\bibnamefont {Wang}}, \bibinfo {author} {\bibfnamefont {F.}~\bibnamefont
  {Tang}}, \bibinfo {author} {\bibfnamefont {Z.}~\bibnamefont {Gao}}, \bibinfo
  {author} {\bibfnamefont {L.}~\bibnamefont {Zhang}}, \bibinfo {author}
  {\bibfnamefont {W.}~\bibnamefont {E}}, \bibinfo {author} {\bibfnamefont
  {Z.-Q.}\ \bibnamefont {Tian}},\ and\ \bibinfo {author} {\bibfnamefont
  {J.}~\bibnamefont {Cheng}},\ }\bibfield  {title} {\bibinfo {title} {Toward a
  unified benchmark and framework for deep learning-based prediction of nuclear
  magnetic resonance chemical shifts},\ }\href
  {https://doi.org/10.1038/s43588-025-00783-z} {\bibfield  {journal} {\bibinfo
  {journal} {Nature Computational Science}\ }\textbf {\bibinfo {volume} {5}},\
  \bibinfo {pages} {292} (\bibinfo {year} {2025})}\BibitemShut {NoStop}%
\bibitem [{\citenamefont
  {Charpentier}(2025)}]{charpentierFirstprinciplesNMROxide2025}%
  \BibitemOpen
  \bibfield  {author} {\bibinfo {author} {\bibfnamefont {T.}~\bibnamefont
  {Charpentier}},\ }\bibfield  {title} {\bibinfo {title} {First-principles
  {{NMR}} of oxide glasses boosted by machine learning},\ }\href
  {https://doi.org/10.1039/d4fd00129j} {\bibfield  {journal} {\bibinfo
  {journal} {Faraday Discussions}\ }\textbf {\bibinfo {volume} {255}},\
  \bibinfo {pages} {370} (\bibinfo {year} {2025})}\BibitemShut {NoStop}%
\bibitem [{\citenamefont {Cuny}\ \emph {et~al.}(2016)\citenamefont {Cuny},
  \citenamefont {Xie}, \citenamefont {Pickard},\ and\ \citenamefont
  {Hassanali}}]{cunyInitioQualityNMR2016}%
  \BibitemOpen
  \bibfield  {author} {\bibinfo {author} {\bibfnamefont {J.}~\bibnamefont
  {Cuny}}, \bibinfo {author} {\bibfnamefont {Y.}~\bibnamefont {Xie}}, \bibinfo
  {author} {\bibfnamefont {C.~J.}\ \bibnamefont {Pickard}},\ and\ \bibinfo
  {author} {\bibfnamefont {A.~A.}\ \bibnamefont {Hassanali}},\ }\bibfield
  {title} {\bibinfo {title} {{\emph{Ab }}{{{\emph{Initio}}}} {{Quality NMR
  Parameters}} in {{Solid-State Materials Using}} a {{High-Dimensional
  Neural-Network Representation}}},\ }\href
  {https://doi.org/10.1021/acs.jctc.5b01006} {\bibfield  {journal} {\bibinfo
  {journal} {Journal of Chemical Theory and Computation}\ }\textbf {\bibinfo
  {volume} {12}},\ \bibinfo {pages} {765} (\bibinfo {year} {2016})}\BibitemShut
  {NoStop}%
\bibitem [{\citenamefont {Venetos}\ \emph {et~al.}(2023)\citenamefont
  {Venetos}, \citenamefont {Wen},\ and\ \citenamefont
  {Persson}}]{venetosMachineLearningFull2023}%
  \BibitemOpen
  \bibfield  {author} {\bibinfo {author} {\bibfnamefont {M.~C.}\ \bibnamefont
  {Venetos}}, \bibinfo {author} {\bibfnamefont {M.}~\bibnamefont {Wen}},\ and\
  \bibinfo {author} {\bibfnamefont {K.~A.}\ \bibnamefont {Persson}},\
  }\bibfield  {title} {\bibinfo {title} {Machine {{Learning Full NMR Chemical
  Shift Tensors}} of {{Silicon Oxides}} with {{Equivariant Graph Neural
  Networks}}},\ }\href {https://doi.org/10.1021/acs.jpca.2c07530} {\bibfield
  {journal} {\bibinfo  {journal} {The Journal of Physical Chemistry A}\
  }\textbf {\bibinfo {volume} {127}},\ \bibinfo {pages} {2388} (\bibinfo {year}
  {2023})}\BibitemShut {NoStop}%
\bibitem [{\citenamefont {Ben~Mahmoud}\ \emph {et~al.}(2025)\citenamefont
  {Ben~Mahmoud}, \citenamefont {Rosset}, \citenamefont {Yates},\ and\
  \citenamefont
  {Deringer}}]{benmahmoudGraphneuralnetworkPredictionsSolidstate2025}%
  \BibitemOpen
  \bibfield  {author} {\bibinfo {author} {\bibfnamefont {C.}~\bibnamefont
  {Ben~Mahmoud}}, \bibinfo {author} {\bibfnamefont {L.~A.~M.}\ \bibnamefont
  {Rosset}}, \bibinfo {author} {\bibfnamefont {J.~R.}\ \bibnamefont {Yates}},\
  and\ \bibinfo {author} {\bibfnamefont {V.~L.}\ \bibnamefont {Deringer}},\
  }\bibfield  {title} {\bibinfo {title} {Graph-neural-network predictions of
  solid-state {{NMR}} parameters in silica from spherical tensor
  decomposition},\ }\href {https://doi.org/10.1063/5.0274240} {\bibfield
  {journal} {\bibinfo  {journal} {The Journal of Chemical Physics}\ }\textbf
  {\bibinfo {volume} {163}},\ \bibinfo {pages} {024118} (\bibinfo {year}
  {2025})}\BibitemShut {NoStop}%
\bibitem [{\citenamefont {Bornes}\ \emph {et~al.}(2026)\citenamefont {Bornes},
  \citenamefont {Mahmoud}, \citenamefont {Deringer}, \citenamefont {Heard},\
  and\ \citenamefont {Grajciar}}]{bornesAccurateTensorialModel2026}%
  \BibitemOpen
  \bibfield  {author} {\bibinfo {author} {\bibfnamefont {C.}~\bibnamefont
  {Bornes}}, \bibinfo {author} {\bibfnamefont {C.~B.}\ \bibnamefont {Mahmoud}},
  \bibinfo {author} {\bibfnamefont {V.~L.}\ \bibnamefont {Deringer}}, \bibinfo
  {author} {\bibfnamefont {C.~J.}\ \bibnamefont {Heard}},\ and\ \bibinfo
  {author} {\bibfnamefont {L.}~\bibnamefont {Grajciar}},\ }\bibfield  {title}
  {\bibinfo {title} {An {{Accurate Tensorial Model}} for {{Prediction}} of
  {{Full Zeolite NMR Spectra}}},\ }\bibfield  {journal} {\bibinfo  {journal}
  {arXiv}\ }\href {https://doi.org/10.48550/ARXIV.2603.22268}
  {10.48550/ARXIV.2603.22268} (\bibinfo {year} {2026})\BibitemShut {NoStop}%
\bibitem [{\citenamefont {Zaverkin}\ \emph {et~al.}(2022)\citenamefont
  {Zaverkin}, \citenamefont {Netz}, \citenamefont {Zills}, \citenamefont
  {K{\"o}hn},\ and\ \citenamefont
  {K{\"a}stner}}]{zaverkinThermallyAveragedMagnetic2022c}%
  \BibitemOpen
  \bibfield  {author} {\bibinfo {author} {\bibfnamefont {V.}~\bibnamefont
  {Zaverkin}}, \bibinfo {author} {\bibfnamefont {J.}~\bibnamefont {Netz}},
  \bibinfo {author} {\bibfnamefont {F.}~\bibnamefont {Zills}}, \bibinfo
  {author} {\bibfnamefont {A.}~\bibnamefont {K{\"o}hn}},\ and\ \bibinfo
  {author} {\bibfnamefont {J.}~\bibnamefont {K{\"a}stner}},\ }\bibfield
  {title} {\bibinfo {title} {Thermally {{Averaged Magnetic Anisotropy Tensors}}
  via {{Machine Learning Based}} on {{Gaussian Moments}}},\ }\href
  {https://doi.org/10.1021/acs.jctc.1c00853} {\bibfield  {journal} {\bibinfo
  {journal} {Journal of Chemical Theory and Computation}\ }\textbf {\bibinfo
  {volume} {18}},\ \bibinfo {pages} {1} (\bibinfo {year} {2022})}\BibitemShut
  {NoStop}%
\bibitem [{\citenamefont {F.~Harper}\ \emph {et~al.}(2025)\citenamefont
  {F.~Harper}, \citenamefont {Huss}, \citenamefont {S.~K{\"o}cher},\ and\
  \citenamefont {Scheurer}}]{f.harperTrackingLiAtoms2025a}%
  \BibitemOpen
  \bibfield  {author} {\bibinfo {author} {\bibfnamefont {A.}~\bibnamefont
  {F.~Harper}}, \bibinfo {author} {\bibfnamefont {T.}~\bibnamefont {Huss}},
  \bibinfo {author} {\bibfnamefont {S.}~\bibnamefont {S.~K{\"o}cher}},\ and\
  \bibinfo {author} {\bibfnamefont {C.}~\bibnamefont {Scheurer}},\ }\bibfield
  {title} {\bibinfo {title} {Tracking {{Li}} atoms in real-time with ultra-fast
  {{NMR}} simulations},\ }\href {https://doi.org/10.1039/D4FD00074A} {\bibfield
   {journal} {\bibinfo  {journal} {Faraday Discussions}\ }\textbf {\bibinfo
  {volume} {255}},\ \bibinfo {pages} {411} (\bibinfo {year}
  {2025})}\BibitemShut {NoStop}%
\bibitem [{\citenamefont {Huss}\ \emph {et~al.}(2026)\citenamefont {Huss},
  \citenamefont {Civaia}, \citenamefont {K{\"o}cher}, \citenamefont {Reuter},
  \citenamefont {Granwehr},\ and\ \citenamefont
  {Scheurer}}]{hussSimulatingQuadrupolarNMR2026}%
  \BibitemOpen
  \bibfield  {author} {\bibinfo {author} {\bibfnamefont {T.}~\bibnamefont
  {Huss}}, \bibinfo {author} {\bibfnamefont {F.}~\bibnamefont {Civaia}},
  \bibinfo {author} {\bibfnamefont {S.~S.}\ \bibnamefont {K{\"o}cher}},
  \bibinfo {author} {\bibfnamefont {K.}~\bibnamefont {Reuter}}, \bibinfo
  {author} {\bibfnamefont {J.}~\bibnamefont {Granwehr}},\ and\ \bibinfo
  {author} {\bibfnamefont {C.}~\bibnamefont {Scheurer}},\ }\bibfield  {title}
  {\bibinfo {title} {Simulating quadrupolar {{NMR}} dynamics in solid
  electrolyte {{Li10GeP2S12}}},\ }\href {https://doi.org/10.1063/5.0308803}
  {\bibfield  {journal} {\bibinfo  {journal} {The Journal of Chemical Physics}\
  }\textbf {\bibinfo {volume} {164}},\ \bibinfo {pages} {084116} (\bibinfo
  {year} {2026})}\BibitemShut {NoStop}%
\bibitem [{\citenamefont {Unzueta}\ \emph {et~al.}(2021)\citenamefont
  {Unzueta}, \citenamefont {Greenwell},\ and\ \citenamefont
  {Beran}}]{unzueta_predicting_2021}%
  \BibitemOpen
  \bibfield  {author} {\bibinfo {author} {\bibfnamefont {P.~A.}\ \bibnamefont
  {Unzueta}}, \bibinfo {author} {\bibfnamefont {C.~S.}\ \bibnamefont
  {Greenwell}},\ and\ \bibinfo {author} {\bibfnamefont {G.~J.~O.}\ \bibnamefont
  {Beran}},\ }\bibfield  {title} {\bibinfo {title} {Predicting {{Density
  Functional Theory-Quality Nuclear Magnetic Resonance Chemical Shifts}} via
  {{$\Delta$-Machine Learning}}},\ }\href
  {https://doi.org/10.1021/acs.jctc.0c00979} {\bibfield  {journal} {\bibinfo
  {journal} {Journal of Chemical Theory and Computation}\ }\textbf {\bibinfo
  {volume} {17}},\ \bibinfo {pages} {826} (\bibinfo {year} {2021})}\BibitemShut
  {NoStop}%
\bibitem [{\citenamefont {Kleine~B{\"u}ning}\ and\ \citenamefont
  {Grimme}(2023)}]{kleinebuningComputationCCSDTQualityNMR2023}%
  \BibitemOpen
  \bibfield  {author} {\bibinfo {author} {\bibfnamefont {J.~B.}\ \bibnamefont
  {Kleine~B{\"u}ning}}\ and\ \bibinfo {author} {\bibfnamefont {S.}~\bibnamefont
  {Grimme}},\ }\bibfield  {title} {\bibinfo {title} {Computation of
  {{CCSD}}({{T}})-{{Quality NMR Chemical Shifts}} via {{$\Delta$-Machine
  Learning}} from {{DFT}}},\ }\href {https://doi.org/10.1021/acs.jctc.3c00165}
  {\bibfield  {journal} {\bibinfo  {journal} {Journal of Chemical Theory and
  Computation}\ }\textbf {\bibinfo {volume} {19}},\ \bibinfo {pages} {3601}
  (\bibinfo {year} {2023})}\BibitemShut {NoStop}%
\bibitem [{\citenamefont {Cordova}\ \emph {et~al.}(2022)\citenamefont
  {Cordova}, \citenamefont {Engel}, \citenamefont {Stefaniuk}, \citenamefont
  {Paruzzo}, \citenamefont {Hofstetter}, \citenamefont {Ceriotti},\ and\
  \citenamefont {Emsley}}]{cordova_machine_2022}%
  \BibitemOpen
  \bibfield  {author} {\bibinfo {author} {\bibfnamefont {M.}~\bibnamefont
  {Cordova}}, \bibinfo {author} {\bibfnamefont {E.~A.}\ \bibnamefont {Engel}},
  \bibinfo {author} {\bibfnamefont {A.}~\bibnamefont {Stefaniuk}}, \bibinfo
  {author} {\bibfnamefont {F.}~\bibnamefont {Paruzzo}}, \bibinfo {author}
  {\bibfnamefont {A.}~\bibnamefont {Hofstetter}}, \bibinfo {author}
  {\bibfnamefont {M.}~\bibnamefont {Ceriotti}},\ and\ \bibinfo {author}
  {\bibfnamefont {L.}~\bibnamefont {Emsley}},\ }\bibfield  {title}
  {{\selectlanguage {en}\bibinfo {title} {A {Machine} {Learning} {Model} of
  {Chemical} {Shifts} for {Chemically} and {Structurally} {Diverse} {Molecular}
  {Solids}}},\ }\href {https://doi.org/10.1021/acs.jpcc.2c03854} {\bibfield
  {journal} {\bibinfo  {journal} {The Journal of Physical Chemistry C}\
  }\textbf {\bibinfo {volume} {126}},\ \bibinfo {pages} {16710} (\bibinfo
  {year} {2022})}\BibitemShut {NoStop}%
\bibitem [{\citenamefont {Kellner}\ \emph {et~al.}(2025)\citenamefont
  {Kellner}, \citenamefont {Holmes}, \citenamefont {{Rodriguez-Madrid}},
  \citenamefont {Viscosi}, \citenamefont {Zhang}, \citenamefont {Emsley},\ and\
  \citenamefont {Ceriotti}}]{kellnerDeepLearningModel2025b}%
  \BibitemOpen
  \bibfield  {author} {\bibinfo {author} {\bibfnamefont {M.}~\bibnamefont
  {Kellner}}, \bibinfo {author} {\bibfnamefont {J.~B.}\ \bibnamefont {Holmes}},
  \bibinfo {author} {\bibfnamefont {R.}~\bibnamefont {{Rodriguez-Madrid}}},
  \bibinfo {author} {\bibfnamefont {F.}~\bibnamefont {Viscosi}}, \bibinfo
  {author} {\bibfnamefont {Y.}~\bibnamefont {Zhang}}, \bibinfo {author}
  {\bibfnamefont {L.}~\bibnamefont {Emsley}},\ and\ \bibinfo {author}
  {\bibfnamefont {M.}~\bibnamefont {Ceriotti}},\ }\bibfield  {title} {\bibinfo
  {title} {A {{Deep Learning Model}} for {{Chemical Shieldings}} in {{Molecular
  Organic Solids Including Anisotropy}}},\ }\href
  {https://doi.org/10.1021/acs.jpclett.5c01819} {\bibfield  {journal} {\bibinfo
   {journal} {The Journal of Physical Chemistry Letters}\ }\textbf {\bibinfo
  {volume} {16}},\ \bibinfo {pages} {8714} (\bibinfo {year}
  {2025})}\BibitemShut {NoStop}%
\bibitem [{\citenamefont {Gunaga}\ \emph {et~al.}(2026)\citenamefont {Gunaga},
  \citenamefont {Schurko}, \citenamefont {Holmes},\ and\ \citenamefont
  {{Mentink-Vigier}}}]{gunagaAccessibleHybridDFTquality}%
  \BibitemOpen
  \bibfield  {author} {\bibinfo {author} {\bibfnamefont {S.~S.}\ \bibnamefont
  {Gunaga}}, \bibinfo {author} {\bibfnamefont {R.~W.}\ \bibnamefont {Schurko}},
  \bibinfo {author} {\bibfnamefont {S.~T.}\ \bibnamefont {Holmes}},\ and\
  \bibinfo {author} {\bibfnamefont {F.}~\bibnamefont {{Mentink-Vigier}}},\
  }\bibfield  {title} {\bibinfo {title} {Accessible hybrid {{DFT-quality NMR}}
  crystallography via gas-phase {{Machine Learning Interatomic Potentials}}},\
  }\bibfield  {journal} {\bibinfo  {journal} {ChemRxiv}\ }\textbf {\bibinfo
  {volume} {2026}},\ \href {https://doi.org/10.26434/chemrxiv.15000992/v2}
  {10.26434/chemrxiv.15000992/v2} (\bibinfo {year} {2026})\BibitemShut
  {NoStop}%
\bibitem [{\citenamefont {Czernek}\ \emph {et~al.}(2026)\citenamefont
  {Czernek}, \citenamefont {Brus},\ and\ \citenamefont
  {Potrzebowski}}]{czernekAssessingReliabilityShiftML32026}%
  \BibitemOpen
  \bibfield  {author} {\bibinfo {author} {\bibfnamefont {J.}~\bibnamefont
  {Czernek}}, \bibinfo {author} {\bibfnamefont {J.}~\bibnamefont {Brus}},\ and\
  \bibinfo {author} {\bibfnamefont {M.~J.}\ \bibnamefont {Potrzebowski}},\
  }\bibfield  {title} {\bibinfo {title} {Assessing the reliability of the
  {{ShiftML3}} model for the {{13C NMR}} chemical shielding tensors of
  crystalline solids},\ }\href {https://doi.org/10.1016/j.cplett.2026.142953}
  {\bibfield  {journal} {\bibinfo  {journal} {Chemical Physics Letters}\
  }\textbf {\bibinfo {volume} {897}},\ \bibinfo {pages} {142953} (\bibinfo
  {year} {2026})}\BibitemShut {NoStop}%
\bibitem [{\citenamefont {Kellner}\ \emph {et~al.}(2026)\citenamefont
  {Kellner}, \citenamefont {{Rodriguez-Madrid}}, \citenamefont {Holmes},
  \citenamefont {Principe}, \citenamefont {Emsley},\ and\ \citenamefont
  {Ceriotti}}]{kellnerQuantumcorrectedNMRCrystallography2026}%
  \BibitemOpen
  \bibfield  {author} {\bibinfo {author} {\bibfnamefont {M.}~\bibnamefont
  {Kellner}}, \bibinfo {author} {\bibfnamefont {R.}~\bibnamefont
  {{Rodriguez-Madrid}}}, \bibinfo {author} {\bibfnamefont {J.~B.}\ \bibnamefont
  {Holmes}}, \bibinfo {author} {\bibfnamefont {V.~P.}\ \bibnamefont
  {Principe}}, \bibinfo {author} {\bibfnamefont {L.}~\bibnamefont {Emsley}},\
  and\ \bibinfo {author} {\bibfnamefont {M.}~\bibnamefont {Ceriotti}},\ }\href
  {https://doi.org/10.48550/arXiv.2603.06236} {\bibinfo {title}
  {Quantum-corrected {{NMR}} crystallography at scale}} (\bibinfo {year}
  {2026}),\ \Eprint {https://arxiv.org/abs/2603.06236} {arXiv:2603.06236
  [physics.chem-ph]} \BibitemShut {NoStop}%
\bibitem [{\citenamefont {Socha}\ \emph {et~al.}(2026)\citenamefont {Socha},
  \citenamefont {Pavli{\v s}ov{\'a}}, \citenamefont {Manna},\ and\
  \citenamefont {Dra{\v
  c}{\'i}nsk{\'y}}}]{sochaQuantitativePredictionExchangeable2026}%
  \BibitemOpen
  \bibfield  {author} {\bibinfo {author} {\bibfnamefont {O.}~\bibnamefont
  {Socha}}, \bibinfo {author} {\bibfnamefont {J.}~\bibnamefont {Pavli{\v
  s}ov{\'a}}}, \bibinfo {author} {\bibfnamefont {D.}~\bibnamefont {Manna}},\
  and\ \bibinfo {author} {\bibfnamefont {M.}~\bibnamefont {Dra{\v
  c}{\'i}nsk{\'y}}},\ }\bibfield  {title} {\bibinfo {title} {Quantitative
  {{Prediction}} of {{Exchangeable Proton Chemical Shifts}}},\ }\bibfield
  {journal} {\bibinfo  {journal} {Nature Communications}\ }\href
  {https://doi.org/10.1038/s41467-026-75743-w} {10.1038/s41467-026-75743-w}
  (\bibinfo {year} {2026})\BibitemShut {NoStop}%
\bibitem [{\citenamefont {Gauss}\ and\ \citenamefont
  {Stanton}(1995)}]{gaussCoupledclusterCalculationsNuclear1995}%
  \BibitemOpen
  \bibfield  {author} {\bibinfo {author} {\bibfnamefont {J.}~\bibnamefont
  {Gauss}}\ and\ \bibinfo {author} {\bibfnamefont {J.~F.}\ \bibnamefont
  {Stanton}},\ }\bibfield  {title} {\bibinfo {title} {Coupled-cluster
  calculations of nuclear magnetic resonance chemical shifts},\ }\href
  {https://doi.org/10.1063/1.470240} {\bibfield  {journal} {\bibinfo  {journal}
  {The Journal of Chemical Physics}\ }\textbf {\bibinfo {volume} {103}},\
  \bibinfo {pages} {3561} (\bibinfo {year} {1995})}\BibitemShut {NoStop}%
\bibitem [{\citenamefont {Hartman}\ \emph {et~al.}(2015)\citenamefont
  {Hartman}, \citenamefont {Monaco}, \citenamefont {Schatschneider},\ and\
  \citenamefont {Beran}}]{hartmanFragmentbased13CNuclear2015}%
  \BibitemOpen
  \bibfield  {author} {\bibinfo {author} {\bibfnamefont {J.~D.}\ \bibnamefont
  {Hartman}}, \bibinfo {author} {\bibfnamefont {S.}~\bibnamefont {Monaco}},
  \bibinfo {author} {\bibfnamefont {B.}~\bibnamefont {Schatschneider}},\ and\
  \bibinfo {author} {\bibfnamefont {G.~J.~O.}\ \bibnamefont {Beran}},\
  }\bibfield  {title} {\bibinfo {title} {Fragment-based {{13C}} nuclear
  magnetic resonance chemical shift predictions in molecular crystals: {{An}}
  alternative to planewave methods},\ }\href
  {https://doi.org/10.1063/1.4922649} {\bibfield  {journal} {\bibinfo
  {journal} {The Journal of Chemical Physics}\ }\textbf {\bibinfo {volume}
  {143}},\ \bibinfo {pages} {102809} (\bibinfo {year} {2015})}\BibitemShut
  {NoStop}%
\bibitem [{\citenamefont {Dittmer}\ \emph {et~al.}(2020)\citenamefont
  {Dittmer}, \citenamefont {Stoychev}, \citenamefont {Maganas}, \citenamefont
  {Auer},\ and\ \citenamefont {Neese}}]{dittmerComputationNMRShielding2020}%
  \BibitemOpen
  \bibfield  {author} {\bibinfo {author} {\bibfnamefont {A.}~\bibnamefont
  {Dittmer}}, \bibinfo {author} {\bibfnamefont {G.~L.}\ \bibnamefont
  {Stoychev}}, \bibinfo {author} {\bibfnamefont {D.}~\bibnamefont {Maganas}},
  \bibinfo {author} {\bibfnamefont {A.~A.}\ \bibnamefont {Auer}},\ and\
  \bibinfo {author} {\bibfnamefont {F.}~\bibnamefont {Neese}},\ }\bibfield
  {title} {\bibinfo {title} {Computation of {{NMR Shielding Constants}} for
  {{Solids Using}} an {{Embedded Cluster Approach}} with {{DFT}},
  {{Double-Hybrid DFT}}, and {{MP2}}},\ }\href
  {https://doi.org/10.1021/acs.jctc.0c00067} {\bibfield  {journal} {\bibinfo
  {journal} {Journal of Chemical Theory and Computation}\ }\textbf {\bibinfo
  {volume} {16}},\ \bibinfo {pages} {6950} (\bibinfo {year}
  {2020})}\BibitemShut {NoStop}%
\bibitem [{\citenamefont {Humbel}\ \emph {et~al.}(1996)\citenamefont {Humbel},
  \citenamefont {Sieber},\ and\ \citenamefont
  {Morokuma}}]{humbelIMOMOMethodIntegration1996}%
  \BibitemOpen
  \bibfield  {author} {\bibinfo {author} {\bibfnamefont {S.}~\bibnamefont
  {Humbel}}, \bibinfo {author} {\bibfnamefont {S.}~\bibnamefont {Sieber}},\
  and\ \bibinfo {author} {\bibfnamefont {K.}~\bibnamefont {Morokuma}},\
  }\bibfield  {title} {\bibinfo {title} {The {{IMOMO}} method: {{Integration}}
  of different levels of molecular orbital approximations for geometry
  optimization of large systems: {{Test}} for n-butane conformation and {{SN2}}
  reaction: {{RCl}}+{{Cl}}-},\ }\href {https://doi.org/10.1063/1.472065}
  {\bibfield  {journal} {\bibinfo  {journal} {The Journal of Chemical Physics}\
  }\textbf {\bibinfo {volume} {105}},\ \bibinfo {pages} {1959} (\bibinfo {year}
  {1996})}\BibitemShut {NoStop}%
\bibitem [{\citenamefont {Dapprich}\ \emph {et~al.}(1999)\citenamefont
  {Dapprich}, \citenamefont {Kom{\'a}romi}, \citenamefont {Byun}, \citenamefont
  {Morokuma},\ and\ \citenamefont
  {Frisch}}]{dapprichNewONIOMImplementation1999}%
  \BibitemOpen
  \bibfield  {author} {\bibinfo {author} {\bibfnamefont {S.}~\bibnamefont
  {Dapprich}}, \bibinfo {author} {\bibfnamefont {I.}~\bibnamefont
  {Kom{\'a}romi}}, \bibinfo {author} {\bibfnamefont {K.~S.}\ \bibnamefont
  {Byun}}, \bibinfo {author} {\bibfnamefont {K.}~\bibnamefont {Morokuma}},\
  and\ \bibinfo {author} {\bibfnamefont {M.~J.}\ \bibnamefont {Frisch}},\
  }\bibfield  {title} {\bibinfo {title} {A new {{ONIOM}} implementation in
  {{Gaussian98}}. {{Part I}}. {{The}} calculation of energies, gradients,
  vibrational frequencies and electric field derivatives1},\ }\href
  {https://doi.org/10.1016/S0166-1280(98)00475-8} {\bibfield  {journal}
  {\bibinfo  {journal} {Journal of Molecular Structure: THEOCHEM}\ }\textbf
  {\bibinfo {volume} {461--462}},\ \bibinfo {pages} {1} (\bibinfo {year}
  {1999})}\BibitemShut {NoStop}%
\bibitem [{\citenamefont
  {Nakajima}(2017)}]{nakajimaExtrapolationSchemeSolidstate2017}%
  \BibitemOpen
  \bibfield  {author} {\bibinfo {author} {\bibfnamefont {T.}~\bibnamefont
  {Nakajima}},\ }\bibfield  {title} {\bibinfo {title} {An extrapolation scheme
  for solid-state {{NMR}} chemical shift calculations},\ }\href
  {https://doi.org/10.1016/j.cplett.2017.04.013} {\bibfield  {journal}
  {\bibinfo  {journal} {Chemical Physics Letters}\ }\textbf {\bibinfo {volume}
  {677}},\ \bibinfo {pages} {99} (\bibinfo {year} {2017})}\BibitemShut
  {NoStop}%
\bibitem [{\citenamefont {Dra{\v c}{\'i}nsk{\'y}}\ \emph
  {et~al.}(2019)\citenamefont {Dra{\v c}{\'i}nsk{\'y}}, \citenamefont
  {Unzueta},\ and\ \citenamefont
  {Beran}}]{dracinskyImprovingAccuracySolidstate2019a}%
  \BibitemOpen
  \bibfield  {author} {\bibinfo {author} {\bibfnamefont {M.}~\bibnamefont
  {Dra{\v c}{\'i}nsk{\'y}}}, \bibinfo {author} {\bibfnamefont {P.}~\bibnamefont
  {Unzueta}},\ and\ \bibinfo {author} {\bibfnamefont {G.~J.~O.}\ \bibnamefont
  {Beran}},\ }\bibfield  {title} {\bibinfo {title} {Improving the accuracy of
  solid-state nuclear magnetic resonance chemical shift prediction with a
  simple molecular correction},\ }\href {https://doi.org/10.1039/C9CP01666J}
  {\bibfield  {journal} {\bibinfo  {journal} {Physical Chemistry Chemical
  Physics}\ }\textbf {\bibinfo {volume} {21}},\ \bibinfo {pages} {14992}
  (\bibinfo {year} {2019})}\BibitemShut {NoStop}%
\bibitem [{\citenamefont
  {Ditchfield}(1974)}]{ditchfieldSelfconsistentPerturbationTheory1974}%
  \BibitemOpen
  \bibfield  {author} {\bibinfo {author} {\bibfnamefont {R.}~\bibnamefont
  {Ditchfield}},\ }\bibfield  {title} {\bibinfo {title} {Self-consistent
  perturbation theory of diamagnetism: {{I}}. {{A}} gauge-invariant {{LCAO}}
  method for {{N}}.{{M}}.{{R}}. chemical shifts},\ }\href
  {https://doi.org/10.1080/00268977400100711} {\bibfield  {journal} {\bibinfo
  {journal} {Molecular Physics}\ }\textbf {\bibinfo {volume} {27}},\ \bibinfo
  {pages} {789} (\bibinfo {year} {1974})}\BibitemShut {NoStop}%
\bibitem [{\citenamefont {Dra{\v c}{\'i}nsk{\'y}}\ \emph
  {et~al.}(2020)\citenamefont {Dra{\v c}{\'i}nsk{\'y}}, \citenamefont
  {V{\'i}cha}, \citenamefont {B{\'a}rtov{\'a}},\ and\ \citenamefont
  {Hodgkinson}}]{dracinskyAccuratePredictionsProton2020}%
  \BibitemOpen
  \bibfield  {author} {\bibinfo {author} {\bibfnamefont {M.}~\bibnamefont
  {Dra{\v c}{\'i}nsk{\'y}}}, \bibinfo {author} {\bibfnamefont {J.}~\bibnamefont
  {V{\'i}cha}}, \bibinfo {author} {\bibfnamefont {K.}~\bibnamefont
  {B{\'a}rtov{\'a}}},\ and\ \bibinfo {author} {\bibfnamefont {P.}~\bibnamefont
  {Hodgkinson}},\ }\bibfield  {title} {\bibinfo {title} {Towards {{Accurate
  Predictions}} of {{Proton NMR Spectroscopic Parameters}} in {{Molecular
  Solids}}},\ }\href {https://doi.org/10.1002/cphc.202000629} {\bibfield
  {journal} {\bibinfo  {journal} {ChemPhysChem}\ }\textbf {\bibinfo {volume}
  {21}},\ \bibinfo {pages} {2075} (\bibinfo {year} {2020})}\BibitemShut
  {NoStop}%
\bibitem [{\citenamefont {Chaloupeck{\'a}}\ \emph {et~al.}(2024)\citenamefont
  {Chaloupeck{\'a}}, \citenamefont {Tyrpekl}, \citenamefont {B{\'a}rtov{\'a}},
  \citenamefont {Nishiyama},\ and\ \citenamefont {Dra{\v
  c}{\'i}nsk{\'y}}}]{chaloupeckaNMRCrystallographyAmino2024}%
  \BibitemOpen
  \bibfield  {author} {\bibinfo {author} {\bibfnamefont {E.}~\bibnamefont
  {Chaloupeck{\'a}}}, \bibinfo {author} {\bibfnamefont {V.}~\bibnamefont
  {Tyrpekl}}, \bibinfo {author} {\bibfnamefont {K.}~\bibnamefont
  {B{\'a}rtov{\'a}}}, \bibinfo {author} {\bibfnamefont {Y.}~\bibnamefont
  {Nishiyama}},\ and\ \bibinfo {author} {\bibfnamefont {M.}~\bibnamefont
  {Dra{\v c}{\'i}nsk{\'y}}},\ }\bibfield  {title} {\bibinfo {title} {{{NMR}}
  crystallography of amino acids},\ }\href
  {https://doi.org/10.1016/j.ssnmr.2024.101921} {\bibfield  {journal} {\bibinfo
   {journal} {Solid State Nuclear Magnetic Resonance}\ }\textbf {\bibinfo
  {volume} {130}},\ \bibinfo {pages} {101921} (\bibinfo {year}
  {2024})}\BibitemShut {NoStop}%
\bibitem [{\citenamefont {Hartman}\ and\ \citenamefont
  {Harper}(2022)}]{hartmanImprovingAccuracyGIPAW2022}%
  \BibitemOpen
  \bibfield  {author} {\bibinfo {author} {\bibfnamefont {J.~D.}\ \bibnamefont
  {Hartman}}\ and\ \bibinfo {author} {\bibfnamefont {J.~K.}\ \bibnamefont
  {Harper}},\ }\bibfield  {title} {\bibinfo {title} {Improving the accuracy of
  {{GIPAW}} chemical shielding calculations with cluster and fragment
  corrections},\ }\href {https://doi.org/10.1016/j.ssnmr.2022.101832}
  {\bibfield  {journal} {\bibinfo  {journal} {Solid State Nuclear Magnetic
  Resonance}\ }\textbf {\bibinfo {volume} {122}},\ \bibinfo {pages} {101832}
  (\bibinfo {year} {2022})}\BibitemShut {NoStop}%
\bibitem [{\citenamefont {Iuliucci}\ \emph {et~al.}(2023)\citenamefont
  {Iuliucci}, \citenamefont {Hartman},\ and\ \citenamefont
  {Beran}}]{iuliucciModelsHybridDensity2023}%
  \BibitemOpen
  \bibfield  {author} {\bibinfo {author} {\bibfnamefont {R.~J.}\ \bibnamefont
  {Iuliucci}}, \bibinfo {author} {\bibfnamefont {J.~D.}\ \bibnamefont
  {Hartman}},\ and\ \bibinfo {author} {\bibfnamefont {G.~J.~O.}\ \bibnamefont
  {Beran}},\ }\bibfield  {title} {\bibinfo {title} {Do {{Models}} beyond
  {{Hybrid Density Functionals Increase}} the {{Agreement}} with {{Experiment}}
  for {{Predicted NMR Chemical Shifts}} or {{Electric Field Gradient Tensors}}
  in {{Organic Solids}}?},\ }\href {https://doi.org/10.1021/acs.jpca.2c07657}
  {\bibfield  {journal} {\bibinfo  {journal} {The Journal of Physical Chemistry
  A}\ }\textbf {\bibinfo {volume} {127}},\ \bibinfo {pages} {2846} (\bibinfo
  {year} {2023})}\BibitemShut {NoStop}%
\bibitem [{\citenamefont {Ramos}\ \emph {et~al.}(2025)\citenamefont {Ramos},
  \citenamefont {Mueller},\ and\ \citenamefont
  {Beran}}]{ramosInterplayDensityFunctional2024}%
  \BibitemOpen
  \bibfield  {author} {\bibinfo {author} {\bibfnamefont {S.~A.}\ \bibnamefont
  {Ramos}}, \bibinfo {author} {\bibfnamefont {L.~J.}\ \bibnamefont {Mueller}},\
  and\ \bibinfo {author} {\bibfnamefont {G.~J.~O.}\ \bibnamefont {Beran}},\
  }\bibfield  {title} {\bibinfo {title} {The interplay of density functional
  selection and crystal structure for accurate {{NMR}} chemical shift
  predictions},\ }\href {https://doi.org/10.1039/D4FD00072B} {\bibfield
  {journal} {\bibinfo  {journal} {Faraday Discussions}\ }\textbf {\bibinfo
  {volume} {255}},\ \bibinfo {pages} {119} (\bibinfo {year}
  {2025})}\BibitemShut {NoStop}%
\bibitem [{\citenamefont {Dra{\v
  c}{\'i}nsk{\'y}}(2021)}]{dracinskyAnalyzingDiscrepanciesChemicalShift2021}%
  \BibitemOpen
  \bibfield  {author} {\bibinfo {author} {\bibfnamefont {M.}~\bibnamefont
  {Dra{\v c}{\'i}nsk{\'y}}},\ }\bibfield  {title} {\bibinfo {title} {Analyzing
  {{Discrepancies}} in {{Chemical-Shift Predictions}} of {{Solid Pyridinium
  Fumarates}}},\ }\href {https://doi.org/10.3390/molecules26133857} {\bibfield
  {journal} {\bibinfo  {journal} {Molecules}\ }\textbf {\bibinfo {volume}
  {26}},\ \bibinfo {pages} {3857} (\bibinfo {year} {2021})}\BibitemShut
  {NoStop}%
\bibitem [{\citenamefont {Holmes}\ \emph {et~al.}(2020)\citenamefont {Holmes},
  \citenamefont {Engl}, \citenamefont {Srnec}, \citenamefont {Madura},
  \citenamefont {Qui{\~n}ones}, \citenamefont {Harper}, \citenamefont
  {Schurko},\ and\ \citenamefont {Iuliucci}}]{holmesChemicalShiftTensors2020}%
  \BibitemOpen
  \bibfield  {author} {\bibinfo {author} {\bibfnamefont {S.~T.}\ \bibnamefont
  {Holmes}}, \bibinfo {author} {\bibfnamefont {O.~G.}\ \bibnamefont {Engl}},
  \bibinfo {author} {\bibfnamefont {M.~N.}\ \bibnamefont {Srnec}}, \bibinfo
  {author} {\bibfnamefont {J.~D.}\ \bibnamefont {Madura}}, \bibinfo {author}
  {\bibfnamefont {R.}~\bibnamefont {Qui{\~n}ones}}, \bibinfo {author}
  {\bibfnamefont {J.~K.}\ \bibnamefont {Harper}}, \bibinfo {author}
  {\bibfnamefont {R.~W.}\ \bibnamefont {Schurko}},\ and\ \bibinfo {author}
  {\bibfnamefont {R.~J.}\ \bibnamefont {Iuliucci}},\ }\bibfield  {title}
  {\bibinfo {title} {Chemical {{Shift Tensors}} of {{Cimetidine Form A
  Modeled}} with {{Density Functional Theory Calculations}}: {{Implications}}
  for {{NMR Crystallography}}},\ }\href
  {https://doi.org/10.1021/acs.jpca.0c00421} {\bibfield  {journal} {\bibinfo
  {journal} {The Journal of Physical Chemistry A}\ }\textbf {\bibinfo {volume}
  {124}},\ \bibinfo {pages} {3109} (\bibinfo {year} {2020})}\BibitemShut
  {NoStop}%
\bibitem [{\citenamefont {Toomey}\ \emph {et~al.}(2024)\citenamefont {Toomey},
  \citenamefont {Wang}, \citenamefont {Heider}, \citenamefont {Hartman},
  \citenamefont {Nichols}, \citenamefont {Myles}, \citenamefont {Gardberg},
  \citenamefont {McIntyre}, \citenamefont {Zeller}, \citenamefont {Mehta},\
  and\ \citenamefont {Harper}}]{toomeyNMRguidedRefinementCrystal2024}%
  \BibitemOpen
  \bibfield  {author} {\bibinfo {author} {\bibfnamefont {R.}~\bibnamefont
  {Toomey}}, \bibinfo {author} {\bibfnamefont {L.}~\bibnamefont {Wang}},
  \bibinfo {author} {\bibfnamefont {E.~C.}\ \bibnamefont {Heider}}, \bibinfo
  {author} {\bibfnamefont {J.~D.}\ \bibnamefont {Hartman}}, \bibinfo {author}
  {\bibfnamefont {A.~J.}\ \bibnamefont {Nichols}}, \bibinfo {author}
  {\bibfnamefont {D.~A.~A.}\ \bibnamefont {Myles}}, \bibinfo {author}
  {\bibfnamefont {A.~S.}\ \bibnamefont {Gardberg}}, \bibinfo {author}
  {\bibfnamefont {G.~J.}\ \bibnamefont {McIntyre}}, \bibinfo {author}
  {\bibfnamefont {M.}~\bibnamefont {Zeller}}, \bibinfo {author} {\bibfnamefont
  {M.~A.}\ \bibnamefont {Mehta}},\ and\ \bibinfo {author} {\bibfnamefont
  {J.~K.}\ \bibnamefont {Harper}},\ }\bibfield  {title} {\bibinfo {title}
  {{{NMR-guided}} refinement of crystal structures using{\textsuperscript{15}}
  {{N}} chemical shift tensors},\ }\href {https://doi.org/10.1039/D4CE00237G}
  {\bibfield  {journal} {\bibinfo  {journal} {CrystEngComm}\ }\textbf {\bibinfo
  {volume} {26}},\ \bibinfo {pages} {3289} (\bibinfo {year}
  {2024})}\BibitemShut {NoStop}%
\bibitem [{\citenamefont {Mathews}\ and\ \citenamefont
  {Hartman}(2021)}]{mathewsAccurateFragmentbased51V2021}%
  \BibitemOpen
  \bibfield  {author} {\bibinfo {author} {\bibfnamefont {A.}~\bibnamefont
  {Mathews}}\ and\ \bibinfo {author} {\bibfnamefont {J.~D.}\ \bibnamefont
  {Hartman}},\ }\bibfield  {title} {\bibinfo {title} {Accurate fragment-based
  51-{{V}} chemical shift predictions in molecular crystals},\ }\href
  {https://doi.org/10.1016/j.ssnmr.2021.101733} {\bibfield  {journal} {\bibinfo
   {journal} {Solid State Nuclear Magnetic Resonance}\ }\textbf {\bibinfo
  {volume} {114}},\ \bibinfo {pages} {101733} (\bibinfo {year}
  {2021})}\BibitemShut {NoStop}%
\bibitem [{\citenamefont {Chaloupeck{\'a}}\ \emph {et~al.}(2025)\citenamefont
  {Chaloupeck{\'a}}, \citenamefont {Socha},\ and\ \citenamefont {Dra{\v
  c}{\'i}nsk{\'y}}}]{chaloupeckaDivergingErrorsComparison2025}%
  \BibitemOpen
  \bibfield  {author} {\bibinfo {author} {\bibfnamefont {E.}~\bibnamefont
  {Chaloupeck{\'a}}}, \bibinfo {author} {\bibfnamefont {O.}~\bibnamefont
  {Socha}},\ and\ \bibinfo {author} {\bibfnamefont {M.}~\bibnamefont {Dra{\v
  c}{\'i}nsk{\'y}}},\ }\bibfield  {title} {\bibinfo {title} {Diverging errors:
  {{A}} comparison of {{DFT}} and machine-learning predictions of {{NMR}}
  shieldings},\ }\href {https://doi.org/10.1016/j.ssnmr.2025.102019} {\bibfield
   {journal} {\bibinfo  {journal} {Solid State Nuclear Magnetic Resonance}\
  }\textbf {\bibinfo {volume} {138}},\ \bibinfo {pages} {102019} (\bibinfo
  {year} {2025})}\BibitemShut {NoStop}%
\bibitem [{\citenamefont {Chaloupecká}(2025)}]{chaloupecka2025nmr}%
  \BibitemOpen
  \bibfield  {author} {\bibinfo {author} {\bibfnamefont {E.}~\bibnamefont
  {Chaloupecká}},\ }\emph {\bibinfo {title} {Study of Structure and
  Interactions of Biologically Active Molecules by {NMR} Spectroscopy}},\
  \href@noop {} {\bibinfo {type} {Doctoral thesis}},\ \bibinfo  {school}
  {Charles University, Faculty of Science}, \bibinfo {address} {Prague, Czech
  Republic} (\bibinfo {year} {2025})\BibitemShut {NoStop}%
\bibitem [{\citenamefont
  {Wiberg}(1968)}]{wibergApplicationPoplesantrysegalCNDO1968}%
  \BibitemOpen
  \bibfield  {author} {\bibinfo {author} {\bibfnamefont {K.~B.}\ \bibnamefont
  {Wiberg}},\ }\bibfield  {title} {\bibinfo {title} {Application of the
  pople-santry-segal {{CNDO}} method to the cyclopropylcarbinyl and cyclobutyl
  cation and to bicyclobutane},\ }\href
  {https://doi.org/10.1016/0040-4020(68)88057-3} {\bibfield  {journal}
  {\bibinfo  {journal} {Tetrahedron}\ }\textbf {\bibinfo {volume} {24}},\
  \bibinfo {pages} {1083} (\bibinfo {year} {1968})}\BibitemShut {NoStop}%
\bibitem [{\citenamefont {Mayer}(1983)}]{mayerChargeBondOrder1983}%
  \BibitemOpen
  \bibfield  {author} {\bibinfo {author} {\bibfnamefont {I.}~\bibnamefont
  {Mayer}},\ }\bibfield  {title} {\bibinfo {title} {Charge, bond order and
  valence in the {{AB}} initio {{SCF}} theory},\ }\href
  {https://doi.org/10.1016/0009-2614(83)80005-0} {\bibfield  {journal}
  {\bibinfo  {journal} {Chemical Physics Letters}\ }\textbf {\bibinfo {volume}
  {97}},\ \bibinfo {pages} {270} (\bibinfo {year} {1983})}\BibitemShut
  {NoStop}%
\bibitem [{\citenamefont {Bannwarth}\ \emph {et~al.}(2019)\citenamefont
  {Bannwarth}, \citenamefont {Ehlert},\ and\ \citenamefont
  {Grimme}}]{bannwarthGFN2xTBAnAccurateBroadly2019}%
  \BibitemOpen
  \bibfield  {author} {\bibinfo {author} {\bibfnamefont {C.}~\bibnamefont
  {Bannwarth}}, \bibinfo {author} {\bibfnamefont {S.}~\bibnamefont {Ehlert}},\
  and\ \bibinfo {author} {\bibfnamefont {S.}~\bibnamefont {Grimme}},\
  }\bibfield  {title} {\bibinfo {title} {{{GFN2-xTB}}---{{An Accurate}} and
  {{Broadly Parametrized Self-Consistent Tight-Binding Quantum Chemical
  Method}} with {{Multipole Electrostatics}} and {{Density-Dependent Dispersion
  Contributions}}},\ }\href {https://doi.org/10.1021/acs.jctc.8b01176}
  {\bibfield  {journal} {\bibinfo  {journal} {Journal of Chemical Theory and
  Computation}\ }\textbf {\bibinfo {volume} {15}},\ \bibinfo {pages} {1652}
  (\bibinfo {year} {2019})}\BibitemShut {NoStop}%
\bibitem [{\citenamefont {Ong}\ \emph {et~al.}(2013)\citenamefont {Ong},
  \citenamefont {Richards}, \citenamefont {Jain}, \citenamefont {Hautier},
  \citenamefont {Kocher}, \citenamefont {Cholia}, \citenamefont {Gunter},
  \citenamefont {Chevrier}, \citenamefont {Persson},\ and\ \citenamefont
  {Ceder}}]{ongPythonMaterialsGenomics2013}%
  \BibitemOpen
  \bibfield  {author} {\bibinfo {author} {\bibfnamefont {S.~P.}\ \bibnamefont
  {Ong}}, \bibinfo {author} {\bibfnamefont {W.~D.}\ \bibnamefont {Richards}},
  \bibinfo {author} {\bibfnamefont {A.}~\bibnamefont {Jain}}, \bibinfo {author}
  {\bibfnamefont {G.}~\bibnamefont {Hautier}}, \bibinfo {author} {\bibfnamefont
  {M.}~\bibnamefont {Kocher}}, \bibinfo {author} {\bibfnamefont
  {S.}~\bibnamefont {Cholia}}, \bibinfo {author} {\bibfnamefont
  {D.}~\bibnamefont {Gunter}}, \bibinfo {author} {\bibfnamefont {V.~L.}\
  \bibnamefont {Chevrier}}, \bibinfo {author} {\bibfnamefont {K.~A.}\
  \bibnamefont {Persson}},\ and\ \bibinfo {author} {\bibfnamefont
  {G.}~\bibnamefont {Ceder}},\ }\bibfield  {title} {\bibinfo {title} {Python
  {{Materials Genomics}} (pymatgen): {{A}} robust, open-source python library
  for materials analysis},\ }\href
  {https://doi.org/10.1016/j.commatsci.2012.10.028} {\bibfield  {journal}
  {\bibinfo  {journal} {Computational Materials Science}\ }\textbf {\bibinfo
  {volume} {68}},\ \bibinfo {pages} {314} (\bibinfo {year} {2013})}\BibitemShut
  {NoStop}%
\bibitem [{\citenamefont {Hjorth~Larsen}\ \emph {et~al.}(2017)\citenamefont
  {Hjorth~Larsen}, \citenamefont {J{\o}rgen~Mortensen}, \citenamefont
  {Blomqvist}, \citenamefont {Castelli}, \citenamefont {Christensen},
  \citenamefont {Du{\l}ak}, \citenamefont {Friis}, \citenamefont {Groves},
  \citenamefont {Hammer}, \citenamefont {Hargus}, \citenamefont {Hermes},
  \citenamefont {Jennings}, \citenamefont {Bjerre~Jensen}, \citenamefont
  {Kermode}, \citenamefont {Kitchin}, \citenamefont {Leonhard~Kolsbjerg},
  \citenamefont {Kubal}, \citenamefont {Kaasbjerg}, \citenamefont {Lysgaard},
  \citenamefont {Bergmann~Maronsson}, \citenamefont {Maxson}, \citenamefont
  {Olsen}, \citenamefont {Pastewka}, \citenamefont {Peterson}, \citenamefont
  {Rostgaard}, \citenamefont {Schi{\o}tz}, \citenamefont {Sch{\"u}tt},
  \citenamefont {Strange}, \citenamefont {Thygesen}, \citenamefont {Vegge},
  \citenamefont {Vilhelmsen}, \citenamefont {Walter}, \citenamefont {Zeng},\
  and\ \citenamefont {Jacobsen}}]{hjorthlarsenAtomicSimulationEnvironment2017}%
  \BibitemOpen
  \bibfield  {author} {\bibinfo {author} {\bibfnamefont {A.}~\bibnamefont
  {Hjorth~Larsen}}, \bibinfo {author} {\bibfnamefont {J.}~\bibnamefont
  {J{\o}rgen~Mortensen}}, \bibinfo {author} {\bibfnamefont {J.}~\bibnamefont
  {Blomqvist}}, \bibinfo {author} {\bibfnamefont {I.~E.}\ \bibnamefont
  {Castelli}}, \bibinfo {author} {\bibfnamefont {R.}~\bibnamefont
  {Christensen}}, \bibinfo {author} {\bibfnamefont {M.}~\bibnamefont
  {Du{\l}ak}}, \bibinfo {author} {\bibfnamefont {J.}~\bibnamefont {Friis}},
  \bibinfo {author} {\bibfnamefont {M.~N.}\ \bibnamefont {Groves}}, \bibinfo
  {author} {\bibfnamefont {B.}~\bibnamefont {Hammer}}, \bibinfo {author}
  {\bibfnamefont {C.}~\bibnamefont {Hargus}}, \bibinfo {author} {\bibfnamefont
  {E.~D.}\ \bibnamefont {Hermes}}, \bibinfo {author} {\bibfnamefont {P.~C.}\
  \bibnamefont {Jennings}}, \bibinfo {author} {\bibfnamefont {P.}~\bibnamefont
  {Bjerre~Jensen}}, \bibinfo {author} {\bibfnamefont {J.}~\bibnamefont
  {Kermode}}, \bibinfo {author} {\bibfnamefont {J.~R.}\ \bibnamefont
  {Kitchin}}, \bibinfo {author} {\bibfnamefont {E.}~\bibnamefont
  {Leonhard~Kolsbjerg}}, \bibinfo {author} {\bibfnamefont {J.}~\bibnamefont
  {Kubal}}, \bibinfo {author} {\bibfnamefont {K.}~\bibnamefont {Kaasbjerg}},
  \bibinfo {author} {\bibfnamefont {S.}~\bibnamefont {Lysgaard}}, \bibinfo
  {author} {\bibfnamefont {J.}~\bibnamefont {Bergmann~Maronsson}}, \bibinfo
  {author} {\bibfnamefont {T.}~\bibnamefont {Maxson}}, \bibinfo {author}
  {\bibfnamefont {T.}~\bibnamefont {Olsen}}, \bibinfo {author} {\bibfnamefont
  {L.}~\bibnamefont {Pastewka}}, \bibinfo {author} {\bibfnamefont
  {A.}~\bibnamefont {Peterson}}, \bibinfo {author} {\bibfnamefont
  {C.}~\bibnamefont {Rostgaard}}, \bibinfo {author} {\bibfnamefont
  {J.}~\bibnamefont {Schi{\o}tz}}, \bibinfo {author} {\bibfnamefont
  {O.}~\bibnamefont {Sch{\"u}tt}}, \bibinfo {author} {\bibfnamefont
  {M.}~\bibnamefont {Strange}}, \bibinfo {author} {\bibfnamefont {K.~S.}\
  \bibnamefont {Thygesen}}, \bibinfo {author} {\bibfnamefont {T.}~\bibnamefont
  {Vegge}}, \bibinfo {author} {\bibfnamefont {L.}~\bibnamefont {Vilhelmsen}},
  \bibinfo {author} {\bibfnamefont {M.}~\bibnamefont {Walter}}, \bibinfo
  {author} {\bibfnamefont {Z.}~\bibnamefont {Zeng}},\ and\ \bibinfo {author}
  {\bibfnamefont {K.~W.}\ \bibnamefont {Jacobsen}},\ }\bibfield  {title}
  {\bibinfo {title} {The atomic simulation environment---a {{Python}} library
  for working with atoms},\ }\href {https://doi.org/10.1088/1361-648X/aa680e}
  {\bibfield  {journal} {\bibinfo  {journal} {Journal of Physics: Condensed
  Matter}\ }\textbf {\bibinfo {volume} {29}},\ \bibinfo {pages} {273002}
  (\bibinfo {year} {2017})}\BibitemShut {NoStop}%
\bibitem [{\citenamefont {Hagberg}\ \emph {et~al.}(2008)\citenamefont
  {Hagberg}, \citenamefont {Schult},\ and\ \citenamefont
  {Swart}}]{hagbergExploringNetworkStructure2008}%
  \BibitemOpen
  \bibfield  {author} {\bibinfo {author} {\bibfnamefont {A.~A.}\ \bibnamefont
  {Hagberg}}, \bibinfo {author} {\bibfnamefont {D.~A.}\ \bibnamefont
  {Schult}},\ and\ \bibinfo {author} {\bibfnamefont {P.~J.}\ \bibnamefont
  {Swart}},\ }\bibfield  {title} {\bibinfo {title} {Exploring {{Network
  Structure}}, {{Dynamics}}, and {{Function}} using {{NetworkX}}},\ }\href
  {https://doi.org/10.25080/tcwv9851} {\bibfield  {journal} {\bibinfo
  {journal} {Proceedings of the Python in Science Conference}\ ,\ \bibinfo
  {pages} {11}} (\bibinfo {year} {2008})}\BibitemShut {NoStop}%
\bibitem [{\citenamefont
  {Mulliken}(1955)}]{mullikenElectronicPopulationAnalysis1955}%
  \BibitemOpen
  \bibfield  {author} {\bibinfo {author} {\bibfnamefont {R.~S.}\ \bibnamefont
  {Mulliken}},\ }\bibfield  {title} {\bibinfo {title} {Electronic {{Population
  Analysis}} on {{LCAO}}--{{MO Molecular Wave Functions}}. {{I}}},\ }\href
  {https://doi.org/10.1063/1.1740588} {\bibfield  {journal} {\bibinfo
  {journal} {The Journal of Chemical Physics}\ }\textbf {\bibinfo {volume}
  {23}},\ \bibinfo {pages} {1833} (\bibinfo {year} {1955})}\BibitemShut
  {NoStop}%
\bibitem [{Tbl(2026)}]{TbliteTblite2026}%
  \BibitemOpen
  \href@noop {} {\bibinfo {title} {Tblite package,
  https://github.com/tblite/tblite, last visited 23.07.2026}} (\bibinfo {year}
  {2026})\BibitemShut {NoStop}%
\bibitem [{\citenamefont {Stoychev}\ \emph {et~al.}(2018)\citenamefont
  {Stoychev}, \citenamefont {Auer}, \citenamefont {Izs{\'a}k},\ and\
  \citenamefont {Neese}}]{stoychevSelfConsistentFieldCalculation2018}%
  \BibitemOpen
  \bibfield  {author} {\bibinfo {author} {\bibfnamefont {G.~L.}\ \bibnamefont
  {Stoychev}}, \bibinfo {author} {\bibfnamefont {A.~A.}\ \bibnamefont {Auer}},
  \bibinfo {author} {\bibfnamefont {R.}~\bibnamefont {Izs{\'a}k}},\ and\
  \bibinfo {author} {\bibfnamefont {F.}~\bibnamefont {Neese}},\ }\bibfield
  {title} {\bibinfo {title} {Self-{{Consistent Field Calculation}} of {{Nuclear
  Magnetic Resonance Chemical Shielding Constants Using Gauge-Including Atomic
  Orbitals}} and {{Approximate Two-Electron Integrals}}},\ }\href
  {https://doi.org/10.1021/acs.jctc.7b01006} {\bibfield  {journal} {\bibinfo
  {journal} {Journal of Chemical Theory and Computation}\ }\textbf {\bibinfo
  {volume} {14}},\ \bibinfo {pages} {619} (\bibinfo {year} {2018})}\BibitemShut
  {NoStop}%
\bibitem [{\citenamefont {Neese}(2025)}]{neeseSoftwareUpdateORCA2025}%
  \BibitemOpen
  \bibfield  {author} {\bibinfo {author} {\bibfnamefont {F.}~\bibnamefont
  {Neese}},\ }\bibfield  {title} {\bibinfo {title} {Software {{Update}}: {{The
  ORCA Program System}}---{{Version}} 6.0},\ }\href
  {https://doi.org/10.1002/wcms.70019} {\bibfield  {journal} {\bibinfo
  {journal} {WIREs Computational Molecular Science}\ }\textbf {\bibinfo
  {volume} {15}},\ \bibinfo {pages} {e70019} (\bibinfo {year}
  {2025})}\BibitemShut {NoStop}%
\bibitem [{\citenamefont {Tetenberg}\ \emph {et~al.}(2026)\citenamefont
  {Tetenberg}, \citenamefont {Neugebauer}, \citenamefont {Plett}, \citenamefont
  {Santhosh}, \citenamefont {Bursch},\ and\ \citenamefont
  {Riplinger}}]{tetenbergORCAMeetsPythonThe2026}%
  \BibitemOpen
  \bibfield  {author} {\bibinfo {author} {\bibfnamefont {T.}~\bibnamefont
  {Tetenberg}}, \bibinfo {author} {\bibfnamefont {H.}~\bibnamefont
  {Neugebauer}}, \bibinfo {author} {\bibfnamefont {C.}~\bibnamefont {Plett}},
  \bibinfo {author} {\bibfnamefont {N.}~\bibnamefont {Santhosh}}, \bibinfo
  {author} {\bibfnamefont {M.}~\bibnamefont {Bursch}},\ and\ \bibinfo {author}
  {\bibfnamefont {C.}~\bibnamefont {Riplinger}},\ }\bibfield  {title} {\bibinfo
  {title} {{{ORCA Meets Python}}-{{The ORCA Python Interface OPI}}},\ }\href
  {https://doi.org/10.1021/acs.jctc.5c02141} {\bibfield  {journal} {\bibinfo
  {journal} {Journal of Chemical Theory and Computation}\ }\textbf {\bibinfo
  {volume} {22}},\ \bibinfo {pages} {4951} (\bibinfo {year}
  {2026})}\BibitemShut {NoStop}%
\bibitem [{\citenamefont {Perdew}\ \emph {et~al.}(1996)\citenamefont {Perdew},
  \citenamefont {Burke},\ and\ \citenamefont
  {Ernzerhof}}]{perdewGeneralizedGradientApproximation1996}%
  \BibitemOpen
  \bibfield  {author} {\bibinfo {author} {\bibfnamefont {J.~P.}\ \bibnamefont
  {Perdew}}, \bibinfo {author} {\bibfnamefont {K.}~\bibnamefont {Burke}},\ and\
  \bibinfo {author} {\bibfnamefont {M.}~\bibnamefont {Ernzerhof}},\ }\bibfield
  {title} {\bibinfo {title} {Generalized {{Gradient Approximation Made
  Simple}}},\ }\href {https://doi.org/10.1103/PhysRevLett.77.3865} {\bibfield
  {journal} {\bibinfo  {journal} {Physical Review Letters}\ }\textbf {\bibinfo
  {volume} {77}},\ \bibinfo {pages} {3865} (\bibinfo {year}
  {1996})}\BibitemShut {NoStop}%
\bibitem [{\citenamefont {Adamo}\ and\ \citenamefont
  {Barone}(1999)}]{adam-baro99jcp}%
  \BibitemOpen
  \bibfield  {author} {\bibinfo {author} {\bibfnamefont {C.}~\bibnamefont
  {Adamo}}\ and\ \bibinfo {author} {\bibfnamefont {V.}~\bibnamefont {Barone}},\
  }\bibfield  {title} {\bibinfo {title} {{Toward reliable density functional
  methods without adjustable parameters: The PBE0 model}},\ }\href@noop {}
  {\bibfield  {journal} {\bibinfo  {journal} {J. Chem. Phys.}\ }\textbf
  {\bibinfo {volume} {110}},\ \bibinfo {pages} {6158} (\bibinfo {year}
  {1999})}\BibitemShut {NoStop}%
\bibitem [{\citenamefont {Dunning}(1989)}]{dunningGaussianBasisSets1989}%
  \BibitemOpen
  \bibfield  {author} {\bibinfo {author} {\bibfnamefont {T.~H.}\ \bibnamefont
  {Dunning}, \bibfnamefont {Jr.}},\ }\bibfield  {title} {\bibinfo {title}
  {Gaussian basis sets for use in correlated molecular calculations. {{I}}.
  {{The}} atoms boron through neon and hydrogen},\ }\href
  {https://doi.org/10.1063/1.456153} {\bibfield  {journal} {\bibinfo  {journal}
  {The Journal of Chemical Physics}\ }\textbf {\bibinfo {volume} {90}},\
  \bibinfo {pages} {1007} (\bibinfo {year} {1989})}\BibitemShut {NoStop}%
\bibitem [{\citenamefont {Woon}\ and\ \citenamefont
  {Dunning}(1993)}]{woonGaussianBasisSets1993}%
  \BibitemOpen
  \bibfield  {author} {\bibinfo {author} {\bibfnamefont {D.~E.}\ \bibnamefont
  {Woon}}\ and\ \bibinfo {author} {\bibfnamefont {T.~H.}\ \bibnamefont
  {Dunning}, \bibfnamefont {Jr.}},\ }\bibfield  {title} {\bibinfo {title}
  {Gaussian basis sets for use in correlated molecular calculations. {{III}}.
  {{The}} atoms aluminum through argon},\ }\href
  {https://doi.org/10.1063/1.464303} {\bibfield  {journal} {\bibinfo  {journal}
  {The Journal of Chemical Physics}\ }\textbf {\bibinfo {volume} {98}},\
  \bibinfo {pages} {1358} (\bibinfo {year} {1993})}\BibitemShut {NoStop}%
\bibitem [{\citenamefont {Vahtras}\ \emph {et~al.}(1993)\citenamefont
  {Vahtras}, \citenamefont {Alml{\"o}f},\ and\ \citenamefont
  {Feyereisen}}]{vahtrasIntegralApproximationsLCAOSCF1993}%
  \BibitemOpen
  \bibfield  {author} {\bibinfo {author} {\bibfnamefont {O.}~\bibnamefont
  {Vahtras}}, \bibinfo {author} {\bibfnamefont {J.}~\bibnamefont
  {Alml{\"o}f}},\ and\ \bibinfo {author} {\bibfnamefont {M.~W.}\ \bibnamefont
  {Feyereisen}},\ }\bibfield  {title} {\bibinfo {title} {Integral
  approximations for {{LCAO-SCF}} calculations},\ }\href
  {https://doi.org/10.1016/0009-2614(93)89151-7} {\bibfield  {journal}
  {\bibinfo  {journal} {Chemical Physics Letters}\ }\textbf {\bibinfo {volume}
  {213}},\ \bibinfo {pages} {514} (\bibinfo {year} {1993})}\BibitemShut
  {NoStop}%
\bibitem [{\citenamefont
  {Neese}(2003)}]{neeseImprovementResolutionIdentity2003}%
  \BibitemOpen
  \bibfield  {author} {\bibinfo {author} {\bibfnamefont {F.}~\bibnamefont
  {Neese}},\ }\bibfield  {title} {\bibinfo {title} {An improvement of the
  resolution of the identity approximation for the formation of the {{Coulomb}}
  matrix},\ }\href {https://doi.org/10.1002/jcc.10318} {\bibfield  {journal}
  {\bibinfo  {journal} {Journal of Computational Chemistry}\ }\textbf {\bibinfo
  {volume} {24}},\ \bibinfo {pages} {1740} (\bibinfo {year}
  {2003})}\BibitemShut {NoStop}%
\bibitem [{\citenamefont {Neese}\ \emph {et~al.}(2009)\citenamefont {Neese},
  \citenamefont {Wennmohs}, \citenamefont {Hansen},\ and\ \citenamefont
  {Becker}}]{neeseEfficientApproximateParallel2009}%
  \BibitemOpen
  \bibfield  {author} {\bibinfo {author} {\bibfnamefont {F.}~\bibnamefont
  {Neese}}, \bibinfo {author} {\bibfnamefont {F.}~\bibnamefont {Wennmohs}},
  \bibinfo {author} {\bibfnamefont {A.}~\bibnamefont {Hansen}},\ and\ \bibinfo
  {author} {\bibfnamefont {U.}~\bibnamefont {Becker}},\ }\bibfield  {title}
  {\bibinfo {title} {Efficient, approximate and parallel {{Hartree}}--{{Fock}}
  and hybrid {{DFT}} calculations. {{A}} `chain-of-spheres' algorithm for the
  {{Hartree}}--{{Fock}} exchange},\ }\href
  {https://doi.org/10.1016/j.chemphys.2008.10.036} {\bibfield  {journal}
  {\bibinfo  {journal} {Chemical Physics}\ }\bibinfo {series} {Moving
  {{Frontiers}} in {{Quantum Chemistry}}:},\ \textbf {\bibinfo {volume}
  {356}},\ \bibinfo {pages} {98} (\bibinfo {year} {2009})}\BibitemShut
  {NoStop}%
\bibitem [{\citenamefont {Barone}\ and\ \citenamefont
  {Cossi}(1998)}]{baroneQuantumCalculationMolecular1998}%
  \BibitemOpen
  \bibfield  {author} {\bibinfo {author} {\bibfnamefont {V.}~\bibnamefont
  {Barone}}\ and\ \bibinfo {author} {\bibfnamefont {M.}~\bibnamefont {Cossi}},\
  }\bibfield  {title} {\bibinfo {title} {Quantum {{Calculation}} of {{Molecular
  Energies}} and {{Energy Gradients}} in {{Solution}} by a {{Conductor Solvent
  Model}}},\ }\href {https://doi.org/10.1021/jp9716997} {\bibfield  {journal}
  {\bibinfo  {journal} {The Journal of Physical Chemistry A}\ }\textbf
  {\bibinfo {volume} {102}},\ \bibinfo {pages} {1995} (\bibinfo {year}
  {1998})}\BibitemShut {NoStop}%
\bibitem [{\citenamefont {Pozdnyakov}\ and\ \citenamefont
  {Ceriotti}(2023)}]{NEURIPS2023_fb4a7e35}%
  \BibitemOpen
  \bibfield  {author} {\bibinfo {author} {\bibfnamefont {S.}~\bibnamefont
  {Pozdnyakov}}\ and\ \bibinfo {author} {\bibfnamefont {M.}~\bibnamefont
  {Ceriotti}},\ }\bibfield  {title} {\bibinfo {title} {Smooth, exact rotational
  symmetrization for deep learning on point clouds},\ }in\ \href@noop {} {\emph
  {\bibinfo {booktitle} {Advances in Neural Information Processing Systems}}},\
  Vol.~\bibinfo {volume} {36},\ \bibinfo {editor} {edited by\ \bibinfo {editor}
  {\bibfnamefont {A.}~\bibnamefont {Oh}}, \bibinfo {editor} {\bibfnamefont
  {T.}~\bibnamefont {Naumann}}, \bibinfo {editor} {\bibfnamefont
  {A.}~\bibnamefont {Globerson}}, \bibinfo {editor} {\bibfnamefont
  {K.}~\bibnamefont {Saenko}}, \bibinfo {editor} {\bibfnamefont
  {M.}~\bibnamefont {Hardt}},\ and\ \bibinfo {editor} {\bibfnamefont
  {S.}~\bibnamefont {Levine}}}\ (\bibinfo  {publisher} {Curran Associates,
  Inc.},\ \bibinfo {year} {2023})\ pp.\ \bibinfo {pages}
  {79469--79501}\BibitemShut {NoStop}%
\bibitem [{\citenamefont {Bigi}\ \emph {et~al.}(2026)\citenamefont {Bigi},
  \citenamefont {Abbott}, \citenamefont {Loche}, \citenamefont {Mazitov},
  \citenamefont {Tisi}, \citenamefont {Langer}, \citenamefont {Goscinski},
  \citenamefont {Pegolo}, \citenamefont {Chong}, \citenamefont {Goswami},
  \citenamefont {Febrer}, \citenamefont {Chorna}, \citenamefont {Kellner},
  \citenamefont {Ceriotti},\ and\ \citenamefont
  {Fraux}}]{bigiMetatensorMetatomicFoundational2026}%
  \BibitemOpen
  \bibfield  {author} {\bibinfo {author} {\bibfnamefont {F.}~\bibnamefont
  {Bigi}}, \bibinfo {author} {\bibfnamefont {J.~W.}\ \bibnamefont {Abbott}},
  \bibinfo {author} {\bibfnamefont {P.}~\bibnamefont {Loche}}, \bibinfo
  {author} {\bibfnamefont {A.}~\bibnamefont {Mazitov}}, \bibinfo {author}
  {\bibfnamefont {D.}~\bibnamefont {Tisi}}, \bibinfo {author} {\bibfnamefont
  {M.~F.}\ \bibnamefont {Langer}}, \bibinfo {author} {\bibfnamefont
  {A.}~\bibnamefont {Goscinski}}, \bibinfo {author} {\bibfnamefont
  {P.}~\bibnamefont {Pegolo}}, \bibinfo {author} {\bibfnamefont
  {S.}~\bibnamefont {Chong}}, \bibinfo {author} {\bibfnamefont
  {R.}~\bibnamefont {Goswami}}, \bibinfo {author} {\bibfnamefont
  {P.}~\bibnamefont {Febrer}}, \bibinfo {author} {\bibfnamefont
  {S.}~\bibnamefont {Chorna}}, \bibinfo {author} {\bibfnamefont
  {M.}~\bibnamefont {Kellner}}, \bibinfo {author} {\bibfnamefont
  {M.}~\bibnamefont {Ceriotti}},\ and\ \bibinfo {author} {\bibfnamefont
  {G.}~\bibnamefont {Fraux}},\ }\bibfield  {title} {\bibinfo {title}
  {Metatensor and metatomic: {{Foundational}} libraries for interoperable
  atomistic machine learning},\ }\href {https://doi.org/10.1063/5.0304911}
  {\bibfield  {journal} {\bibinfo  {journal} {The Journal of Chemical Physics}\
  }\textbf {\bibinfo {volume} {164}},\ \bibinfo {pages} {064113} (\bibinfo
  {year} {2026})}\BibitemShut {NoStop}%
\bibitem [{\citenamefont {Kingma}\ and\ \citenamefont
  {Ba}(2017)}]{kingmaAdamMethodStochastic2017a}%
  \BibitemOpen
  \bibfield  {author} {\bibinfo {author} {\bibfnamefont {D.~P.}\ \bibnamefont
  {Kingma}}\ and\ \bibinfo {author} {\bibfnamefont {J.}~\bibnamefont {Ba}},\
  }\href {https://doi.org/10.48550/arXiv.1412.6980} {\bibinfo {title} {Adam: A
  method for stochastic optimization}},\ \bibinfo {howpublished} {Preprint,
  arXiv:1412.6980 [cs]. \url{https://arxiv.org/abs/1412.6980}} (\bibinfo {year}
  {2017}),\ \bibinfo {note} {prepublished: 2017-01-30; accessed
  2025-04-06}\BibitemShut {NoStop}%
\bibitem [{\citenamefont {Batatia}\ \emph {et~al.}(2026)\citenamefont
  {Batatia}, \citenamefont {Baldwin}, \citenamefont {Kuryla}, \citenamefont
  {Hart}, \citenamefont {Kasoar}, \citenamefont {Elena}, \citenamefont {Moore},
  \citenamefont {Gawkowski}, \citenamefont {Shi}, \citenamefont {Kapil},
  \citenamefont {Kourtis}, \citenamefont {Magd{\u a}u},\ and\ \citenamefont
  {Cs{\'a}nyi}}]{batatiaMACEPOLAR1PolarisableElectrostatic2026}%
  \BibitemOpen
  \bibfield  {author} {\bibinfo {author} {\bibfnamefont {I.}~\bibnamefont
  {Batatia}}, \bibinfo {author} {\bibfnamefont {W.~J.}\ \bibnamefont
  {Baldwin}}, \bibinfo {author} {\bibfnamefont {D.}~\bibnamefont {Kuryla}},
  \bibinfo {author} {\bibfnamefont {J.}~\bibnamefont {Hart}}, \bibinfo {author}
  {\bibfnamefont {E.}~\bibnamefont {Kasoar}}, \bibinfo {author} {\bibfnamefont
  {A.~M.}\ \bibnamefont {Elena}}, \bibinfo {author} {\bibfnamefont
  {H.}~\bibnamefont {Moore}}, \bibinfo {author} {\bibfnamefont {M.~J.}\
  \bibnamefont {Gawkowski}}, \bibinfo {author} {\bibfnamefont {B.~X.}\
  \bibnamefont {Shi}}, \bibinfo {author} {\bibfnamefont {V.}~\bibnamefont
  {Kapil}}, \bibinfo {author} {\bibfnamefont {P.}~\bibnamefont {Kourtis}},
  \bibinfo {author} {\bibfnamefont {I.-B.}\ \bibnamefont {Magd{\u a}u}},\ and\
  \bibinfo {author} {\bibfnamefont {G.}~\bibnamefont {Cs{\'a}nyi}},\ }\href
  {https://doi.org/10.48550/arXiv.2602.19411} {\bibinfo {title}
  {{{MACE-POLAR-1}}: {{A Polarisable Electrostatic Foundation Model}} for
  {{Molecular Chemistry}}}} (\bibinfo {year} {2026}),\ \Eprint
  {https://arxiv.org/abs/2602.19411} {arXiv:2602.19411 [physics.chem-ph]}
  \BibitemShut {NoStop}%
\bibitem [{\citenamefont {Wood}\ \emph {et~al.}(2026)\citenamefont {Wood},
  \citenamefont {Dzamba}, \citenamefont {Fu}, \citenamefont {Gao},
  \citenamefont {Shuaibi}, \citenamefont {{Barroso-Luque}}, \citenamefont
  {Abdelmaqsoud}, \citenamefont {Gharakhanyan}, \citenamefont {Kitchin},
  \citenamefont {Levine}, \citenamefont {Michel}, \citenamefont {Sriram},
  \citenamefont {Cohen}, \citenamefont {Das}, \citenamefont {Rizvi},
  \citenamefont {Sahoo}, \citenamefont {Ulissi},\ and\ \citenamefont
  {Zitnick}}]{woodUMAFamilyUniversal2026a}%
  \BibitemOpen
  \bibfield  {author} {\bibinfo {author} {\bibfnamefont {B.~M.}\ \bibnamefont
  {Wood}}, \bibinfo {author} {\bibfnamefont {M.}~\bibnamefont {Dzamba}},
  \bibinfo {author} {\bibfnamefont {X.}~\bibnamefont {Fu}}, \bibinfo {author}
  {\bibfnamefont {M.}~\bibnamefont {Gao}}, \bibinfo {author} {\bibfnamefont
  {M.}~\bibnamefont {Shuaibi}}, \bibinfo {author} {\bibfnamefont
  {L.}~\bibnamefont {{Barroso-Luque}}}, \bibinfo {author} {\bibfnamefont
  {K.}~\bibnamefont {Abdelmaqsoud}}, \bibinfo {author} {\bibfnamefont
  {V.}~\bibnamefont {Gharakhanyan}}, \bibinfo {author} {\bibfnamefont {J.~R.}\
  \bibnamefont {Kitchin}}, \bibinfo {author} {\bibfnamefont {D.~S.}\
  \bibnamefont {Levine}}, \bibinfo {author} {\bibfnamefont {K.}~\bibnamefont
  {Michel}}, \bibinfo {author} {\bibfnamefont {A.}~\bibnamefont {Sriram}},
  \bibinfo {author} {\bibfnamefont {T.}~\bibnamefont {Cohen}}, \bibinfo
  {author} {\bibfnamefont {A.}~\bibnamefont {Das}}, \bibinfo {author}
  {\bibfnamefont {A.}~\bibnamefont {Rizvi}}, \bibinfo {author} {\bibfnamefont
  {S.~J.}\ \bibnamefont {Sahoo}}, \bibinfo {author} {\bibfnamefont {Z.~W.}\
  \bibnamefont {Ulissi}},\ and\ \bibinfo {author} {\bibfnamefont {C.~L.}\
  \bibnamefont {Zitnick}},\ }\href {https://doi.org/10.48550/arXiv.2506.23971}
  {\bibinfo {title} {{{UMA}}: {{A Family}} of {{Universal Models}} for
  {{Atoms}}}} (\bibinfo {year} {2026}),\ \Eprint
  {https://arxiv.org/abs/2506.23971} {arXiv:2506.23971 [cs.LG]} \BibitemShut
  {NoStop}%
\bibitem [{\citenamefont {Engel}\ \emph {et~al.}(2021)\citenamefont {Engel},
  \citenamefont {Kapil},\ and\ \citenamefont {Ceriotti}}]{enge+21jpcl}%
  \BibitemOpen
  \bibfield  {author} {\bibinfo {author} {\bibfnamefont {E.~A.}\ \bibnamefont
  {Engel}}, \bibinfo {author} {\bibfnamefont {V.}~\bibnamefont {Kapil}},\ and\
  \bibinfo {author} {\bibfnamefont {M.}~\bibnamefont {Ceriotti}},\ }\bibfield
  {title} {\bibinfo {title} {Importance of {{Nuclear Quantum Effects}} for
  {{NMR Crystallography}}},\ }\href
  {https://doi.org/10.1021/acs.jpclett.1c01987} {\bibfield  {journal} {\bibinfo
   {journal} {The Journal of Physical Chemistry Letters}\ }\textbf {\bibinfo
  {volume} {12}},\ \bibinfo {pages} {7701} (\bibinfo {year}
  {2021})}\BibitemShut {NoStop}%
\bibitem [{\citenamefont {Kalakewich}\ \emph {et~al.}(2015)\citenamefont
  {Kalakewich}, \citenamefont {Iuliucci}, \citenamefont {Mueller},
  \citenamefont {Eloranta},\ and\ \citenamefont
  {Harper}}]{kalakewichMonitoringRefinementCrystal2015}%
  \BibitemOpen
  \bibfield  {author} {\bibinfo {author} {\bibfnamefont {K.}~\bibnamefont
  {Kalakewich}}, \bibinfo {author} {\bibfnamefont {R.}~\bibnamefont
  {Iuliucci}}, \bibinfo {author} {\bibfnamefont {K.~T.}\ \bibnamefont
  {Mueller}}, \bibinfo {author} {\bibfnamefont {H.}~\bibnamefont {Eloranta}},\
  and\ \bibinfo {author} {\bibfnamefont {J.~K.}\ \bibnamefont {Harper}},\
  }\bibfield  {title} {\bibinfo {title} {Monitoring the refinement of crystal
  structures with {{15N}} solid-state {{NMR}} shift tensor data},\ }\href
  {https://doi.org/10.1063/1.4935367} {\bibfield  {journal} {\bibinfo
  {journal} {The Journal of Chemical Physics}\ }\textbf {\bibinfo {volume}
  {143}},\ \bibinfo {pages} {194702} (\bibinfo {year} {2015})}\BibitemShut
  {NoStop}%
\bibitem [{\citenamefont {Hartman}\ and\ \citenamefont
  {Beran}(2018)}]{hartmanAccurate13C15N2018}%
  \BibitemOpen
  \bibfield  {author} {\bibinfo {author} {\bibfnamefont {J.~D.}\ \bibnamefont
  {Hartman}}\ and\ \bibinfo {author} {\bibfnamefont {G.~J.~O.}\ \bibnamefont
  {Beran}},\ }\bibfield  {title} {\bibinfo {title} {Accurate 13-{{C}} and
  15-{{N}} molecular crystal chemical shielding tensors from fragment-based
  electronic structure theory},\ }\href
  {https://doi.org/10.1016/j.ssnmr.2018.09.003} {\bibfield  {journal} {\bibinfo
   {journal} {Solid State Nuclear Magnetic Resonance}\ }\textbf {\bibinfo
  {volume} {96}},\ \bibinfo {pages} {10} (\bibinfo {year} {2018})}\BibitemShut
  {NoStop}%
\bibitem [{\citenamefont {Guo}\ \emph {et~al.}(2017)\citenamefont {Guo},
  \citenamefont {Pleiss}, \citenamefont {Sun},\ and\ \citenamefont
  {Weinberger}}]{guoCalibrationModernNeural2017a}%
  \BibitemOpen
  \bibfield  {author} {\bibinfo {author} {\bibfnamefont {C.}~\bibnamefont
  {Guo}}, \bibinfo {author} {\bibfnamefont {G.}~\bibnamefont {Pleiss}},
  \bibinfo {author} {\bibfnamefont {Y.}~\bibnamefont {Sun}},\ and\ \bibinfo
  {author} {\bibfnamefont {K.~Q.}\ \bibnamefont {Weinberger}},\ }\bibfield
  {title} {\bibinfo {title} {On {{Calibration}} of {{Modern Neural
  Networks}}},\ }in\ \href@noop {} {\emph {\bibinfo {booktitle} {Proceedings of
  the 34th {{International Conference}} on {{Machine Learning}}}}}\ (\bibinfo
  {publisher} {PMLR},\ \bibinfo {year} {2017})\ pp.\ \bibinfo {pages}
  {1321--1330}\BibitemShut {NoStop}%
\bibitem [{\citenamefont {Kuleshov}\ \emph {et~al.}(2018)\citenamefont
  {Kuleshov}, \citenamefont {Fenner},\ and\ \citenamefont
  {Ermon}}]{kuleshovAccurateUncertaintiesDeep2018c}%
  \BibitemOpen
  \bibfield  {author} {\bibinfo {author} {\bibfnamefont {V.}~\bibnamefont
  {Kuleshov}}, \bibinfo {author} {\bibfnamefont {N.}~\bibnamefont {Fenner}},\
  and\ \bibinfo {author} {\bibfnamefont {S.}~\bibnamefont {Ermon}},\ }\bibfield
   {title} {\bibinfo {title} {Accurate {{Uncertainties}} for {{Deep Learning
  Using Calibrated Regression}}},\ }in\ \href@noop {} {\emph {\bibinfo
  {booktitle} {Proceedings of the 35th {{International Conference}} on
  {{Machine Learning}}}}}\ (\bibinfo  {publisher} {PMLR},\ \bibinfo {year}
  {2018})\ pp.\ \bibinfo {pages} {2796--2804}\BibitemShut {NoStop}%
\bibitem [{\citenamefont {Musil}\ \emph {et~al.}(2019)\citenamefont {Musil},
  \citenamefont {Willatt}, \citenamefont {Langovoy},\ and\ \citenamefont
  {Ceriotti}}]{musi+19jctc}%
  \BibitemOpen
  \bibfield  {author} {\bibinfo {author} {\bibfnamefont {F.}~\bibnamefont
  {Musil}}, \bibinfo {author} {\bibfnamefont {M.~J.}\ \bibnamefont {Willatt}},
  \bibinfo {author} {\bibfnamefont {M.~A.}\ \bibnamefont {Langovoy}},\ and\
  \bibinfo {author} {\bibfnamefont {M.}~\bibnamefont {Ceriotti}},\ }\bibfield
  {title} {\bibinfo {title} {{Fast and Accurate Uncertainty Estimation in
  Chemical Machine Learning}},\ }\href@noop {} {\bibfield  {journal} {\bibinfo
  {journal} {J. Chem. Theory Comput.}\ }\textbf {\bibinfo {volume} {15}},\
  \bibinfo {pages} {906} (\bibinfo {year} {2019})}\BibitemShut {NoStop}%
\bibitem [{\citenamefont {Imbalzano}\ \emph {et~al.}(2021)\citenamefont
  {Imbalzano}, \citenamefont {Zhuang}, \citenamefont {Kapil}, \citenamefont
  {Rossi}, \citenamefont {Engel}, \citenamefont {Grasselli},\ and\
  \citenamefont {Ceriotti}}]{imbalzanoUncertaintyEstimationMolecular2021}%
  \BibitemOpen
  \bibfield  {author} {\bibinfo {author} {\bibfnamefont {G.}~\bibnamefont
  {Imbalzano}}, \bibinfo {author} {\bibfnamefont {Y.}~\bibnamefont {Zhuang}},
  \bibinfo {author} {\bibfnamefont {V.}~\bibnamefont {Kapil}}, \bibinfo
  {author} {\bibfnamefont {K.}~\bibnamefont {Rossi}}, \bibinfo {author}
  {\bibfnamefont {E.~A.}\ \bibnamefont {Engel}}, \bibinfo {author}
  {\bibfnamefont {F.}~\bibnamefont {Grasselli}},\ and\ \bibinfo {author}
  {\bibfnamefont {M.}~\bibnamefont {Ceriotti}},\ }\bibfield  {title} {\bibinfo
  {title} {Uncertainty estimation for molecular dynamics and sampling},\ }\href
  {https://doi.org/10.1063/5.0036522} {\bibfield  {journal} {\bibinfo
  {journal} {The Journal of Chemical Physics}\ }\textbf {\bibinfo {volume}
  {154}},\ \bibinfo {pages} {074102} (\bibinfo {year} {2021})}\BibitemShut
  {NoStop}%
\end{thebibliography}
\end{document}